\documentclass[doublespace,a4paper,12pt]{thesis}
\usepackage[top=1.2in,bottom=1in,left=1.5in,right=1.1in,includehead]{geometry}

\usepackage{amssymb,amsthm,amscd,geometry,fancyhdr,hyperref,bbold}
\usepackage{amsmath,amsfonts,setspace,indentfirst,epsfig,slashed}
\usepackage[english]{babel}
\def\be{\begin{equation}}
\def\ee{\end{equation}}
\def\ben{\begin{equation*}}
\def\een{\end{equation*}}
\def\ba{\begin{array}}
\def\ea{\end{array}}
\def\bn{\begin{aligned}}
\def\en{\end{aligned}}
\def\bnn{\begin{eqnarray*}}
\def\enn{\end{eqnarray*}}
\def\bsub{\begin{subequations}}
\def\esub{\end{subequations}}

\def\a{{\alpha}}
\def\b{{\beta}}
\def\G{{\Gamma}}
\def\g{{\gamma}}
\def\d{{\delta}}
\def\e{{\epsilon}}
\def\eh{{\hat\epsilon}}
\def\eb{{\bar\epsilon}}

\def\h{{\eta}}
\def\th{{\theta}}

\def\k{{\kappa}}
\def\l{{\lambda}}
\def\L{{\Lambda}}
\def\m{{\mu}}
\def\n{{\nu}}
\def\x{{\xi}}

\def\r{{\rho}}
\def\s{{\sigma}}
\def\S{{\Sigma}}
\def\t{{\tau}}

\def\f{{\phi}}
\def\c{{\chi}}
\def\ps{{\psi}}
\def\w{{\omega}}
\def\W{{\Omega}}
\def\vare{{\varepsilon}}

\def\p{{\partial}}
\def\pb{{\bar{\partial}}}
\def\F{{\mathcal{F}}}
\def\C{{\mathcal{C}}}

\def\Cl{{\mathit{Cl}}}
\def\Spin{{\mathit{Spin}}}
\def\4{AdS_4 \times \mathbb{C}P^3}
\def\5{AdS_5 \times \mathbb{S}^5}
\def\cp3{\mathbb{C}P^3}
\def\zb{{\bar{z}}}
\def\wb{{\bar{w}}}

\title{Fermionic T-duality and U-duality in type II supergravity}
\author{Ilya Bakhmatov}
\department{Physics}
\division{Centre for research in string theory}
\date{October 2011}

\begin{document}
\maketitle

\topmatter{Abstract}
This thesis deals with the two duality symmetries of $\mathcal{N}=2$ $D=10$ supergravity theories that are descendant from the full superstring theory: fermionic T-duality and U-duality.

The fermionic T-duality transformation is applied to the D-brane and pp-wave solutions of type IIB supergravity. New supersymmetric solutions of complexified supergravity are generated. We show that the pp-wave yields a purely imaginary background after two dualities, undergoes a geometric transformation after four dualities, and is self-dual after eight dualities. 

Next we apply six bosonic and six fermionic T-dualities to the $\4$ background of type IIA supergravity, which is relevant to the current research in the amplitude physics. This helps to elucidate the potential obstacles in establishing the self-duality, and quite independently from that shows us that fermionic T-dualities may be degenerate under some circumstances.

Finally, we make a step towards constructing a manifestly U-duality covariant action for $D=10$ supergravities by deriving the generalized metric for a D1-brane. This is a single structure that treats brane wrapping coordinates on the same footing as spacetime coordinates. It turns out that the generalized metric of a D-string results from that of the fundamental string if one replaces the spacetime metric with the open string metric. We also find an antisymmetric contribution to the generalized metric that can be interpreted as a noncommutativity parameter.

\topmatter{Declaration}
The work presented in this thesis is the original research of the author, except where explicitly acknowledged. Much of the research presented here has appeared in the publications \cite{Bakhmatov:2009be, Bakhmatov:2010fp}. This dissertation has not been submitted before, in whole or in part, for a degree at this or any other institution.

\topmatter{Acknowledgements}
Firstly I wish to thank all the people who contributed to making this stay in London easy and fruitful, in particular the administrative staff at Queen Mary and seminar organizers in various London string theory groups.

I am particularly grateful to David Berman, for being a perfect supervisor, providing valuable guidance in course of this work, and giving clear and vivid explanations whenever needed.

I have benefitted a lot from discussions with all the fellow students at Queen Mary string theory group, most notably Will Black, Andrew Low, Moritz\linebreak McGarrie, Edvard Musaev, Jurgis Pa\v{s}ukonis, Gianni Tallarita, Daniel Thomp\nolinebreak son, and David Turton.

Finally, I thank the people at the Physics Department of Kazan University, who taught me physics and with whom I worked, particularly Nail Khusnutdinov. Special thanks are due to Emil Akhmedov for introducing me to the world of theoretical physics and string theory, and to Yevgeniy Patrin for doing the same for mathematics.

\tableofcontents

\mainmatter
\renewcommand{\chaptermark}[1]{\markboth{#1}{}}

\chapter{Introduction}
\label{ch:intro}

\section{String theory dualities}

\begin{figure}[t]
\begin{center}
\includegraphics[scale=0.31]{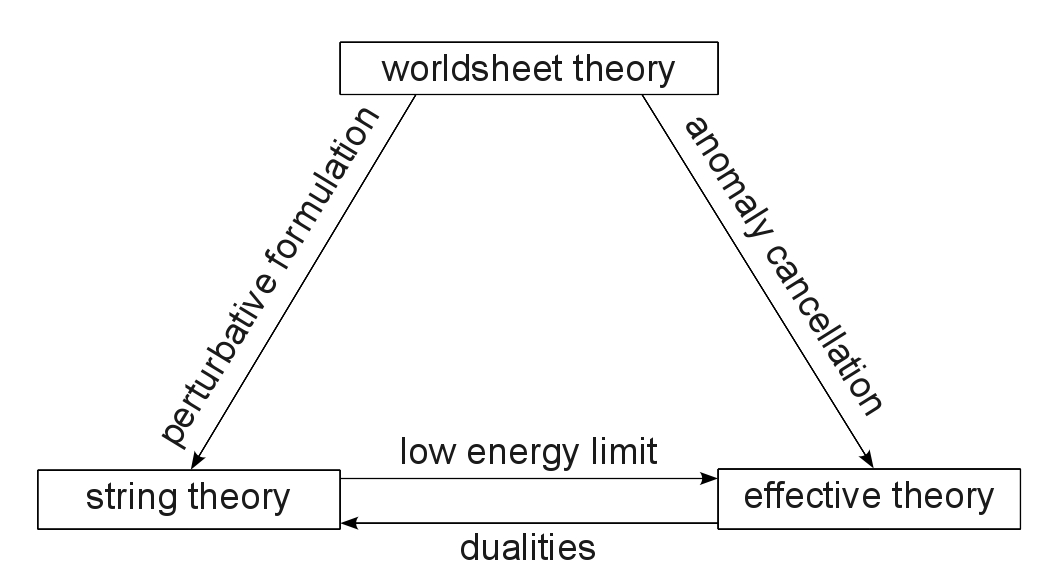}
\caption{Relations between the worldsheet actions, string theory, and effective theory.}
\label{sxema}
\end{center}
\end{figure}

The development of string theory during and after what is commonly referred to as ``second superstring revolution'' is marked by an increasing role played by dualities \cite{Hull:1994ys, Witten:1995ex, Vafa:1997pm} and D-branes \cite{Polchinski:1995mt, Johnson:2003gi}. It is the shift of paradigm from the traditional methodology centred around worldsheet techniques \cite{Green:1987sp, Green:1987mn} to the newer ``spacetime approach'' that has promoted string dualities to the important position they occupy nowadays. ``Spacetime approach'' here stands for the methods and objectives dictated by the greater role played by the effective low energy theories in string theory research. The two approaches are of course interdependent, and indeed the approach based on effective theories has been made possible in the first place by the worldsheet derivation of dynamics of the effective theories. Namely, one obtains the dynamical equations of the effective supergravity theories by imposing consistency constraints on the quantum worldsheet theory. The scheme in figure \ref{sxema} summarizes these relationships and highlights the place occupied by string theory dualities.

On the one hand, the basic worldsheet action of a string gives rise to the perturbative formulation of string theory. Some information regarding the nonperturbative states (D-branes) can be extracted from the worldsheet formalism as well. On the other hand, low energy effective theories also arise from the worldsheet theory, as mentioned above. String theory dualities enter the scene as one turns to the interplay between the two: the common interpretation is that string dualities are reflected by the global symmetries of the corresponding low energy effective theories \cite{Hull:1994ys}, or by the symmetries between distinct low energy theories. Historically this interpretation was established by moving from the right to the left on the above scheme (along the $\leftarrow$ arrow). For example, T-duality \cite{Giveon:1994fu}, which is arguably the oldest string duality known, was originally discovered as a symmetry of the effective potential for the compactification radius in the toroidal compactification with respect to the inversion $R \rightarrow \frac{\a'}{R}$ \cite{Kikkawa:1984cp, Sakai:1985cs}. It was soon realized that this radial inversion $\mathbb{Z}_2$ symmetry is in fact embedded into a larger $\mathrm{O}(d,d;\mathbb{Z})$ group for string theory compactified on a torus $\mathbb{T}^d$ \cite{Narain:1985jj, Narain:1986am, Giveon:1988tt, Shapere:1988zv}.

T-duality holds order by order in string perturbation theory: as we will see, the string coupling is just scaled under this transformation, $g'_s \propto g_s$. One can derive T-duality from the worldsheet approach to string theory (i.e. going in the~$\swarrow$ direction on the scheme \ref{sxema}), and it is the existence of well-developed worldsheet technique for T-duality (so-called Buscher procedure \cite{Buscher:1985kb, Buscher:1987sk, Buscher:1987qj}, to be reviewed later) that has made possible the discovery of fermionic T-duality, which is the main subject of this thesis. 

Alternatively, evading worldsheet approach, one can conjecture that T-duality is in fact the symmetry not only of the effective action, but also of string theory by following the aforementioned logic: to an effective action symmetry may correspond a string theory duality. This logic is represented by the horizontal $\leftarrow$ arrow on Fig.\ref{sxema}. Such logic is possible due to the perturbative nature of T-duality: one can observe it in the perturbative string spectrum. This approach is hindered in the case of nonperturbative dualities, that generally go by the name S-dualities. These are characterized by the fact that they relate weakly and strongly coupled regimes, $g'_s \propto g_s^{-1}$. The prototypical example of S-duality, already displaying the characteristic $\mathrm{SL}(2,\mathbb{R})$ symmetry group, was found in the effective theory of the heterotic string compactified to four dimensions \cite{Rey:1989xj, Font:1990gx, Sen:1992fr} (which is just $d=4$ $\mathcal{N}=4$ supergravity). The same symmetry group, and the same inversion of the string coupling also appear in the effective theory of type IIB superstring, $d=10$ $\mathcal{N}=2$ supergravity (type IIB) \cite{Schwarz:1983qr, Bergshoeff:1995as}. The $\mathrm{SL}(2,\mathbb{R})$ S-duality group in this case has an M-theoretic explanation as a modular group of the torus in a $\mathbb{T}^2$ compactification of M-theory (this is due to the fact that this compactification is dual to a circle compactification of IIB superstring).

M-theory arguments play a crucial role in the unification of T- and S-dualities. For type II theories, T- and S-dualities are in fact subgroups of bigger nonperturbative duality groups, called U-dualities \cite{Hull:1994ys}. These also contain transformations, that are neither T- nor S-dualities. Studying the global supergravity symmetries that correspond to nonperturbative string dualities in general can give us some hints to the nonperturbative spectrum of string theory (this again corresponds to proceeding along the $\leftarrow$ in the figure \ref{sxema}). Worldsheet approach {\it to string theory} is of little use in this case, as it is essentially perturbative. However, it may be possible to find the {\it worldvolume} motivation for U-dualities in M-theory framework. Some low-dimensional U-duality groups have been reproduced starting from the M2-brane worldvolume theory \cite{Berman:2010is}. Developments related to this are reviewed in the last chapter of the thesis.

Fermionic T-duality, to which most of the thesis is devoted, is a new nonperturbative symmetry, which so far has been formulated for type II superstring theories. It has been discovered by extending the worldsheet techniques of standard T-duality to the superspace setup. Fermionic T-duality is only valid at tree level in string perturbation theory, and is in this sense nonperturbative (although the behaviour of the string coupling is qualitatively the same as in the case of perturbative bosonic T-duality, $g'_s \propto g_s$). Most of this thesis will be devoted to following the $\rightarrow$ arrow on the figure \ref{sxema}, in order to study the implications of fermionic T-duality for type II supergravities. From the supergravity point of view the transformation looks rather strange and has many unexpected consequences.

\subsection{Structure of thesis}

In the remaining sections of this introductory chapter we review the way in which effective field theories emerge from string theory, briefly describe the bosonic field content and the actions of $d=10$ supergravities, and overview the derivation of traditional, bosonic T-duality. This is done mainly in the worldsheet approach (so called Buscher's procedure), but a general overview of the perturbative spectrum symmetry is also given.

Chapter \ref{ch:techintro} provides a thorough introduction into the derivation, properties and some of the applications of fermionic T-duality. We derive the transformations of the supergravity background fields by means of the fermionic generalization of the Buscher's procedure. As this is accomplished in the pure spinor formalism for the worldsheet superstring action, a brief review of the formalism is included.

In chapter \ref{ch:paper1} we apply the fermionic T-duality transformation to the D1-brane and pp-wave backgrounds of type IIB supergravity and discuss various properties of the transformed solutions. The pp-wave is shown to be self-dual under a certain combination of dualities, and some other combination is shown to be equivalent to a geometric transformation of the pp-wave background.

Following this, in chapter \ref{ch:paper2} we consider the action of combined bosonic and fermionic T-dualities on the $\4$ background of type IIA supergravity. This is done in pursuit of a so far unsolved problem of current interest in the field of gauge theory scattering amplitudes. The set of T-dualities was expected \cite{Bargheer:2010hn} to produce self-duality of the background based on AdS/CFT considerations, but is shown to fail due to degeneracy of the transformation. We discuss the possible ways out and comment on some of the explanations found in the literature.

Finally, we turn to the issues of worldvolume derivation of some U-duality aspects. Using the generalized geometry approach, in chapter \ref{ch:branes} we derive the generalized metric description of D1-brane in type IIB superstring theory. This is proposed as a building block for the ultimate goal of reformulation of type II supergravity in a U-duality covariant manner.

We conclude (chapter \ref{ch:conc}) with an overview of achievements presented in the main body of the thesis and discuss some possible ways to extend this work. There are several appendices (\ref{app:2b}, \ref{gamma}, \ref{app-spinors}, \ref{actions}) with technical details and conventions relevant to various parts of the thesis.

\section{Supergravity}

Since in this thesis we will be considering fermionic T-duality within the supergravity approximation to string theory, let us look at how does supergravity arise from sting theory. Dynamics of a bosonic string is encoded in the Polyakov action
\be\label{polyakov}
S = \frac{1}{4\pi\a'}\int d^2 \x \sqrt{-h} h^{\a\b} \p_\a x^m \p_\b x^n g_{mn}(x), 
\ee
where the functions $x^m(\x)$ describe the spacetime embedding of the string worldsheet, which is parameterized by the two coordinates $\x^\a = (\t,\s)$, and $h^{\a\b}$ is an auxilliary worldsheet metric. This theory is nonlinear (commonly referred to as a nonlinar sigma-model for historical reasons) because of the spacetime dependence of the background metric $g_{mn}(x)$. The action has the only dimensionful parameter $\a'$, which is a square of the fundamental string scale. Using the symmetries of the action one can fix the conformal gauge $\sqrt{-h} h^{\a\b} = \d^{\a\b}$. In order to study theory (\ref{polyakov}) perturbatively, consider quantum fluctuations around a classical solution $x_0(\x)$ \cite{Polchinski:1998rq}, so that
\be
x^m = x_0^m + \sqrt{\a'} y^m,
\ee
where $y^m \ll 1$ are dimensionless fields. Expanding the integrand in a series around $x_0$
\be\label{expan}\bn
g_{mn}(x) \p_\a x^m \p_\b x^n = \a' & \left[ g_{mn}(x_0) + \sqrt{\a'} g_{mn,r}(x_0) y^r + \right. \\ + & \left. \a'\frac12 g_{mn,rs}(x_0) y^r y^s + \ldots \right] \p_\a y^m \p_\b y^n 
\en\ee
we see explicitly the infinite series of coulping constants for the vertices with ever-increasing number of fields $y^m(\x)$ in each. If we introduce a characteristic curvature radius $R_c$ that controls the spacetime variation of the metric according to
\be
\frac{\p g}{\p x} \propto \frac{1}{R_c},
\ee
then it is obvious that the effective dimensionless coupling that controls the expansion (\ref{expan}) is 
\be
\frac{\sqrt{\a'}}{R_c}.
\ee
One can study the theory (\ref{polyakov}) perturbatively for big curvature radii, $R_c \gg \sqrt{\a'}$. In this very limit it is also appropriate to restrict the choice of possible sigma-model couplings in (\ref{polyakov}) to massless fields only (since for big enough wavelengths massive states are not excited), and to neglect the finite string size, studying a low energy effective {\it field} theory. This field theory is the theory of supergravity.

The supergravity action can be obtained from string theory by requiring that conformal symmetry is kept at a quantum level \cite{Friedan:1980jm, Callan:1985ia}. To this end, one considers the beta-functions of the quantum string theory and imposes that they vanish, so that no renormalization scale is introduced. This leads to the constraints for the sigma-model couplings, which can also be thought of as the field equations for the background fields (the spacetime metric $g_{mn}$ in the above example). One then reconstructs a spacetime action for the background fields, which would lead to these equations. If the beta-functions have been computed at one-loop order in the $\a'$-perturbation theory, then one talks of supergravity effective action; higher order corrections to the beta-functions correspond to the stringy corrections to supergravity (which are of course higher order in $\a'$).

The one-loop order beta-functions for the theory (\ref{polyakov}) with the only background field $g_{mn}$ are given by
\be\label{renorm}
\b^g_{mn} = \a' R_{mn} + O(\a'^2),
\ee
which obviously gives us the vacuum Einstein equations. If one includes the other two massless bosonic string fields, the antisymmetric gauge field potential $b_{mn}$ and the dilaton $\f$, the beta-functions give the field equations of the Einstein theory coupled to the dilaton and $b_{mn}$ as matter fields \cite{Callan:1985ia, Polchinski:1998rq}:
\be\label{nsns}
S = \frac{1}{2\k^2} \int d^d x \sqrt{|g|} \,e^{-2\f} \left[R+4(\p\f)^2-\frac12\frac{1}{3!}{\tilde H}^2\right],
\ee
where $d=26$ in the case of bosonic string, and $d=10$ for the superstring. 3-form $\tilde H$ is just the field strength of the Neveu-Schwarz potential $H = db$ with some modifications in the heterotic superstring case. The action (\ref{nsns}) represents the dynamics of a common supergravity sector; it arises in the low energy limit of any type of superstring theory, which all have a common Neveu-Schwarz--Neveu-Schwarz (NSNS) sector: $g_{mn}, b_{mn}$, and $\f$. Field content of this sector is the same as that of the bosonic string. If we consider the beta-functions for all the background fields of a superstring, then of course there will be more equations of motion and corresponding extra terms in the effective action. The extra fields are either gauge potentials from the Ramond-Ramond (RR) sector in type II theories, or nonabelian gauge fields for heterotic string theories. Coupling of the latter to $H$ comprises the difference between $H$ and $\tilde H$ (see below for the details).

We will now give a brief overview of different $d=10$ supergravity actions \cite{Polchinski:1998rr, Ortin:2004ms}. The actions naturally decompose into a sum of the action for the common sector (\ref{nsns}), the action for extra theory-specific bosonic fields, and finally the action for massless fermions. Omitting the fermionic parts of the actions, we will only look at the theory-specific bosonic contributions. For the practical applications in this thesis we will need the actions and field equations of type II theories, so let us begin with these.\\
\begin{itemize}
\item {\bf Type IIA}. $\tilde H$ in the NSNS part of the action (\ref{nsns}) is just $\tilde H = H = db$ in this case. The RR field potentials are a 1-form $C_m$ and a 3-form $C_{mnr}$; we denote their 2- and 4-form field strengths as $F_2$ and $F_4$. Their action that one needs to add to (\ref{nsns}) is
\be
-\frac{1}{4\k^2}\int d^{10}x \sqrt{|g|} \left[\frac{1}{2} {F_{(2)}}^2 + \frac{1}{4!} \tilde F_{(4)}\!^2 \right] - \frac{1}{4\k^2}\int b_{(2)} \wedge F_{(4)} \wedge F_{(4)}.
\ee
The modified RR 4-form field strength is $\tilde F_{(4)} = dC_{(3)} - C_{(1)} \wedge H_{(3)}$.
\item {\bf Type IIB}. Again $\tilde H = H$ with no modification. The RR field content is $C_{(0)}, C_{(2)}$, and $C_{(4)}$, such that the modified field strength of the latter is self dual, $\tilde F_{(5)} = \star \tilde F_{(5)}$. The extra terms in the action are given by
\be
-\frac{1}{4\k^2}\int d^{10}x \sqrt{|g|} \left[{F_{(1)}}^2 + \frac{1}{3!}\tilde F_{(3)}\!^2 +  \frac12\frac{1}{5!} \tilde F_{(5)}\!^2 \right] - \frac{1}{4\k^2}\int C_{(4)}\wedge H_{(3)}\wedge F_{(4)},
\ee
with the modified RR field strengths being 
\be\bn
\tilde F_{(3)} &= dC_{(2)} - C_{(0)} \wedge H_{(3)}, \\
\tilde F_{(5)} &= dC_{(4)} - \frac12 C_{(2)} \wedge H_{(3)} + \frac12 b_{(2)} \wedge F_{(3)}.
\en\ee
Note that the self-duality of $\tilde F_{(5)}$ field strength does not follow from the action and needs to be imposed independently.

In the appendix \ref{app:2b} one can find a more detailed account of type IIB supergravity conventions together with the field equations, as this will be used later on in the chapter \ref{ch:paper1}.
\item {\bf Heterotic}. In the two heterotic theories there are no RR fields, and the only massless bosonic field not from the common sector is the gauge field strength $F_{(2)}$, taking values in the Lie algebra of either $\mathrm{SO}(32)$ or $\mathrm{E}_8 \times \mathrm{E}_8$ (which are the gauge groups of the two heterotic theories). One should supplement the action of the common sector (\ref{nsns}) with the standard Yang-Mills action for $F_{(2)}$,
\be
\frac{\a'}{8\k^2}\int d^{10}x \sqrt{|g|}\, \mathrm{Tr}\, {F_{(2)}}^2.
\ee
It is the heterotic supergravity case where the NSNS 2-form field strength gets modified in the common sector action (\ref{nsns}): $\tilde H = db - \frac{\a'}{4} \w$, where the 3-form $\w$ is the Chern-Simons form for the gauge potential $A$ of $F_{(2)}$, 
\be
\w = \mathrm{Tr} \left( A \wedge dA + \frac{2}{3} A \wedge A \wedge A \right).
\ee
\end{itemize}

\section{Bosonic T-duality}

T-duality is one of the remarkable features of string theory \cite{Giveon:1994fu}. It is a map between different string backgrounds that leaves the partition function of the string sigma model invariant. From the point of view of the worldsheet theory one may interpret it as an abelian two-dimensional S-duality: as will be shown shortly, the most characteristic T-duality transformation consists of inverting the spacetime metric component, which acts as a coupling in the worldsheet theory:
\be
g'_{11} \propto g_{11}^{-1}.
\ee
From the spacetime viewpoint T-duality is somewhat mysterious since it provides an equivalence between completely different geometries. A key application of T-duality is to use this symmetry as a solution generating mechanism in supergravity \cite{Bergshoeff:1995as} where one begins with a particular solution and then through application of the T-duality rules produces a new set of solutions. This technique has proved particularly useful in constructing solutions deformed by NS flux such as for the gravity duals of noncommutative theories \cite{Maldacena:1999mh, Alishahiha:1999ci, Cai:1999aw}, beta-deformed Yang-Mills \cite{Lunin:2005jy} and so-called dipole deformed theories \cite{Gursoy:2005cn} (similar techniques have also been used for deformation of M-theory geometries \cite{Berman:2007tf}).

\subsection{Radial inversion symmetry}

Let us now give a more detailed account of the traditional bosonic T-duality, which will serve as a preparation to the overview of the fermionic version in the chapter \ref{ch:techintro}. There exist several alternative ways leading to the duality laws. Perhaps the most straightforward and simple one is to consider a closed bosonic string in $\mathbb{S}^1 \times \mathbb{R}^{1,24}$ spacetime (i.e. a compactification on a circle of radius $R$) and find its energy spectrum. Simple calculation \cite{Lust:1989tj, Tong:2009np} shows that the masses of the quantum states take the values
\be\label{mass}
M^2 = \frac{m^2}{R^2} + \frac{n^2 R^2}{\a'^2} + \frac{2}{\a'} \left( N + \tilde N - 2 \right),
\ee
where $N$ and $\tilde N$ are the number operators for left and right-moving oscillation modes of the string. The possibility that the string centre of mass may have momentum in the compactified direction leads to the appearance of the Kaluza-Klein contribution to mass squared, which is governed by the Kaluza-Klein momentum quantum number $m$. This effect is common to the standard Kaluza-Klein theory of a relativistic particle, as opposed to the possibility of winding on the compactification circle, which is only possible in the case of a string. The potential energy of a wound string also contributes to the total energy, and this contribution is controlled by the winding mode $n$.

One can immediately notice that the mass squared (\ref{mass}) is invariant under the $\mathbb{Z}_2$ transformation
\be
m\leftrightarrow n,\quad R\leftrightarrow\frac{\a'}{R},
\ee
which is the famous compactification radius inversion symmetry. This transformation has a clear physical meaning: in the decompactification limit $R\rightarrow\infty$ the spectrum of Kaluza-Klein states becomes continuous, while the winding modes get infinitely heavy and cannot be excited. In the opposite limit $R\rightarrow 0$ the situation is reversed, with the momentum modes becoming heavy and winding modes tending to a continuum. The two strings compactified on circles of T-dual radii $R$ and $\frac{\a'}{R}$ thus have identical spectra, with the roles of winding and Kaluza-Klein momentum reversed. Note that there exists the self-dual compactification radius $R_0 = \sqrt{\a'}$, which coincides with the string scale. This motivates the intuitively natural idea that strings may only be useful in probing distances bigger than the string scale.

Furthermore, one can consider the full string theory partition function, which includes contributions from worldsheets of all genera. In this way it can be shown that the spectra of T-dual theories coincide at any order of the string perturbation theory \cite{Giveon:1994fu}. We will not review this derivation here since it is irrelevant to fermionic T-duality: as will be shown in the chapter \ref{ch:techintro}, the latter is only a symmetry of tree-level string theory. 

Finally, T-duality may be viewed as a canonical transformation in phase space. A simple change of variables in the Hamiltonian formalism for the string sigma-model leaves the Hamiltonian invariant, if one transforms the background fields according to the T-duality rules. This approach, first proposed in \cite{Alvarez:1994wj, Curtright:1994be}, has been recently extended to include fermionic T-duality \cite{Sfetsos:2010xa, Thompson:2010sr}.

We will now concentrate on the most traditional approach to T-duality, formulated in \cite{Buscher:1985kb, Buscher:1987sk, Buscher:1987qj}, which we will use later to derive fermionic T-duality.

\subsection{Buscher's procedure}

Although classical in essence, Buscher's approach to T-duality can be easily incorporated into the path integral treatment of the quantum string. As a starting point we take the Polyakov action of a bosonic string in conformal gauge \cite{Polyakov:1981rd}:
\be\label{sigma}
S = \int d^2 z \left[g_{mn}(x) + b_{mn}(x)\right] \p x^m \pb x^n.
\ee
This is written in terms of complex worldsheet coordinate $z = \frac{1}{\sqrt 2} (\t+i\s)$. The spacetime metric tensor $g_{mn}$ and its antisymmetric counterpart $b_{mn}$ play the role of the sigma-model coupling constants.

Assume that the background is invariant under shifts generated by a spacetime vector field $k^m(x)$. This means, that $k^m$ is a Killing vector
\be
\nabla_m k_n + \nabla_n k_m = 0,
\ee
and that Lie derivatives of any other background fields (such as the field strength of $b_{mn}$) with respect to $k^m$ vanish. After choosing coordinates $\{x^1,x^i\}$, $i>1$ in such a way that the symmetry acts by shifting $x^1$ the action may be rewritten as
\be\label{dej'}
S' = \int d^2 z \left[ g_{11} A \bar A + l_{1i} A\, \pb x^i + l_{i1} \p x^i \bar A + l_{ij} \p x^i \pb x^j + \tilde x^1 (\p \bar A - \pb A) \right], 
\ee
where $l_{mn} = g_{mn} + b_{mn}$, and the background fields are independent of $x^1$. We have also made a replacement
\be\label{zamena}
(\p x^1, \pb x^1) \rightarrow (A, \bar A)
\ee
where $(A,\bar A)$ is an auxilliary worldsheet vector field. This replacement may be interpreted \cite{Giveon:1994fu} as gauging the shift symmetry of the original sigma-model by a minimal coupling to the gauge field $A$:
\be
\p x^1 \rightarrow D x^1 = \p x^1 + A.
\ee

The last term in (\ref{dej'}) imposes the constraint $F=dA = 0$ via the field equation of the Lagrange multiplier $\tilde x^1$. This constraint can be solved (on a topologically trivial worldsheet) by setting $A$ to a differential of a scalar. This has the effect of reversing the arrow in (\ref{zamena}), and one recovers the initial sigma-model (\ref{sigma}). On the other hand, eliminating the gauge field via its field equations
\be\label{dej''}\bn
     A &=  g_{11}^{-1} \left( \p\tilde x^1 - l_{i1} \p x^i \right),\\
\bar A &= -g_{11}^{-1} \left( \pb\tilde x^1 + l_{1i} \pb x^i \right),
\en\ee
one obtains the dual theory whose action
\be
S'' = \int d^2 z \left[\tilde g_{mn}(x) + \tilde b_{mn}(x)\right] \p y^m \pb y^n
\ee
is written in terms of the coordinates $\{y^m\} = \{\tilde x^1, x^i\}$. The Lagrange multiplier from (\ref{dej'}) acts as a dual coordinate, and the dual theory is again isometric in the $\tilde x^1$ direction. The dual background fields are related to the original ones by:
\be\label{T}\bn
&\tilde g_{11} = (g_{11})^{-1}, \quad \tilde g_{1i} = (g_{11})^{-1} b_{1i}, \quad \tilde b_{1i} = -(g_{11})^{-1} g_{1i},\\
&\tilde g_{ij} = g_{ij} - (g_{11})^{-1} (g_{i1} g_{1j} + b_{i1} b_{1j}), \quad \tilde b_{ij} = b_{ij} - (g_{11})^{-1} (g_{i1} b_{1j} + b_{i1} g_{1j}).
\en\ee

This procedure may also be carried out in a covariant manner, without going to the adapted coordinate system. The expressions for the dual background fields are then written in terms of the Killing vector field $k^m$ \cite{Alvarez:1993qi}. Furthermore, at a quantum level the above manipulations are carried out in the same manner. One eliminates the gauge field $A$ by completing the square with respect to $A$ in the path integral
\be
\int \mathcal{D}A \,\mathcal{D}\bar A \,\mathcal{D} x^i \,\mathcal{D} \tilde x^1 e^{-S'\left[ \tilde x, x, A \right]}
\ee
and performing Gaussian integral. The result is of course the same (\ref{T}), but integration over the vector field brings in a Jacobian factor in the path integral, which is interpreted as a rescaling of the string coupling, i.e. the shift of the dilaton \cite{Buscher:1987qj, Schwarz:1992te, DeJaegher:1998pp}:
\be\label{fi}
\f' = \f - \frac12 \log g_{11}
\ee
(assuming that the dilaton coupling has been included in the original action by means of the Fradkin-Tseytlin term \cite{Fradkin:1984pq}). This transformation of the dilaton agrees with the result of the partition function approach to the duality transformation \cite{Giveon:1994fu}. It should be noted, however, that subtleties arise in the path integral treatment if the string background is not conformally invariant: one would need to define the path integral carefully to take care of the renormalization of the metric and other sigma-model couplings according to (\ref{renorm}), to include the higher order $\a'$ corrections.

There are subtleties in proving that the Buscher's procedure holds on the worldsheets of higher genera \cite{Rocek:1991ps, Giveon:1994fu}. We will review this later in the context of fermionic T-duality transformation, where it prevents one from extending the duality beyond tree level in string perturbation theory.

In this overview we have completely omitted the aspects of T-duality that are specific to the superstring theory (as opposed to the bosonic string). Most importantly, this refers to the transformation laws of RR and fermionic fields. Treatment of the superstring case reveals that the chirality of one of the supersymmetry generators is reversed by T-duality, which thus maps type IIA and IIB string theories to one another, with the corresponding interchange between the D-branes of the two theories. Several alternative derivations of the T-duality transformation of RR fields have appeared \cite{Bergshoeff:1995as, Hassan:1999bv, Cvetic:1999zs, Kulik:2000nr, Bandos:2003bz, Benichou:2008it}.

\chapter{Introduction to fermionic T-duality}
\label{ch:techintro}

\section{Overview}

Bosonic T-duality is crucial in establishing the connection between the different branes of type II string theory and has been a central pillar in string duality for many years. It relies on using an isometry of the background to generate the T-duality transformation.

Fermionic T-duality is a tree-level symmetry of type II string theory that can be viewed as extending this idea to the superspace setup. If one has a Green-Schwarz-type sigma-model that describes the embedding of a string worldsheet in type II superspace, then a fermionic analog of the classic Buscher procedure can be carried out, resulting in the redefinition of the sigma-model couplings. The necessary condition for the duality is that the background preserves a supersymmetry, parameterized by some Killing spinors $(\e,\eh)$ (we are considering an $\mathcal{N}=2$ theory, hence a couple of supersymmetry parameters) that generate an Abelian subgroup of the symmetry supergroup. 

Initially the fermionic T-duality transformation was introduced as an ingredient of string theory interpretation of the amplitude/Wilson loop correspondence, which is a symmetry of the scattering amplitudes in $\mathcal{N}=4$ supersymmetric Yang-Mills theory. From string theory point of view, this correspondence (together with closely related dual superconformal invariance of $\mathcal{N}=4$ SYM) manifests itself as self-duality of the $\5$ background under a certain set of T-duality transformations that map a string configuration corresponding to an amplitude to a configuration corresponding to a Wilson loop \cite{Alday:2007hr}. It was required to supplement the bosonic T-dualities employed in \cite{Alday:2007hr} with fermionic ones in order to achieve the exact self-duality \cite{Berkovits:2008ic, Beisert:2008iq}. 

Let us note various aspects of this transformation. Firstly, it is not a full symmetry of string theory like bosonic T-duality since it is broken at one loop in $g_s$. This is because of the presence of fermionic zero modes in the path integral over topologically nontrivial worldsheets, which make the path integral vanish. It is interesting to consider if one could extend the duality beyond tree level by soaking up these zero modes and making sense of such a path integral including the fermionic insertion. Some of the quantum aspects of fermionic T-duality have been considered recently in \cite{Grassi:2011zf}.

The background field transformation laws that results from the fermionic Buscher procedure are quite different from the ordinary T-duality transformation. In fact the entire NSNS sector is not modified, except for the dilaton that gets an additive contribution
\be
\f' = \f + \frac12 \log C,
\ee
where $C$ is determined by the Killing spinors $(\e,\eh)$ that parameterize the fermionic isometries. This transformation law is very similar to the way dilaton changes under ordinary T-duality, but the sign of the logarithm term is opposite. This difference turns out to be crucial in establishing self-duality of the $\5$ background, which was the original motivation for developing the formalism of fermionic T-duality. As for the bosonic fields of the RR sector, their transformation can be written concisely in terms of the bispinor $F^{\a\b}$:
\be
e^{\f'}F'^{\a\hat\b} = e^{\f}F^{\a\hat\b} + 16 i \, \e^\a \eh^{\hat\b} C^{-1}.
\ee
The bispinor $F^{\a\hat\b}$ is formed by contracting all the RR forms of the theory with appropriate antisymmetrized products of gamma-matrices. We will show how these formulae can be derived later in this chapter.

An important feature of the fermionic T-duality transformation is that it can only be done with complexified Killing spinors, which means that the resulting target space background will generically be a solution to complexified supergravity, as we will demonstrate explicitly in chapter \ref{ch:paper1}. Paper \cite{Godazgar:2010ph} deals with the extension of fermionic T-duality to a larger class of fermionic symmetries in supergravity, which also include some real transformations.

A crucial ingredient in the proper theoretical understanding of fermionic T-duality would be to formulate it as a group symmetry \cite{Fre:2009ki}, in analogy with the $\mathrm{O}(d,d)$ group representation of the ordinary T-duality.

Some applications of fermionic T-duality to spacetime noncommutativity have appeared recently \cite{ChangYoung:2011rs, Nikolic:2011ps}.

\section{Fermionic Buscher's procedure}
In this section we will review in detail the fermionic T-duality transformation procedure formulated in \cite{Berkovits:2008ic}. To begin with, one needs a spacetime supersymmetric sigma-model describing a string propagating in superspace with coordinates $(x^m,\th^\a)$, where $x^m$ are bosonic and $\th^\a$ fermionic coordinates. We assume that the worldsheet action is invariant under the shifts of a particular fermionic coordinate $\th^1$ by a constant fermionic parameter $\r$:
\be\label{sdvig}
\th^1 \rightarrow \th^1 + \r,\quad x^m \rightarrow x^m,\quad \th^\a \rightarrow \th^\a \quad (\a\neq 1).
\ee
Such invariance implies that $\th^1$ enters the action only in the form of derivatives. Without specifying a particular form of the action (such as Green-Schwarz or pure spinor action), we can represent it in the following prototypical form:
\be\label{act}\bn
S = \int d^2 z \left[ B_{11}(Z)\p\th^1\pb\th^1 \right. + L_{1M} \p\th^1\pb Z^M & +  L_{M1}(Z) \p Z^M \pb\th^1  \\ &\left. + L_{MN}(Z) \p Z^M \pb Z^N \right],
\en\ee
where $Z^M = (x^\m,\th^\a),\; \a\neq 1$, and the sigma-model couplings form a superfield $L_{MN}(Z) = G_{MN}(Z)+B_{MN}(Z)$. The summands are a graded symmetric and a graded antisymmetric tensors, respectively:
\be
G_{MN} = (-)^{MN} G_{NM},\quad B_{MN} = -(-)^{MN} B_{NM},
\ee
\begin{equation*}
(-)^{MN} = \left\{\begin{array}{ll} -1 & M,N\; \mathrm{fermionic}, \\ +1 & \mathrm{otherwise}, \end{array}\right.
\end{equation*}
and they contain all the background fields as their components.

In the superspace formulation the shift (\ref{sdvig}) is seen as being generated by the supercharges $Q^\a$ for a particular choice of the fermionic displacement (which is essentially the supersymmetry parameter) $\e^\a = \r \d^{\a 1}$:
\be\label{ieq}
\exp\, i\eb Q:\quad (x^m, \th^\a) \rightarrow (x^m - \bar\th\g^m\e, \th^\a+\e^\a).
\ee
This very transformation, when acting upon the superfield $L_{MN}$, is known to generate the supersymmetry transformations of the component fields with a supersymmetry parameter (Killing spinor) $\e^\a$ \cite{Wess:1992cp, FigueroaO'Farrill:2001tr}. We can therefore think of a supergravity background, given by some solution of the field equations for the component fields of $L_{MN}$, such as those presented in the appendix \ref{app:2b}, with the corresponding Killing spinors $\e^\a$. Such a setup would be a starting point for the fermionic analog of the Buscher procedure, and for the corresponding fermionic T-duality transformation. 

Note that in the above we have assumed that the supersymmetry acts by simple shifts in a certain fermionic direction in superspace (\ref{sdvig}), which means that the Killing spinor is constant. However, this is not the case for most nontrivial supergravity backgrounds. In order to fully justify the above derivation one needs to provide a proof that such `adapted' superspace coordinates can be chosen for a more complicated Killing spinor as well. Alternatively, a fermionic Buscher procedure needs to be carried out with a generic Killing spinor, in the same way as an arbitrary Killing vector has been used to derive the bosonic T-duality transformation in \cite{Alvarez:1993qi}.

For a nonzero $B_{11}$ in (\ref{act}) we can use the Buscher procedure to T-dualize the fermionic direction $\th^1$ in a manner identical to the case of ordinary T-duality (as demonstrated in the previous chapter \ref{ch:intro}). We introduce two extra worldvolume fields: a vector field $(A,\bar A)$ and a scalar $\tilde\th^1$. The latter acts as a Lagrange multiplier enforcing that the field strenght of $A$ vanishes:
\be\label{act'}\bn
S' = \int d^2 z &\left[ B_{11}(Z) A\bar A \right. + L_{1M} A\, \pb Z^M +  L_{M1}(Z) \p Z^M \bar A  \\ &\left. + L_{MN}(Z) \p Z^M \pb Z^N + \tilde\th^1 (\p\bar A - \pb A)\right],
\en\ee
where we have also replaced the derivatives of $\th^1$ with the vector field, as in the bosonic case. This clearly requires that $A$ is fermionic.

Integrating out the Lagrange multiplier we establish the equivalence of $S'$ and $S$, because on a topologically trivial Riemann surface $dA = 0$ implies that $A = d\th^1$ (by topologically trivial we mean that no non-trivial non-contractible cycles exist on a surface, so that this refers to tree level in string perturbation theory). Treating worldsheets of higher genera is more subtle, as in the bosonic case, but the resolution here is problematic, leading to the fermionic T-duality being ill-defined beyond tree level in string coupling. This will be discussed separately below.

We can instead integrate out the fermionic vector field $A$, which will produce the same sigma-model as in (\ref{act}), but with the dual fermionic coordinate $\th^1 \rightarrow \tilde\th^1$:
\be\label{act''}\bn
S'' = \int d^2 z &\left[ B'_{11}(Z)\p\tilde\th^1\pb\tilde\th^1 \right. + L'_{1M} \p\tilde\th^1\pb Z^M +  L'_{M1}(Z) \p Z^M \pb\tilde\th^1  \\ &\left. + L'_{MN}(Z) \p Z^M \pb Z^N \right]
\en\ee
and with fermionic T-dual couplings:
\be\label{dual1}
\bn
&B'_{11} = -(B_{11})^{-1},\quad L'_{1M} = (B_{11})^{-1} L_{1M},\quad L'_{M1} = (B_{11})^{-1} L_{M1},\\
&L'_{MN} = L_{MN} - (B_{11})^{-1} L_{1N} L_{1M}.
\en\ee
These formulae look much like the ordinary T-duality transformation (\ref{T}), but they are now written for the superfields rather than just for the metric and the $b$-field. The transformation will thus look quite different when rewritten in terms of the component fields. This of course depends crucially on a particular sigma-model action one works with, and in a further subsection we will do this for the pure spinor superstring.

Several important distinctions from the bosonic T-duality case arise due to the fermionic nature of the auxilliary vector field and the coordinate being dualized. Firstly, the transformation of the dilaton now emerges with an opposite sign:
\be\label{dual1fi}
\f' = \f + \frac12 \log (B_{11})|_{\th=0}
\ee
(compare to (\ref{fi})). This is a crucial point that makes the self-duality of the $\5$ background possible, which was the original motivation to introduce the fermionic T-duality transformation. One has the dilaton shifts coming from a series of bosonic T-dualities cancelling precisely with those coming from fermionic T-dualities \cite{Berkovits:2008ic, Beisert:2008iq}.

Secondly, there is an important sign difference in the equations of motion for $A$ that follow from (\ref{act'}):
\be\bn
\p\tilde\th^1 &= B_{11} A + L_{M1} \p Z^M,\\
\pb\tilde\th^1 &= B_{11} \bar A - (-1)^{s(M)} L_{1M} \pb Z^M,
\en\ee
where the sign exponent $s(M)$ is zero when $M$ is a bosonic index and one if it is fermionic. If one makes the substitutions $A = \p\th^1$, $\bar A = \pb \th^1$ in these equations, then it is easy to see that there is no relative minus sign between $\p\tilde\th^1 / \p\th^1$ and $\pb\tilde\th^1 / \pb\th^1$, as opposed to the case of bosonic T-duality, where one has 
\be\label{2.10}\bn
\p\tilde x^1 &= g_{11} \p x^1 + l_{i1} \p x^i,\\
\pb\tilde x^1 &= -g_{11} \pb x^1 - l_{1i} \pb x^i
\en\ee
(see (\ref{dej''})). After we specialize to the pure spinor string action in the next section, we will see the important implications of this sign mismatch. Namely, it implies that in contrast to bosonic T-duality the fermionic version does not interchange type IIA and type IIB theories, and does not affect the D-brane dimensions.

\section{Fermionic T-duality in pure spinor formalism}
\label{sect}

The choice of the pure spinor superstring action to derive the fermionic T-duality rules of type II supergravity is due to two simplifications that this choice leads to, despite the apparent complexity of the pure spinor action itself. Firstly, as we shall see shortly, the pure spinor action includes all the supergravity background fields in an explicit manner, as opposed to the Green-Schwarz formalism, where for example only the bosonic components of the supervielbein are present, and one has to invoke the supergravity constraints in order to derive the duality transformations of the fermionic part as well. Furthermore, the pure spinor formulation possesses BRST symmetry generated by the operators
\be\label{brst}
Q = \int dz \l^\a d_\a,\quad \hat Q = \int d\bar z \hat\l^{\hat\a} \hat d_{\hat\a},
\ee
and it is known that nilpotency and (anti)holomorphicity of these operators imply the superspace equations of motion of the background superfields \cite{Berkovits:2001ue}. As we shall see, the form of the BRST operators does not change under the duality transformation, which means that the transformed background is still on-shell.

The pure spinor action of type II superstring in a curved supergravity background is written in terms of the $\mathbb{R}^{10|32}$ superspace coordinates $Z^M = (x^m, \th^\a, \hat\th^{\hat\a})$, where $\th^\a$ and $\hat\th^{\hat\a}$ denote the $\mathrm{SO}(9,1)$ Majorana-Weyl spinors of opposite chiralities for type IIA or of the same chirality for type IIB supergravity. There are also extra worldvolume fields $(d_\a, \l^\a, w_\a)$ and $(\hat d_{\hat\a}, \hat \l^{\hat\a}, \hat w_{\hat\a})$, holomorphic and antiholomorphic, respectively, which not only define the BRST operators (\ref{brst}), but also appear in the action explicitly. In the simpler flat superspace formulation $d_\a$ and $\hat d_{\hat\a}$ were representing the sypersymmetric Green-Schwarz constraints satisfied by $(p_\a, \hat p_{\hat\a})$, the conjugate momenta for spinorial coordinates $(\th^\a, \hat\th^{\hat\a})$:
\be
d_\a = p_\a - \frac12 \g^m_{\a\b} \th^\b \p x_m - \frac18 \g^m_{\a\b} {\g_m}_{\g\d} \th^\b \th^\g \p\th^\d,
\ee
and similarly for $\hat d_{\hat\a}$. Nowever, now that we are in a curved background, $(d_\a, \hat d_{\hat\a})$ are independent fermionic worldvolume fields. $(\l^\a, w_\a)$ and $(\hat \l^{\hat\a}, \hat w_{\hat\a})$ are bosonic ghosts, subject to the pure spinor constraints $\l^\a\g^m_{\a\b}\l^\b = 0$, and similarly for $w_\a, \hat \l^{\hat\a}$, and $\hat w_{\hat\a}$. $\l$~and~$w$ are canonically conjugate variables. With this set of worldvolume fields, the action takes the form
\be\label{pure}\bn
\frac{1}{2\pi\a'}&\int d^2 z \left[ L_{MN}(Z) \p Z^M \pb Z^N + P^{\a\hat\b}(Z) d_\a \hat d_{\hat\b} + E_M^\a(Z) d_\a \pb Z^M \right. \\&+ E_M^{\hat\a}(Z)\p Z^M \hat d_{\hat\a} + \W_{M\a}^\b(Z) \l^\a w_\b \pb Z^M + \hat\W_{M\hat\a}^{\hat\b}(Z) \p Z^M \hat\l^{\hat\a} \hat w_{\hat\b} \\&+ C_\a^{\b\hat\g}(Z) \l^\a w_\b \hat d_{\hat\g} + \hat C_{\hat\a}^{\hat\b\g}(Z) d_\g \hat\l^{\hat\a} \hat w_{\hat\b} + S_{\a\hat\g}^{\b\hat\d} \l^\a w_\b \hat\l^{\hat\g} \hat w_{\hat\d} \\&+ \left. w_\a \pb \l^\a + \hat w_{\hat\a} \p\hat\l^{\hat\a} \right] + \frac{1}{4\pi} \int d^2 z\,\Phi(Z) \mathcal{R}.
\en\ee
As before, $L_{MN}$ stands for the sum of graded-symmetric and graded-antisymmetric tensors $G_{MN}$ and $B_{MN}$, which have the metric and the $b$-field as their lowest-order component fields in the $(\th,\hat\th)$-expansion. The superfield $P^{\a\hat\b}$ takes care of the RR fluxes:
\bsub
\be\label{Pab}
P^{\a\hat\b}|_{\th=\hat\th=0} = \frac{i}{16} e^\f F^{\a\hat\b},
\ee
\be\label{2a}
F_{IIA}^{\a\hat\b} = m + \frac{1}{2} (\g^{m_1 m_2})^{\a\b} F_{m_1 m_2} + \frac{1}{4!} (\g^{m_1 \hdots m_4})^{\a\b} F_{m_1 \hdots m_4},
\ee
\be\label{2b}
F_{IIB}^{\a\hat\b} = (\g^m)^{\a\b} F_m + \frac{1}{3!} (\g^{m_1 m_2 m_3})^{\a\b} F_{m_1 m_2 m_3} + \frac12 \frac{1}{5!} (\g^{m_1 \hdots m_5})^{\a\b} F_{m_1 \hdots m_5}.
\ee
\esub
The numerical coefficient in (\ref{Pab}) may be different depending on the supergravity conventions. In the above formulae $\mathrm{SO}(8)$ ($16\times 16$) gamma-matrices are used, see appendix \ref{gamma10d}. The scalar $m$ in (\ref{2a}) is Romans' mass parameter. In fact, these expressions are only correct for backgrounds with trivial NSNS 2-form. If there is a nontrivial $b$-field, then instead of just the RR field strengths one should use the modified RR field strengths that are invariant under the supergravity gauge transformations as given in (\ref{mod}). This correction is beyond the first order in component fields and thus was omitted from the original derivation.

$E^\a_M$ and $E^{\hat\a}_M$ are parts of supervielbein, containing ordinary vielbein and (for $M$ spinorial) $N=2$ $d=10$ gravitinos $\psi^\a_m$, $\psi^{\hat\a}_m$. The $\th=\hat\th=0$ components of $\W$, $C$, and $S$ are, respectively, the spin connection mixed with NSNS three-form $H=db$, gravitino field strengths, and Riemann tensor again mixed with $H$. For details of the pure spinor formalism see \cite{Berkovits:2002zk, Oz:2008zz, Bedoya:2009np, Mazzucato:2011jt}. 

Starting with the action (\ref{pure}) we can carry out Buscher procedure as described in the previous section, effectively replacing the fermionic isometry coordinate $\th^1$ with dual $\tilde\th^1$, and all the background superfields with their fermionic T-duals. The dual fields $B'_{11}, L'_{1M}, L'_{M1}, L'_{MN}, \Phi'$ are the same as in the example of the previous section (\ref{dual1}), (\ref{dual1fi}), and for the rest of the superfields that are present in (\ref{pure}) we get:
\be\label{Dual}\bn
P'^{\a\hat\b} = P^{\a\hat\b} - (B_{11})^{-1} E^\a_1 E^{\hat\b}_1, \quad & E'^\a_1 = (B_{11})^{-1} E^\a_1, \quad E'^{\hat\a}_1 = (B_{11})^{-1} E^{\hat\a}_1,\\
E'^\a_M = E^\a_M - (B_{11})^{-1} L_{1M} E^\a_1, \quad & E'^{\hat\a}_M = E^{\hat\a}_M - (B_{11})^{-1} E^{\hat\a}_1 L_{M1},
\en\ee
etc. (for the complete list of background superfield transformations, as well as the proof that the supersymmetry is preserved see \cite{Berkovits:2008ic}). The supervielbein index $1$ in these formulae is spinorial, corresponding to the isometry coordinate $\th^1$. Taking $\th=\hat\th=0$ components one can establish that fermionic T-duality transformation leaves invariant the NSNS tensor fields $g_{mn}$ and $b_{mn}$. What does transform are the RR fluxes and, of course, the dilaton:
\be\label{dual}
\frac{i}{16} e^{\f'} F'^{\a\hat\b} = \frac{i}{16} e^\f F^{\a\hat\b} - \e^\a \hat\e^{\hat\b} C^{-1}, \quad \f' = \f + \frac12 \log C.
\ee
The dilaton transformation law comes about in precisely the same manner as (\ref{dual1fi}), while the RR bispinor transformation is encoded in the first of the equations (\ref{Dual}). We denote
\be
C = B_{11}|_{\th=\hat\th=0}, \quad (\e^\a, \hat\e^{\hat\a}) = (E^\a_1, E^{\hat\a}_1)|_{\th=\hat\th=0}.
\ee
Furthermore, the superspace torsion constraints help to find an expression for $C$ in terms of $\e^\a, \hat\e^{\hat\a}$ \cite{Berkovits:2008ic}:
\be\label{c}
\p_m C = i \left( \bar\e \,\G_m \e - \bar\eh \,\G_m \eh \right) = \left\{ 
\begin{array}{l}
	\mathrm{IIB}:\quad i\left[(\e c)^\a (\g_m)_{\a\b} \e^\b - (\eh c)^{\hat\a} (\g_m)_{\hat\a\hat\b} \eh^{\hat\b}\right], \\
	\mathrm{IIA}:\quad i\left[(\e c)^\a (\g_m)_{\a\b} \e^\b - (\eh \bar{c})_{\hat\a} (\g_m)^{\hat\a\hat\b} \eh_{\hat\b}\right].
\end{array}
\right.
\ee
The IIA expression can be rewritten concisely in terms of the Majorana spinor $\mathrm{E} = \e + \eh$:
\be
\label{2a-c}
\p_m C = i\bar{\mathrm{E}} \,\G_m \G^{11}\mathrm{E}.
\ee
The above formulae are written in terms of a Majorana-Weyl representation of the $\mathrm{SO}(1,9)$ gamma-matrices $\G^\m$, such that
\be\label{rep}
\G^m =
	\left(
        \begin{array}{cc}
            0 & (\g^m)^{\a\b}\\
            \g^m_{\a\b} & 0\\
        \end{array}
    \right), \quad
C = 
	\left(
        \begin{array}{cc}
            0 & {c_\a}^\b\\
            \bar{c}^\a{}_\b & 0\\
        \end{array}
    \right), \quad
\G^{11} =
	\left(
        \begin{array}{cc}
            1 & 0\\
            0 & -1\\
        \end{array}
    \right).
\ee
The indices $\a, \b$ here take values $1\ldots 16$. Different properties of this class of representations are considered in \cite{Grassi:2003cm}. We use Majorana conjugation for covariant spinors $\bar\ps = \ps^T C$. More details regarding the spinorial conventions are gathered in appendix \ref{gamma10d}. In particular, in the appendix we introduce the representation of the class (\ref{rep}), where the charge conjugation matrix $C=\G^0$, and $c = 1$, $\bar c = -1$. The relation (\ref{c}) in such a basis takes the form (for type IIB)
\be\label{2b-c}
\p_m C = i \e^\a (\g_m)_{\a\b}\e^\b - i \hat\e^{\hat\a} (\g_m)_{\hat\a\hat\b}\hat\e^{\hat\b}.
\ee
In order to clarify the meaning of $\e^\a,\hat\e^{\hat\a}$, which play the role of the parameters of the fermionic T-duality transformation, recall that in curved superspace the supersymmetry parameters can be written as \cite{Wess:1992cp}
\be
(E^\a_M \d Z^M)|_{\th=\hat\th=0},\quad (E^{\hat\a}_M \d Z^M)|_{\th=\hat\th=0}
\ee
(rather than just $(\d\th^\a, \d\hat\th^{\hat\a})$ in the flat case). If we take $\d Z^M$ as in (\ref{sdvig}), then we see that the supersymmetry parameters can be written in terms of the lowest-order components of the supervielbeins as
\be
\e^\a \r,\quad \hat\e^{\hat\a} \r.
\ee
This leads us to conclude that the parameters $\e^\a,\hat\e^{\hat\a}$ of the fermionic T-duality transformation (\ref{dual}) are the Killing spinors of the initial supergravity background. This pair of Kililng spinors describes a supersymmetry preserved by the background, whose existence manifests itself in the shift isometry (\ref{sdvig}). Note that the spinors $\e^\a,\hat\e^{\hat\a}$ are commuting since the dependence on an anticommutative parameter $\r$ has been made explicit, and that we are talking about a single supersymmetry parameterized by a couple $\e^\a,\hat\e^{\hat\a}$ since it is $N=2$ supersymmetric theory.

These Killing spinors cannot be arbitrary, though. It is natural to require that the isometries being dualized form an abelian subalgebra of the symmetry superalgebra of the background, simply to have the result of the transformation well-defined. In the case of bosonic T-duality this requirement is obviously trivial for a single isometry, since it always commutes with itself. The situation is different when we require that multiple supersymmetries that we fermionically T-dualize anticommute. In particular, even for a single supersymmetry one should require that it squares to zero. Making use of the supersymmetry algebra we find the following constraint on the Killing spinors:
\be\label{a-comm}\bn
0 =\left( \bar\e \,Q + \bar\eh \,\hat{Q} \right)^2 &= - \left( \bar\e \,\G^m \e + \bar\eh \,\G^m \eh \right) P_m 
\\
  &= \left\{ 
		\begin{array}{l}
			\mathrm{IIB}:\quad \left[(\e c)^\a \g^m_{\a\b} \e^\b + (\eh c)^{\hat\a} \g^m_{\hat\a\hat\b} \eh^{\hat\b}\right] P_m, \\
			\mathrm{IIA}:\quad \left[(\e c)^\a \g^m_{\a\b} \e^\b + (\eh \bar{c})_{\hat\a} (\g^m)^{\hat\a\hat\b} \eh_{\hat\b}\right] P_m.
		\end{array}
	\right.
\en\ee
Again we can rewrite the IIA expression succinctly as
\be
\label{2a-constraint}
\bar{\mathrm{E}} \,\G^m \mathrm{E} = 0.
\ee
In the gamma-matrix representation given in appendix \ref{gamma10d} we can write simply
\be\label{cons}
\e\g_m\e + \eh\g_m\eh = 0
\ee
for the case of type IIB theory. We will be using the constraint in this form in chapter \ref{ch:paper1}, where the focus will be on IIB supergravity. In the chapter \ref{ch:paper2}, which deals with IIA theory, the form (\ref{2a-constraint}) will be preferred.

The constraint (\ref{a-comm}) has far-reaching consequences for the duality transformation. In the standard representation of the gamma-matrices mentioned above, $\g_0$ is a unit matrix. Therefore for $m=0$ (\ref{cons}) cannot be satisfied for any real spinor. This is the reason why in general fermionic T-duality does not preserve the reality of background.

Strictly speaking, imposing the constraint (\ref{a-comm}) is not necessary for the Buscher procedure to hold. One can speculate that whether or not the Killing spinor satisfies this constraint may be related to the possibility to introduce the adapted superspace coordinates for a Killing spinor, so that it acts by simple shifts (this issue has been discussed above, after the equation (\ref{ieq})).

Note that nonabelian T-duality can be formulated consistently in the bosonic case \cite{delaOssa:1992vc, Giveon:1993ai, Alvarez:1994zr, Lozano:1995jx, Sfetsos:2010uq, Lozano:2011kb}. The corresponding research on fermionic T-duality has not appeared yet. It would be interesting to consider such a possibility in order to evade the need to complexify the Killing spinor and the background.

We can now return to the issue mentioned after the equations (\ref{2.10}) of the previous section, namely, that there is an important sign difference between the transformations of bosonic and of fermionic T-duality. What in that example manifested itself as the same sign of $\p\tilde\th^1 / \p\th^1$ and $\pb\tilde\th^1 / \pb\th^1$ is now the coincidence of the signs of $E'^\a_1$ and $\hat E'^{\hat\a}_1$ in the transformation law (\ref{Dual}). This is to be contrasted with the bosonic T-duality case, where the signs are different (\ref{2.10}). Note that there exists a derivation of bosonic T-duality transformation in the pure spinor formalism \cite{Benichou:2008it}, which shows clearly that in the bosonic case the relative minus sign is present in the transformation law of the supervielbein:
\be
E'^\a_1 = (G_{11})^{-1} E^\a_1,\quad \hat E'^{\hat\a}_1 = - (G_{11})^{-1} \hat E^{\hat\a}_1.
\ee
Here we assume that a bosonic coordinate $x^1$ has been T-dualized, and the corresponding supervielbein index is therefore a 'bosonic 1', not fermionic as in (\ref{Dual}). This difference implies that whereas it is crucial to change the chirality of either $\th$ or $\hat\th$ in order to keep the standard supergravity constraints after bosonic T-duality has been done, there is no need to do this after fermionic T-duality. Thus type IIA and Type IIB string theories are not interchanged under fermionic T-duality, and the D-brane dimension is also preserved. For more details of this dissimilarity we refer the reader to the original paper \cite{Berkovits:2008ic}.

Finally let us give a brief discussion of the problems that arise if we try to define the fermionic T-duality transformation beyond tree level in string perturbation theory. As mentioned earlier, strictly speaking, the simplified description of the Buscher procedure given above is only valid for the string worldsheets with topology of a disk or a sphere. Global aspects of the Buscher procedure become important on the worldsheets with handles \cite{Rocek:1991ps, Giveon:1993ai, Alvarez:1993qi}. 

Think of the transition from the intermediate action that relates the two bosonic T-dual sigma-models, to the original action, which is achieved by integrating out the Lagrange multiplier $\tilde x$ in (\ref{dej'}). The main obstacle is that on a nontrivial Riemann surface the condition that field strength of a vector $F=dA$ is zero does not imply that the 1-form $A$ is exact. Integral curves of the vector field may wind around the noncontractible cycles on the worldsheet, with the vector field having no well-defined potential. If the original isometry coordinate $x$ was compact, then one could only get the correct periodicity in $x$ after integrating out $\tilde x$ if $A$ had integer-valued circulations around the noncontractible cycles $\mathcal{C}_a$ of the Riemann surface
\be
\oint_{\mathcal{C}_a} A = n_a,\quad n_a \in \mathbb{Z}.
\ee
In this case one can interpret $A$ as differential of a periodic scalar $x$ and the original sigma-model is recovered. The integer-valuedness of the circulations of $A$ must be imposed by inserting a delta-function (or actually a Dirac comb to account for all integer values for $\oint A$) in the string path integral
\be
\sum_n \prod_{a=1}^g \d \left(n_a = \oint_{\mathcal{C}_a} A\right) = \sum_n \exp \left(i \sum_{a=1}^g n_a \oint_{\mathcal{C}_a} A\right),
\ee
where the product/sum over $a$ corresponds to taking into account all the nontrivial cycles $\mathcal{C}_a$ on a genus $g$ surface, and we also sum over all possible values of $n_a$. The latter now may be interpreted as the winding modes of the dual coordinate $\tilde x$ on the cycles $\mathcal{C}_a$:
\be
\tilde{x}(z+\mathcal{C}_a) = \tilde{x}(z) + n_a.
\ee
Note that the dual coordinate is thus non-periodic, and one must include the integration over its winding modes $n_a$ in the path integral.

This discussion applies to the fermionic Buscher procedure with minimal modifications that are due to the fact that the fermionic variables cannot be compact. This leads to some subtleties in the treatment of the zero modes of the fermionic field being dualized \cite{Berkovits:2008ic}, even at tree level. On the Riemann surfaces of higher genus, however, the procedure is ill-defined because of the presence of an extra fermionic zero mode $\r_a$ in the path integral. It can be thought of as representing either windings of the dual fermionic coordinate on the noncontractible cycles:
\be
\tilde{\th}(z+\mathcal{C}_a) = \tilde{\th}(z) + \r_a,
\ee
or the circulations of the fermionic vector field around these cycles.

\section{Summary}

Fermionic T-duality is a tree level symmetry of string theory, which preserves supersymmetry. It can be carried out with respect to Killing spinors that belong to the abelian subalgebra of the symmetry superalgebra. These Killing spinors determine the transformed solution as follows.

Take, $\e$, a Killing spinor that parameterizes an unbroken supersymmetry. It is a Majorana-Weyl spinor of $(1+9)$-dimensional spacetime, that is, real with sixteen components. Since type II supergravity is an $\mathcal{N}=2$ theory, there is also another Killing spinor, which is denoted by $\eh$ and has the same or different chirality as $\e$ depending on whether we are in type IIB or type IIA theory. A pair $\varepsilon=(\e,\eh)$ generates one supersymmetry transformation. However, the two spinors within the pair are not independent -- they are related by the Killing spinor equations, and furthermore by the constraint (\ref{a-comm}). This relation cannot hold for real spinors, and they must be artificially complexified. This is a characteristic property of fermionic T-duality, which leads to complex RR fluxes after the transformation. In type IIB we write (\ref{a-comm}) as 
\be\label{BM:constr}
\e\g_m\e + \eh\g_m\eh = 0,
\ee
and in type IIA as
\be
\bar{\mathrm{E}} \,\G^m \mathrm{E} = 0
\ee
for a Majorana spinor $E = \e+\eh$.

After the choice of the Killing spinors satisfying (\ref{a-comm}) has been made, one calculates an auxilliary scalar field $C$ defined by the differential equation (\ref{c}), i.e. 
\be
\p_m C = i \e^\a (\g_m)_{\a\b}\e^\b - i \hat\e^{\hat\a} (\g_m)_{\hat\a\hat\b}\hat\e^{\hat\b}
\ee
for type IIB, and
\be
\p_m C = i\bar{\mathrm{E}} \,\G_m \G^{11}\mathrm{E}
\ee
for IIA.

Note that by using the constraint (\ref{BM:constr}) we can simplify the IIB expression:
\be
\label{BM:C}
\p_m C = 2i \e\g_m\e.
\ee
The transformation of the dilaton is given by
\be
\label{BM:dil}
\f' = \f + \frac12 \log C,
\ee
and the transformation of RR fields can be written succinctly in terms of the bispinor $F^{\a\b}$:
\be
\label{BM:RR}
\frac{i}{16}e^{\f'}F' = \frac{i}{16}e^{\f}F - \frac{\e\otimes\eh}{C}.
\ee

In the case when the fermionic T-duality is performed with respect to several supersymmetries, parameterized by the Killing spinors $\varepsilon_i=(\e_i, \eh_i), i\in\{1,\ldots,n\}$, the formulae (\ref{BM:C}), (\ref{BM:dil}), and (\ref{BM:RR}) are generalized to
\bsub
\label{mult-T-d}
\be
\label{C-mult}
\p_m C_{ij} = 2i \e_i \g_m \e_j,
\ee
\be
\label{dil-mult}
\f' = \f + \frac12 \sum_{i=1}^n (\log C)_{ii},
\ee
\be
\label{RR-mult}
\frac{i}{16} e^{\f'} F' = \frac{i}{16} e^\f F - \sum_{i,j=1}^n(\e_i \otimes \eh_j)\, (C^{-1})_{ij}.
\ee
\esub
The set of the Killing spinors must obey
\be
\label{abel}
\e_i\g_m\e_j + \eh_i\g_m\eh_j = 0
\ee
for all $i,j\in\{1,\ldots,n\}$. One can strightforwarldy modify the type IIA formulae to describe the case of multiple fermionic T-dualities in a similar manner (essentially by promoting $C$ to the matrix-valued function).

\chapter{Exploring fermionic T-duality}
\label{ch:paper1}

\section{Introduction}

In this chapter we will gain some practical familiarity with the way fermionic T-duality works by applying the transformation to several solutions of type IIB supergravity. The choice of backgrounds to be transformed shall be dictated by the necessity to demonstrate the properties of the fermionic T-duality transformation hinted at in the chapter \ref{ch:techintro}.

As shown there, the transformation leaves invariant the NSNS sector (apart from the dilaton shift). Fermionic T-duality is a transformation primarily of the RR fields. This really explains the delay in the study of fermionic T-duality; deriving the transformations of the RR backgrounds in bosonic T-duality from the string worldsheet has only been done recently and required using the pure spinor formulation \cite{Benichou:2008it}. We will look in detail at the supergravity fields that result from applying the fermionic T-duality to the standard backgrounds like a D-brane. By doing this one has ample opportunities to get acquainted with all sorts of unusual supergravity backgrounds that look like a familiar D-brane with respect to the metric and the $b$-field, but have very uncommon dilaton and RR fields (and are supersymmetric, just as the original D-brane).

Secondly, because of the requirement that we deal with commuting supersymmetries (just as one deals with commuting isometries in ordinary T-duality) it is necessary that we deal with complexified Killing spinors and in turn complexified RR-fluxes. Thus the transformed background will be a solution of complexified supergravity. One open and indeed crucial question is to determine when these transformations map back to a real supergravity solution. In fact, one need not map directly to a purely real solution since if there exists a time-like isometry (which is almost certain for a supersymmetric solution) then one can
do bosonic T-duality in the timelike direction. Although timelike compactifications may not be a valid feature for a realistic theory, bosonic T-dualities in timelike directions are known to relate type II supergravities to perfectly valid type II* theories \cite{Hull:1998ym,Hull:1998vg}. The transition II~$\rightarrow$~II* involves, among other things, a continuation of RR fields:
\be
C_{(n)} \rightarrow -i C_{(n)},
\ee
so that all $C_{(n)}$ have wrong signs of their kinetic terms in type II* supergravity action. In this sense, one could think of the type II background with purely imaginary RR fluxes (which may result from fermionic T-duality) as being the type II* background with real fluxes. A timelike T-duality transformation would then map it to some real type II background.

This was precisely the case for the fermionic dual of $AdS_5\times S^5$ described by Berkovits and Maldacena \cite{Berkovits:2008ic} where after eight fermionic T-dualities there remained some imaginary RR flux. This was then made real by application of timelike T-duality. We will see an imaginary RR background appear after fermionic T-dualizing the pp-wave background of IIB supergravity.

In any case, perhaps we should be interested in complexified supergravity in its own right. In quantum field theory (such as Yang-Mills) there has been a great deal of progress made by complexifying the theory and then using the power of complex
analysis. This was the origin of the S-matrix programme which has now seen something of a revival \cite{Britto:2005fq,Alday:2007hr,Drummond:2007aua,Brandhuber:2007yx} with recent works on amplitude physics again relying on an implicit complexification of the theory to achieve results. In fact, the motivation  for studying fermionic T-duality \cite{Berkovits:2008ic,Beisert:2008iq} was to explain the duality between certain amplitudes and Wilson lines in Yang-Mills theory, and the relation of the dual superconformal symmetry of scattering amplitudes to string theory integrability \cite{Ricci:2007eq}. Whether we can learn really more about string theory per se through complexification of backgrounds has yet to be seen but ideas along these lines have appeared before (see for example the discussion in \cite{Hull:1998ym}).

Based on the description of the fermionic T-duality technique given in the chapter \ref{ch:techintro} we can formulate the following recipe to perform fermionic T-duality on a given solution:
\begin{enumerate}
\item Find the Killing spinors of the solution. In IIB supergravity we choose to represent these by pairs $\varepsilon=(\e,\eh)$ of 16-component real spinors of the same chirality. This corresponds to the two Majorana-Weyl supersymmetry parameters of the theory.
\item Choose a complex linear combination of the Killing spinors $\varepsilon'=(\e',\eh')$ that satisfies the commutativity condition (\ref{BM:constr}). This linear combination describes the supersymmetry that we a dualising with respect to.
\item Calculate the auxilliary function $C$ from (\ref{BM:C}). To do this consistently, one should work in world indices (i.e. one should integrate $\p_{\underline \m} C = 2i \e' \left(e^\n_{\underline \m} \g_\n\right) \e'$, where $e^\n_{\underline \m}$ is the vielbein, and world indices are underlined to distinguish them from flat ones).
\item If there are any RR fields in the original background, substitute them into (\ref{2b}) to calculate the matrix $F^{\a\b}$:
\be
\label{BM:bispinor-2}
F^{\a\b} = (\g^\m)^{\a\b} F_\m + \frac{1}{3!} (\g^{\m_1 \m_2 \m_3})^{\a\b} F_{\m_1 \m_2 \m_3} + \frac12 \frac{1}{5!} (\g^{\m_1 \hdots \m_5})^{\a\b} F_{\m_1 \hdots \m_5}.
\ee
\item Use $F^{\a\b}, \e^\a, \eh^\b$, and $C$ to calculate the transformed RR background $F'^{\a\b}$ via (\ref{BM:RR}):
\be
\label{BM:RR-2}
\frac{i}{16}e^{\f'}F' = \frac{i}{16}e^{\f}F - \frac{\e\otimes\eh}{C}.
\ee
\item Use (\ref{BM:bispinor-2}) again, this time to find the contributions of $F_1, F_3$, and $F_5$ to $F'^{\a\b}$ separately.
\item Check that the transformed background is a solution to the field equaitons.
\end{enumerate}

It is obvious that this recipe of doing fermionic T-duality involves a great deal of 16 by 16 matrix manipulations. Its practical implementation can be simplified greatly by using mathematical software capable of analytic computations. In our case we used a simple programme for {\it Mathematica} to perform steps 2, 4, 5, and 6 automatically. The only nontrivial step in the algorithm is number 6, where one starts with a 16 by 16 matrix $F'$, and one needs to find the corresponding 1-, 3-, and 5-form components. This calculation is done by separating the matrices in equation (\ref{BM:bispinor-2}) into their symmetric and antisymmetric parts. On the left-hand side of the equation we have a matrix $F'$, which is the output of (\ref{BM:RR-2}). This should be split into symmetric and antisymmetric parts by brute force. As to the right-hand side of (\ref{BM:bispinor-2}), it is naturally separated into symmetric and antisymmetric parts. Namely, a single $\g$-matrix is symmetric, as well as a product of five $\g$-matrices, whereas a triple product is antisymmetric. This can be verified explicitly by using the matrix representation given in appendix \ref{gamma}.

\section{Fermionic T-duals of the D1-brane}
\label{D1susy}

Firstly, we will perform the transformation on the background of a single D1-brane. This will be a simple nontrivial example, which however clearly shows that the fermionic T-dual fields are typically complex-valued. One can also observe that the transformed background is rather nontrivial, unexpected of a supersymmetric solution (recall that supersymmetry is preserved). Next we shall consider the pp-wave background, and apply multiple fermionic T-dualities to it in order to show other more interesting properies.

The D1-brane background has vanishing $B$-field, and its nontrivial metric is supported by the dilaton and RR 2-form potential. These are given by the following \cite{Horowitz:1991cd}:
\be
\label{D1:phi}
e^{2\f} = 1+\frac{Q}{(\d_{mn} x^m x^n)^6};
\ee
\be
\label{D1:G}
g_{\m\n} = (e^{-\f} \eta_{ij},e^{\f} \d_{mn}),
\ee
\be
\label{D1:C_2}
(C_2)_{01} = e^{-2\f}-1;\quad (F_3)_{01m} = -2 e^{-2\f} \p_m \f,
\ee
and the other RR fields ($C_0, C_4$) vanish everywhere. The notation we use is
$$
\eta_{ij} = \mathrm{diag} (-1,1),\qquad i,j \in \{0,1\},
$$
$$
\delta_{mn} = \mathrm{diag} (1,1,1,1,1,1,1,1),\qquad m,n \in \{2,\ldots,9\}.
$$
All components of $C_2$ and $F_3$, other than specified in (\ref{D1:C_2}), are zero. The indices in (\ref{D1:C_2}) are world indices.

As it is easy to see, the above supergravity background is a solitonic solution in the sense that the fields fall off rapidly as one increases the distance $r=\sqrt{\d_{mn} x^m x^n}$ from the string-like core, located at $x^2 =\ldots = x^9 = 0$, and the metric tends to become flat in this limit. This matches with the supergravity background being a low-energy approximation to the nonperturbative string theory state (a D-brane).

The form of the transformed solution depends on the choice of the Killing spinor used for the transformation. So a few words about D-brane Killing spinors are in order. In this discussion we will closely follow \cite{Ortin:2004ms}. For a generic supersymmetric theory with bosonic fields $B$ and fermionic fields $F$, the supersymmetry transformations with respect to a local parameter $\vare(x)$ can be written schematically as
\begin{align}
\d_\vare B &=\bar{\vare} F,\\
\d_\vare F &=\p\vare + \vare B.
\end{align}
This means that for the solutions with only bosonic fields (which is what we are interested in) we only need to ensure that the variations of the fermions vanish. Requiring this imposes constraints on the supersymmetry parameter, which are the Killing spinor equations. In type IIB supergravity there are two doublets of fermions (dilatini and gravitini), and their supersymmetry variations are given in the appendix (\ref{susy-vars}). It is then straightforward to substitute the D-brane background fields and solve the Killing spinor equations. As a result, one finds that type IIB D$p$-branes in general are invariant under the supersymmetry transformations parameterized by the spinors that satisfy the following condition:
\be\label{proj}
(1\pm\G^{0\ldots p} \mathcal{O}_p) \varepsilon = 0,
\ee
where $\mathcal{O}_p$ is an operator that depends on the dimensionality of the brane in question:
\be
\mathcal{O}_p = \left\{ \begin{array}{cl} \s_1, \quad \frac{p+3}{2} &\mathrm{even},\\ i\s_2, \quad \frac{p+3}{2} &\mathrm{odd}, \end{array} \right.
\ee
and we have included both possible signs of the D-brane charge. One can check that the condition (\ref{proj}) is a projection condition that eliminates half of the degrees of freedom of the spinor. Thus a IIB D-brane in ten dimensions has sixteen unbroken supersymmetries generated by the Killing spinors that satisfy the above constraint. This result in fact holds for IIA D-branes as well, since in the absence of the NSNS $B$-field the supersymmetry variations of the two theories are the same (up to the redefinition of $\mathcal{O}_p$).

Confining our attention to the case of D1-brane we have
\be
\varepsilon = \left(\begin{array}{c} \e \\ \eh \end{array}\right),
\ee
where $\e$ and $\eh$ are the two chiral Majorana-Weyl spinors that are
the supersymmetry parameters of type IIB supergravity.
This is written in the two-component formalism, so that $\vare$
is just a two-component column vector, not a 32-component 10d spinor.
For the D1-brane $\mathcal{O}_1 = \s_1$, so that the Killing spinor constraint takes the form
\be
(1\pm\G^{01} \s_1) \varepsilon = \left(\begin{array}{c} \e \\ \eh \end{array}\right) \pm \left(\begin{array}{c} \G^{01}\eh \\ \G^{01}\e \end{array}\right) = 0,
\ee

Taking the minus sign for definitness we see that, for example, we can
take $\e$ to be arbitrary 16-component MW spinor, in which case $\eh = \G^{01} \e$.

Technically, the above algebraic constraint on the Killing spinor results from the requirement that the supersymmetry
variation of the dilatino vanishes. One then goes on to consider the variation of the gravitino. Since the variation of the gravitino contains derivatives of the supersymmetry parameter, this second constraint leads to a differential equation for $\varepsilon$. Solving this equation introduces coordinate dependence into the Killing spinor (note that so far $\e$ and $\eh$ were constant). Thus, it turns out that
\be
\e = e^{-\frac{\f}{4}} \e_0
\ee
for an arbitrary constant $\e_0$, and $\eh = \G^{01} \e$, as before.
The function $e^\f$ has been defined in (\ref{D1:phi}).

Using the explicit realisation of the gamma-matrices (\ref{gamma-m}), we see that corresponding to an arbitrary
\be\label{sp1}
\e = \left( \e_1,\; \e_2,\; \ldots \e_{16} \right)^T,
\ee
is
\be\label{sp2}
\eh = \left( \e_{16},\; -\e_{15},\; -\e_{14},\; \e_{13},\; -\e_{12},\; \e_{11},\; \e_{10},\; -\e_9,\; -\e_8,\; \e_7,\; \e_6,\; -\e_5,\; \e_4,\; -\e_3,\; -\e_2,\; \e_1 \right)^T,
\ee
where the factors of $e^{-\frac{\f}{4}}$
have been omitted for simplicity ($^T$ means transpose, so that $\e$
and $\eh$ are columns). Setting all $\e_i$ but $\e_1$ to zero,
we get the first basis element, which we call $\vare_1 = (\e_1,\eh_1)$. Repeating this process for all of the sixteen
parameters, we end up with the set of basis elements $\{\vare_i\},\, i\in\{1,\ldots,16\}$.

The next step in our programme is to pick a particular linear
combination of the Killing spinors,
so that it satisfies the condition (\ref{BM:constr}).
As mentioned earlier, this constraint cannot be satisfied by real
Killing spinors.
We consider the simplest possible linear combinations, i.e. those of the form
\be
\label{comb}
\vare' = \vare_a + i \vare_b;\, a,b\in\{1,\ldots,16\}.
\ee
Using the explicit form of gamma-matrices (\ref{gamma-m}) one can check that (\ref{BM:constr}) is satisfied by {\it any} such combination, apart from those of the form $\vare_a + i \vare_{17-a}$ for any $a$.

The result of the fermionic T-duality transformation with respect to $\vare' = \vare_a + i \vare_b$ can be of two types depending on the values of $a$ and $b$:
\begin{itemize}
\item If ($a\leq 8$ and $b\leq 8$), or ($a\geq 9$ and $b\geq 9$), then the result is of the `simple' type. For this choice of the Killing spinors we find that
\be
\e\g_\m\eh = 0 = \eh\g_\m\eh,
\ee
which means that the auxilliary function $C$ in (\ref{BM:dil}) and (\ref{BM:RR}) is just a constant. The commutativity condition (\ref{BM:constr}) is satisfied trivially in this case. The dilaton is shifted by a constant, the RR field components that were present in the original background (\ref{D1:C_2}) are multiplied by a constant, and several new components of $F_3$ and $F_5$ emerge. 
\item If ($a\leq 8$ and $b\geq 9$), or ($a\geq 9$ and $b\leq 8$), then the result is of the `complicated' type. Despite $\e\g_\m\e + \eh\g_\m\eh$ is still zero, as required by (\ref{BM:constr}), $\e\g_\m\e$ is nonzero in this case:
\be
\left\{ \begin{array}{l} \e\g_\m\e\neq 0 \\ \e\g_\m\e + \eh\g_\m\eh = 0. \end{array} \right.
\ee
This means that $C$ is not a constant (in our examples $C$ will be a linear complex-valued function of the coordinates transverse to the brane, see below). The dilaton is shifted by a logarithm of this function, the RR fields are scaled by a power of it, and some new components of $F_3$ and $F_5$ appear again, but also the components that were present in the original solution (\ref{D1:C_2}) get additive terms.
\end{itemize}

Let us give some explicit examples. As a representative of the class of
`simple' fermionic T-duals of D1-branes we will consider the result of
the duality with a Killing spinor parameter $\vare_1 + i \vare_2$. For a
`complicated' class of backgrounds we will use $\vare_1 + i \vare_9$. In both
cases we have the same metric (\ref{D1:G}) and $B$-field (zero), as in
the original D1-brane solution -- this is a general property of
fermionic T-duality. In the particular case of D1-brane and for
Killing spinor combinations of the form (\ref{comb}) it turns out that
RR scalar is also the same before and after the transformation
(zero). Shown below are transformed dilaton and the new RR fields.

\subsection{`Simple' case}
\label{easy}

Taking $a=1, b=2$ we see from (\ref{sp1},\ref{sp2}) that the  Killing spinor parameter of the transformation is
\be
\vare_1 + i \vare_2 =
\left\{
\begin{aligned}
&\{1,i,0,0,0,0,0,0,0,0,0,0,0,0,0,0\}\\
&\{0,0,0,0,0,0,0,0,0,0,0,0,0,0,-i,1\}
\end{aligned}
\right\}.
\ee
We then get from (\ref{BM:C})
\be
\p_\m C = 0,\: \forall \m \quad\Rightarrow\quad C = \mathrm{const}.
\ee
Thus, the dilaton dependence after the duality is
\be
\label{phinew}
e^{2\f'} = C\left(1+\frac{Q}{(\d_{mn} x^m x^n)^6}\right),
\ee
which is a constant rescaling of the string coupling $g_s = e^\f$.
The RR 3-form has the components (world indices are used everywhere)
\be
\label{F_3new_}
(F_3)_{01m} = -2 C^{-1/2} e^{-2\f} \p_m \f
\ee
(compare to (\ref{D1:C_2})) and eight new constant components
\be
\label{F_3new}
\begin{array}{llll}
F_{236} = i, & F_{268} = -1, & F_{356} = 1, & F_{568} = -i, \\
F_{237} = 1, & F_{278} = i, & F_{357} = -i, & F_{578} = -1.
\end{array}
\ee

There also appear 16 constant components of the self-dual RR 5-form:
\begin{subequations}
\label{F_5new}
\be
\label{F_5:1}
\begin{array}{llll}
F_{02369} = -i, & F_{02689} = 1, & F_{03569} = -1, & F_{05689} = i, \\
F_{02379} = -1, & F_{02789} = -i, & F_{03579} = i, & F_{05789} = 1, \\
\end{array}
\ee
\be
\label{F_5:2}
\begin{array}{llll}
F_{14578} = -i, & F_{13457} = -1, & F_{12478} = 1, & F_{12347} = i, \\
F_{14568} = 1, & F_{13456} = -i, & F_{12468} = i, & F_{12346} = -1.
\end{array}
\ee
\end{subequations}
Note that the indices in (\ref{F_5:1}) result from appending 0 and 9
to the indices of the 3-form components in (\ref{F_3new}).
The components in (\ref{F_5:2}) are required by the self-duality.
All the values given in (\ref{F_3new}) and (\ref{F_5new}) must be additionally multiplied by $2 C^{-3/2}$.

\subsection{`Complicated' case}
\label{hard}

As an example of this type of a transformed background let's take the following linear combination of Killing spinors:
\be
\vare_1 + i \vare_9 =
\left\{
\begin{aligned}
&\{1,0,0,0,0,0,0,0,i,0,0,0,0,0,0,0\}\\
&\{0,0,0,0,0,0,0,-i,0,0,0,0,0,0,0,1\}
\end{aligned}
\right\},
\ee
from which it follows that
\be
\label{C}
\p_{0,\ldots,7} C = 0,\, \p_8 C = -4,\, \p_9 C = 4i \quad\Rightarrow\quad C = 4i (x^9 + i x^8).
\ee
We see that the dilaton is now complex-valued, and does not have its characteristic solitonic profile any more:
\be\label{dilatonez}
e^{2\f'} = C e^{2\f} = 4i (x^9 + i x^8)\,\left(1+\frac{Q}{(\d_{mn} x^m x^n)^6}\right).
\ee
The RR fields transform similarly to the simple case with one
important difference: of the eight
newly appearing components of the 3-form only six have truly new indices:
\be
\label{F_3new1}
\begin{array}{lll}
F_{278} = i, & F_{348} = -i, & F_{568} = -i, \\
F_{279} = 1, & F_{349} =-1, & F_{569} = -1,
\end{array}
\ee
whereas the lacking two appear as additive contributions to the $(01m)$ components that were already present before the transformation:
\begin{align}
\label{F_3new2}
&F_{012} = -2 C^{-1/2} e^{-2\f} \p_2 \f, \quad\ldots\quad, F_{017} = -2 C^{-1/2} e^{-2\f} \p_7 \f, \nonumber \\
&F_{018} = -2 C^{-1/2} e^{-2\f} \left( \p_8 \f - C^{-1} \right),\quad F_{019} = -2 C^{-1/2} e^{-2\f} \left( \p_9 \f + i C^{-1} \right).
\end{align}
Again there are sixteen components of self-dual 5-form field strength. These components, as well as those of the 3-form in (\ref{F_3new1}), should be multiplied by $2 C^{-3/2} = 2 \left[ 4i(x^9 + i x^8) \right]^{-3/2}$:
\be
\label{F_5new__}
\begin{array}{llll}
F_{02368} = 1, & F_{02458} = 1, & F_{03578} = -1, & F_{04678} = 1, \\
F_{02369} = -i, & F_{02459} = -i, & F_{03579} = i, & F_{04679} = -i, \\
F_{14579} = -1, & F_{13679} = -1, & F_{12469} = 1, & F_{12359} = -1, \\
F_{14578} = -i, & F_{13678} = -i, & F_{12468} = i, & F_{12358} = -i.
\end{array}
\ee

\subsection{Solution checking the fermionic T-dual}

In the previous section we have seen one of the simplest fermionic T-dual backgrounds for the D1-brane. One can notice that it is quite peculiar in many ways, some of which were hinted at in the introduction: the background is not real, and although it has the common D-brane metric, the dilaton and the RR fluxes are unsusual. Not only are they complex-valued, but the real and imaginary parts of the dilaton (\ref{dilatonez}), for example, grow without bound in the $x^8, x^9$ directions. One can nevertheless verify that the transformed backgrounds are indeed solutions to type IIB supergravity equations of motion (given in our conventions in the appendix \ref{app:2b}).

In the so called, `simple' case, all the equations are trival apart from the Einstein equation (\ref{_G})
\be
R_{\m\n} + 2\nabla_\m\nabla_\n\f = \frac{e^{2\f}}{2} \left[ T_{\m\n}^{(1)} + T_{\m\n}^{(3)} + \frac12 T_{\m\n}^{(5)} \right],
\ee
which is satisfied by the transformed solution because the RR fields' energy-momentum tensors change trivially under the transformation -- being quadratic in RR field strengths that scale as $C^{-1/2}$ (\ref{F_3new_}), they simply get multiplied by $C^{-1}$, which is cancelled by the transformation of the dilaton:
\be
\frac{e^{2\f'}}{2} \left[ {T'}_{\m\n}^{(1)} + {T'}_{\m\n}^{(3)} + \frac12 {T'}_{\m\n}^{(5)} \right] =  \frac{C e^{2\f}}{2} \left[\frac{1}{C} T_{\m\n}^{(1)} + \frac{1}{C} T_{\m\n}^{(3)} + \frac{1}{C} \frac12 T_{\m\n}^{(5)} \right],
\ee
so that the right-hand side of (\ref{_G}) does not change (the left-hand side does not change trivially because the dilaton is shifted by a constant and because the curvature is not affected).

An interesting question, however, is how it so happens that the new
components of the 3- and 5-form do not contribute to the
energy-momentum tensor. The reason is an accurate balance of real and
imaginary units, scattered around (\ref{F_3new}) and
(\ref{F_5new}). 

In the so called, `complicated' case, the auxilliary field $C$ in the
transformation is no longer constant. As a result the function $C = 4i
(x^9 + i x^8)$ (\ref{C}) enters into the expressions for the
transformed fields and the verification of most equations is nontrivial.

To gain a flavour of the cancellations involved we will give an
example of solving the dilaton field equation (\ref{_phi})
\be\label{dil-eqn}
R = 4(\p\f)^2 - 4\nabla^2 \f.
\ee
Using
\be
\f' = \f + \frac12 \log C,
\ee
we calculate
\be
\nabla^2 \f' = \frac{1}{\sqrt{|g|}} \p_m \left( \sqrt{|g|} g^{mn} \p_n \f' \right) = -\frac{e^{-\f}}{2C^2} \d^{mn} \left(\p_m C \p_n C - 2C \p_m \f \p_n C \right),
\ee
\be
(\p\f')^2 = e^{-\f} \d^{mn} \left( \p_m \f \p_n \f + \frac{1}{C} \p_m \f \p_n C + \frac{1}{4C^2} \p_m C \p_n C \right),
\ee
where we have taken into account that for the dilaton in the D1-brane background
\be
\d^{mn} \left( \p_m\p_n\f + 2 \p_m\f\p_n\f \right) \equiv 0,
\ee
and that the second derivatives of the linear function $C$ vanish.

For the function $C = 4i(x^9 + i x^8)$ we get
\begin{align}
&\d^{mn} \p_m C \p_n C = (\p_8 C)^2 + (\p_9 C)^2 = 0,\\
&\d^{mn} \p_m \f\, \p_n C = -4 (\p_8 \f - i\, \p_9 \f),
\end{align}
and substituting this into the dilaton field equation (\ref{dil-eqn}) yields
\begin{align}
R + 4 \nabla^2 \f' - 4(\p\f')^2 &= -5e^{-\f} \d^{mn} \p_m \p_n \f - \frac{16 e^{-\f}}{C} (\p_8 \f - i\, \p_9 \f) \nonumber\\
&- 10 e^{-\f} \d^{mn} \p_m\f \p_n\f + \frac{16e^{-\f}}{C} (\p_8\f - i\, \p_9\f) = 0.
\end{align}

All other field equations have been checked and involve many
complicated cancellations. Carrying out these checks one obtains a
healthy respect for the nontriviality of this duality from the point
of view of the supergravity equations of motion.

\section{pp-wave}

Another type IIB background that is interesting to consider is the pp-wave solution~\cite{Blau:2001ne,Berenstein:2002jq}. This is a maximally supersymmetric solution, and so by dualizing it with respect to any of its Killing spinors we can get another maximally supersymmetric background of (complexified) type IIB supergravity.

In our conventions the pp-wave background is given by
\begin{subequations}
\be
ds^2 = 2 dx^+ dx^- - \l^2 \d_{\m\n} x^\m x^\n dx^+ dx^+ + \d_{\m\n} dx^\m dx^\n,
\ee
\be
\label{ppwave-RR}
F_{+1234} = 4\l = F_{+5678}
\ee
\end{subequations}
(in this section we use the lightcone coordinates $x^\pm =
\frac{1}{\sqrt{2}} (x^9 \pm x^0)$,
and $x^\m = \{x^1,\ldots,x^8\}$).
This solves the supergravity field equations for any constant $\l$:
the dilaton equation is $R=0$, which holds for the above metric, and
the only nontrivial Einstein equation is $R_{++} = \frac14
T^{(5)}_{++}$, which also holds with $R_{++} = 8\l^2$. All the other
equations are trivial due to the vanishing of almost all of the type IIB fields.

The Killing spinors of this background have been derived in~\cite{Blau:2001ne} and in our notation are given by
\be
\label{pp-killing}
\e = \left(\mathbb{1} - i x^\m \mathbb{A}_\m\right) \left(\cos\frac{\l x^+}{2} \mathbb{1} - i \sin\frac{\l x^+}{2}\mathbb{I}\right) \left(\cos\frac{\l x^+}{2} \mathbb{1} - i \sin\frac{\l x^+}{2}\mathbb{J}\right) \e_0,
\ee
for an arbitrary $\e_0$, where $\mathbb{1}$ is a $32 \times 32$ unit matrix, $\mathbb{I} = \G_1\G_2\G_3\G_4$, $\mathbb{J} = \G_5\G_6\G_7\G_8$, and
\be
\mathbb{A}_\m = \left\{
\begin{array}{ll}
8\l\, \G_-\, \mathbb{I}\, \G_\m, & \m = 1,2,3,4,\\
8\l\, \G_-\, \mathbb{J}\, \G_\m, & \m = 5,6,7,8.
\end{array}
\right.
\ee
The formula (\ref{pp-killing}) is written in the complex notation for the supersymmetry tranformations, see appendix~\ref{gamma}. Both $\e$ and $\e_0$ are Weyl spinors, i.e. complex, 16-component. Since full 32 by 32 gamma-matrices $\G_\m$ are used here, rather than 16 by 16 $\g_\m$, half of the components of $\e$ and $\e_0$ are zero.

In order to get the 32 basis elements $\{\vare_k = (\e_k,\eh_k)\}$ we first substitute arbitrary complex constants as the components of $\e_0$:
\be
\label{ab}
(\e_0)_k = \a_k + i \b_k,\qquad k\in\{1,\ldots,16\},\quad \a_k,\b_k \in \mathbb{R},
\ee
the rest 16 components of $\e_0$ being zero. Next we evaluate (\ref{pp-killing}) and get 16 complex components of $\e$.
Now, the real and imaginary parts of this Weyl spinor are our Killing spinors $\vare = (\e,\eh)$ in real notation. There are 32 independent pairs $\vare=(\e,\eh)$, corresponding to the thirty-two real parameters $\a_k$, $\b_k$.

The basis Killing spinor pairs then fall into two groups,
those that depend on $x^+$ only (`group $A$'),
and those that depend on the transverse coordinates $x^1,\ldots,x^8$
(`group $B$'). We get 16 group $A$ Killing spinors by keeping any of $\a_1,\ldots,\a_8$
(which we refer to as `group $A1$') or $\b_1,\ldots,\b_8$ (`group $A2$'),
while setting all other parameters to zero. Spinors that comprise group $B$
result from keeping any of $\a_9,\ldots,\a_{16}$ (`group $B1$') or
$\b_9,\ldots,\b_{16}$ (`group $B2$').

Not all of these Killing spinors satisfy the constraint (\ref{BM:constr}) (or its generalisation
(\ref{abel}), if one wants to perform multiple fermionic T-dualities).
If we pick a pair to construct a complex linear combination $\vare' = \vare_a
+ i \vare_b$  so that $\vare_a$ and $\vare_b$ belong to different groups ($A$ and $B$),
then the condition (\ref{BM:constr}) cannot be satisfied.
Thus, necessarily $\vare_a, \vare_b\in A$ or $\vare_a, \vare_b\in B$.
According to the division into subgroups $A1,A2,B1$, and $B2$,
there are four quite distinct fermionic T-dual backgrounds:
\begin{itemize}
\item $\vare_a,\vare_b \in A1$ or $\vare_a,\vare_b \in A2$;
\item $\vare_a,\vare_b \in B1$ or $\vare_a,\vare_b \in B2$;
\item $\vare_a\in B1$, $\vare_b \in B2$, or the other way round;
\item $\vare_a\in A1$, $\vare_b \in A2$, or the other way round.
\end{itemize}

The first case is much like the `simple' case of the transformed
D1-brane discussed in the section~\ref{easy} above. Namely, the duality parameter $C$ is
just a constant, dilaton is shifted by its logarithm and RR 5-form is
scaled by its power. Twenty-four new RR field components appear, eight
in $F_3$ and sixteen in $F_5$. These look much like those given
in (\ref{F_3new}) and (\ref{F_5new}) multiplied additionally by a sine
or a cosine of $2 \l x^+$. Crucially, these new RR fluxes do not contribute to the stress-energy, precisely as in the D-brane case.

In the second case the transformed background is more complex.
It also has constant $C$, and therefore a constant dilaton and a
constant scaling factor for the 5-form components.
New in this case is that there are four nonvanishing components of RR
1-form, thirty-two components of the 3-form and fifty-six components of
the 5-form. All of these look like $\mathrm{const}\cdot (x^\m + i x^\n)$
for some $\m,\n\in \{1,\ldots,8\}$. Again, their stress-energy vanishes, so that no modification of the Einstein equations occurs.

The third case is interesting, the defining equation for $C$ is
nontrivial. We can proceed however forgetting about the factors of $C$
in all the RR form components. Three points are characteristic of a
dual background in this case:
there is no 3-form,
but all the 1-form and the 5-form components are nonzero;
all of these are either first or (more often) second order polynomials
in the transverse coordinates; and
they have nonvanishing stress-energy. The Einstein equations are still
satisfied due to the nontrivial spacetime dependence of the dilaton, which is proportional to $\log C$.

We will look in detail at the fourth case. This can be also
characterized by nontrivial contribution of the new components to the
stress-energy tensor, and a spacetime-dependent dilaton.

\subsection{Transformed pp-wave}
\label{pp-1}

The linear combination of the Killing spinors that we will use is $\vare' = \vare_1 + i \vare_9$, where $\vare_1$ is what results from keeping only $\a_1 = 1$ in (\ref{ab}) while setting all the other parameters to zero (so this is a group $A1$ element), and $\vare_9$ corresponds to $\b_1 = 1$ (group $A2$). Explicitly this has the following form:
\be
\label{pp-1-k-sp}
\vare' =
    \left\{
        \begin{aligned}
            &\{\cos \l x^+,0,0,i\sin \l x^+,0,0,0,0,0,0,0,0,0,0,0,0\}\\
            &\{i\cos \l x^+,0,0,-\sin \l x^+,0,0,0,0,0,0,0,0,0,0,0,0\}
        \end{aligned}
    \right\},
\ee
where the first line is $\e$ and the second line is $\eh$. This Killing spinor manifestly satisfies the constraint (\ref{BM:constr}), since in this case $\eh = i\e$, and thus
\be
\e\g_\m\e + \eh\g_\m\eh = \e\g_\m\e - \e\g_\m\e \equiv 0.
\ee

The defining equation for $C$ (\ref{BM:C}) takes the form
\be
\p_+ C = 2\sqrt{2}i \cos 2\l x^+ \quad\Rightarrow\quad C = \frac{i\sqrt{2}}{\l} \sin 2\l x^+.
\ee
The dilaton now depends on $x^+$:
\be
\f' = \frac12 \log \left(\frac{i\sqrt{2}}{\l} \sin 2\l x^+\right).
\ee

The RR 5-form components that were nonzero in the original background (\ref{ppwave-RR}) gain the following $x^+$ dependence:
\be
F_{+1234} = F_{+5678} = 3\l \left( \frac{i\sqrt{2}}{\l} \sin 2\l x^+ \right)^{-1/2}.
\ee
The transformed background also has nonzero RR 1-form
\be
F_+ = -\cos 2\l x^+ \left( \frac{i\sqrt{2}}{\l} \sin 2\l x^+ \right)^{-3/2}
\ee
and the following new components of the 5-form:
\bsub
\begin{align}
F_{+1256} = F_{+1368} = F_{+1458} = F_{+2367} &= F_{+2457} = F_{+3478} \\
&= -\l \left( \frac{i\sqrt{2}}{\l} \sin 2\l x^+ \right)^{-1/2};\nonumber\\
F_{+1236} = F_{+1245} = F_{+3678} = F_{+4578} &= \left( \frac{i\sqrt{2}}{\l} \sin 2\l x^+ \right)^{-3/2};\\
F_{+1348} = F_{+1568} = F_{+2347} = F_{+2567} &= -\left( \frac{i\sqrt{2}}{\l} \sin 2\l x^+ \right)^{-3/2};\\
F_{+1278} = F_{+1467} = F_{+2358} = F_{+3456} &= \cos 2\l x^+ \left( \frac{i\sqrt{2}}{\l} \sin 2\l x^+ \right)^{-3/2};\\
F_{+1357} = F_{+2468} &= -\cos 2\l x^+ \left( \frac{i\sqrt{2}}{\l} \sin 2\l x^+ \right)^{-3/2}.
\end{align}
\esub

The only nonvanishing component of the energy-momentum tensors of these RR fields is the $(++)$ component, and this is readily calculated to give
\bsub
\begin{align}
T_{++}^{(1)} &= \frac{i\l^3}{\sqrt{2}} \frac{\cos^2 2\l x^+}{\sin^3 2\l x^+},\\
T_{++}^{(5)} &= 15\sqrt{2}i\l^3 \frac{\cos^2 2\l x^+}{\sin^3 2\l x^+} - 8\sqrt{2} i\l^3 \frac{1}{\sin^3 2\l x^+}.
\end{align}
\esub
The combination that enters the $(++)$ Einstein equation (\ref{_G}) is
\be
\frac{e^{2\f}}{2} \left( T_{++}^{(1)} + \frac12 T_{++}^{(5)} \right) = -8\l^2 \frac{\cos^2 2\l x^+}{\sin^2 2\l x^+} + 4\l^2 \frac{1}{\sin^2 2\l x^+}.
\ee
Recalling that $R_{++} = 8\l^2$ and calculating the second derivative of the dilaton to be
\be
\nabla_+\nabla_+\f = \p_+\p_+\f = -2\l^2 \frac{1}{\sin^2 2\l x^+},
\ee
we see that the Einstein equation (\ref{_G}) is satisfied by the transformed background:
\be
8\l^2 - 4\l^2 \frac{1}{\sin^2 2\l x^+} + 8\l^2 \frac{\cos^2 2\l x^+}{\sin^2 2\l x^+} - 4\l^2 \frac{1}{\sin^2 2\l x^+} \equiv 0.
\ee
All the other field equations are satisfied trivially.

\subsection{Purely imaginary fermionic T-dual background}

All the examples of fermionic T-duals considered up to now were complex. Here we give an example of a solution that one can potentially make sense of within non-complexified supergravity. This is produced by carrying out two independent fermionic T-dualities on the pp-wave. The result of the transformation has purely imaginary RR forms, so that timelike bosonic T-duality~\cite{Hull:1998ym} can make it real.

We begin by picking a second Killing spinor alongside with the one that has been used in the previous subsection:
\bsub
\label{pp-2}
\begin{align}
\vare_1' &=
    \left\{
        \begin{aligned}
            &\{\cos \l x^+,0,0,i\sin \l x^+,0,0,0,0,0,0,0,0,0,0,0,0\}\\
            &\{i\cos \l x^+,0,0,-\sin \l x^+,0,0,0,0,0,0,0,0,0,0,0,0\}
        \end{aligned}
    \right\},\\
\vare_2' &=
    \left\{
        \begin{aligned}
            &\{i\sin \l x^+,0,0,\cos \l x^+,0,0,0,0,0,0,0,0,0,0,0,0\}\\
            &\{-\sin \l x^+,0,0,i\cos \l x^+,0,0,0,0,0,0,0,0,0,0,0,0\}
        \end{aligned}
    \right\}.
\end{align}
\esub
The additional Killing spinor is a sum $\vare_2' = \vare_4 + i \vare_{12}$, where $\vare_4$ is a group $A1$ Killing spinor defined by $\a_4 = 1$ in (\ref{ab}) while setting all the other parameters to zero, and $\vare_{12}$ corresponds to $\b_4 = 1$ (group $A2$). The pair $(\vare_1', \vare_2')$ can be checked to satisfy (\ref{abel}).

The auxilliary function $C$ is a two by two matrix, defined by (\ref{C-mult}):
\be
C_{ij} =
    \left(
        \begin{array}{cc}
            a   & b\\
            b & a\\
        \end{array}
    \right),
\ee
where
\begin{align}
a &= \frac{i\sqrt{2}}{\l} \sin 2\l x^+,\\
b &= \frac{\sqrt{2}}{\l} \cos 2\l x^+.
\end{align}

The matrices $\log C$ and $C^{-1}$, which are needed in order to implement the formulae (\ref{mult-T-d}), have the same structure, but with different values for $a$ and $b$. Namely, we have for the inverse of $C$
\begin{align}
\label{a'}a' &= -\frac{i\l}{\sqrt{2}} \sin 2\l x^+,\\
\label{b'}b' &= \frac{\l}{\sqrt{2}} \cos 2\l x^+,
\end{align}
and for $\log C$:
\begin{align}
\label{a''}a'' &= \frac{i\pi}{2} + \log \frac{\sqrt{2}}{\l},\\
\label{b''}b'' &= -\frac{i\pi}{2} +  i\, 2\l x^+.
\end{align}

Using $\log C$ we can calculate the transformed dilaton:
\be
\f' = \frac12 \mathrm{Tr} \log C = a'' = \frac{i\pi}{2} + \log \frac{\sqrt{2}}{\l}, \qquad e^{\f'} = i\frac{\sqrt{2}}{\l}.
\ee
Thus the string coupling is purely imaginary in this background. From this we can already predict, that the transformed background will necessarily have purely imaginary RR flux, so that the sign of the combination $e^{2\f} F^2$ is invariant.

In order to derive this explicitly we calculate the contribution of the Killing spinors to the RR field strength bispinor, which is represented by the last term in (\ref{RR-mult}):
\be
\begin{aligned}
(\e_i \otimes \eh_j)\, (C^{-1})_{ij} =
                                            &-\frac{i\l}{\sqrt{2}} \sin 2\l x^+ \left[ \e_1 \otimes \eh_1 + \e_2 \otimes \eh_2 \right]\\
                                          &+ \frac{\l}{\sqrt{2}} \cos 2\l x^+ \left[\e_1 \otimes \eh_2 + \e_2 \otimes \eh_1 \right],
\end{aligned}
\ee
where $\e_i$ and $\eh_i$ are components of the pairs $\vare_i' = (\e_i,\eh_i)$. Substituting the values of the Killing spinors as given in (\ref{pp-2}), we arrive at the following background, which is indeed purely imaginary:
\bsub
\begin{align}
F_{+1234} &= F_{+5678} = -i \l^2 \sqrt{2} \\
F_{+1256} &= F_{+1368} = F_{+1458} = F_{+2367} = F_{+2457} = F_{+3478} = i \l^2 \sqrt{2}.
\end{align}
\esub
All other components of RR forms vanish. This background clearly satisfies Einstein equations, because
\be
R_{++} + 2\nabla_+\nabla_+ \f' - \frac{e^{2\f'}}{4} T^{(5)}_{++} = 8\l^2 - \frac14 \left(-\frac{2}{\l^2}\right) (-16 \l^4) \equiv 0.
\ee

\subsection{Self-duality of pp-wave}

We shall now show that the pp-wave background is self-dual under the
fermionic T-duality with respect to eight supersymmetries that we
denote by $\{\vare_1', \ldots, \vare_8'\}$. Corresponding Killing spinors are all
of the same form as those
used to demonstrate how a single or double T-duality is done in the two previous subsections.
Namely, recapitulating the discussion after (\ref{ab}),
we pick sixteen real Killing spinors $\{\vare_1,\ldots,\vare_8\}\in A1$, $\{\vare_9,\ldots,\vare_{16}\} \in A2$.
Then the eight complex Killing spinors, satisfying (\ref{abel}), are given by
\be
\vare_i' = \vare_i + i \vare_{i+8},\qquad i\in\{1,\ldots,8\}.
\ee
In particular, $\vare_1'$ is exactly the same as $\vare'$ that was used in section
\ref{pp-1} and was given by (\ref{pp-1-k-sp}).

With this choice of supersymmetries we get the following matrix $C$:
\be
C =
    \left(
        \begin{array}{cc}
            \begin{array}{cccc}
                a & 0 & 0 & b\\
                0 & a & -b& 0\\
                0 & -b& a & 0\\
                b & 0 & 0 & a\\
            \end{array}
            & 0\\
            0 &
            \begin{array}{cccc}
                a & 0 & 0 &-b\\
                0 & a & b & 0\\
                0 & b & a & 0\\
                -b& 0 & 0 & a\\
            \end{array}
        \end{array}
    \right),
\ee
where $a$ and $b$ are the same as in the previous subsection:
\begin{align}
a &= \frac{i\sqrt{2}}{\l} \sin 2\l x^+,\\
b &= \frac{\sqrt{2}}{\l} \cos 2\l x^+.
\end{align}

The matrices $\log C$ and $C^{-1}$ again have the same structure, but with different values for $a$ and $b$, which coincide with those given in the previous subsection, see eqs. (\ref{a'}) to (\ref{b''}).

The transformed dilaton is then evaluated to be
\be
\label{mult-T-d:new-dil}
\f' = 4 a'' = 2\pi i +4 \log \frac{\sqrt{2}}{\l}, \qquad e^{\f'} = \frac{4}{\l^4}.
\ee
We observe that the phases of $\frac{i\pi}{2}$ in (\ref{a''}) add up precisely in the way required to make $g_s = e^{\f}$ real.
The RR bispinor is modified by
\be
\begin{aligned}
(\e_i \otimes \eh_j)\, (C^{-1})_{ij} =
                                            &-\frac{i\l}{\sqrt{2}} \sin 2\l x^+ \left[ \e_1 \otimes \eh_1 + \ldots + \e_8 \otimes \eh_8 \right]\\
                                          &+ \frac{\l}{\sqrt{2}} \cos 2\l x^+
                                                    \left[
                                                        \begin{aligned}
                                                            &\e_1 \otimes \eh_4 + \e_4 \otimes \eh_1 - \e_2 \otimes \eh_3 - \e_3 \otimes \eh_2\\
                                                            -&\e_5 \otimes \eh_8 - \e_8 \otimes \eh_5 + \e_6 \otimes \eh_7 + \e_7 \otimes \eh_6
                                                        \end{aligned}
                                                    \right],
\end{aligned}
\ee
where $(\e_i,\eh_i)=\vare_i'$.

An important feature of this matrix, which becomes obvious only after explicit substitution of the Killing spinors, is that it is proportional to the first term on the right-hand side of (\ref{RR-mult}). This leads to the RR field bispinor after the transformation being proportional to itself before the transformation. More precisely, we have for the transformed RR background
\be
F_{+1234} = -\l^5 = F_{+5678},
\ee
with all other components vanishing. This is just the original flux that was supporting the pp-wave geometry before we have done fermionic T-duality, multiplied by a constant $-\frac{\l^4}{4}$. Since this constant is equal to $-e^{-\f'}$ (\ref{mult-T-d:new-dil}), the Einstein equations hold for the new background because they involve a product $e^{2\f'} T_{\m\n}^{(5)}$:
\be
R_{++} + 2\nabla_+\nabla_+ \f' - \frac{e^{2\f'}}{4} T^{(5)}_{++} = 8\l^2 - \frac14 \left(\frac{4}{\l^4}\right)^2 (\l^{10} + \l^{10}) \equiv 0.
\ee

This transformation clearly leaves the string spectrum invariant since it is just a field redefinition of the Ramond-Ramond field strength. 

Interestingly, if one splits the eight supersymmetries that were used in this section into two groups $\{\vare_1',\ldots,\vare_4'\}$ and $\{\vare_5',\ldots,\vare_8'\}$ and performs fermionic T-dualities of the original pp-wave background with respect to each of these groups independently, then the resulting background has the dilaton $e^{\f'} = \frac{2}{\l^2}$ in both cases, and the RR forms in the two cases are given by
\be
F_{+1458} = F_{+2367} = \pm 2\l^3.
\ee
Thus each group of four fermionic T-dualities also results in a pp-wave background that has undergone a certain rotation in transverse directions as compared to the original pp-wave.

\section{Summary}

Our aim in this chapter has been to establish some basic familiarity with the fermionic T-duality in supergravity, and to observe the basic features of the transformed backgrounds. Several supersymmetric solutions of type IIB supergravity have been generated, at times displaying peculiar properties. Firstly, one should note that fermionic T-duality does not commute with bosonic T-duality. This has been made evident in the D1-brane case where new Ramond-Ramond fields are produced (such as (\ref{F_3new})) that break the $\mathrm{SO}(1,1) \times \mathrm{SO}(8)$ symmetry of the original D1-brane solution.

We have also checked whether fermionic T-duality maps back to the original background if one applies the same transformation twice (this has not been mentioned in the text), and we see that it is not always so. In the examples carried out above the RR fluxes are mapped back to themselves multiplied by a root of unity. This is undoubtedly a consquence of the fermionic nature of the duality.

One of the main goals of this section was to generate real solutions by means of fermionic T-duality. This has been successful in that we have shown that the pp-wave can be transformed to produce real solutions but in that case the transformed solution is again the pp-wave up to some field redefinitions or rotations.

Now we will shift our approach towards more practical issues and consider a certain application of fermionic T-duality to research in the symmetries of scattering amplitudes in gauge theories.

\chapter{Fermionic T-duality in $\4$ background}
\label{ch:paper2}

\section{Introduction}

In order to further deepen our understanding of the way fermionic T-duality works, we now apply the transformation to an $\4$ background of type IIA string theory. This problem has a rich motivation that comes from the field theory side of the AdS/CFT correspondence. Initially an impressive progress has been made in $\mathcal{N}=4$ super-Yang-Mills theory (which is dual to $\5$ string theory). It was established that a correspondence or duality between planar scattering amplitudes and Wilson loops exists \cite{Alday:2007hr}. Essentially the correspondence relates a certain part of the scattering amplitude to a Wilson loop in momentum space, which is built out of the momenta of the particles participating in the scattering process. The Wilson loop is lightlike since the correspondence is formulated for scattering of massless gluons. For details of the correspondence see reviews \cite{Alday:2008yw, Drummond:2010km}. A closely related development was that the dual superconformal symmetry \cite{Drummond:2008vq, Brandhuber:2008pf} has been proven to exist in $\mathcal{N}=4$ SYM, which from the point of view of the amplitude/Wilson loop correspondence is just ordinary superconformal symmetry acting on Wilson loops. Dual superconformal symmetry can be unified with the conventional one under the framework of Yangian symmetry \cite{Beisert:2010jq}, which provides a new perspective on the integrability that $\mathcal{N}=4$ SYM is known to possess \cite{Ricci:2007eq}.

\begin{figure}[t]
\begin{center}
\includegraphics[scale=0.4]{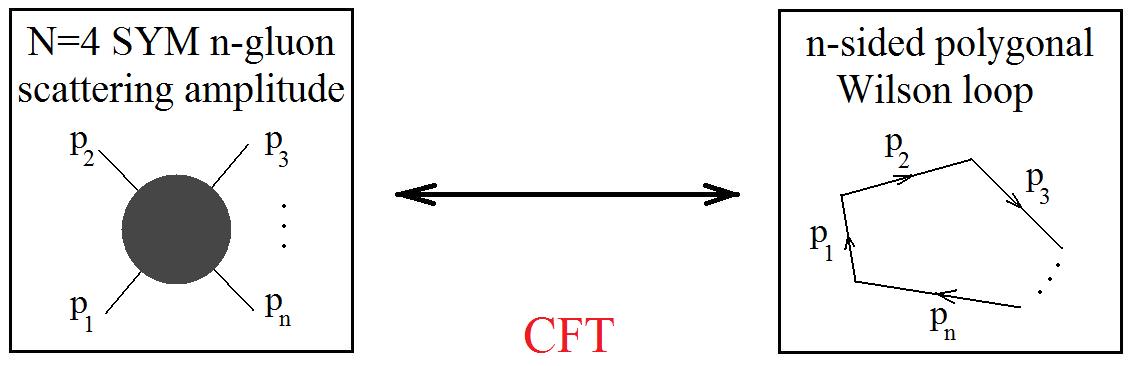}
\includegraphics[scale=0.4]{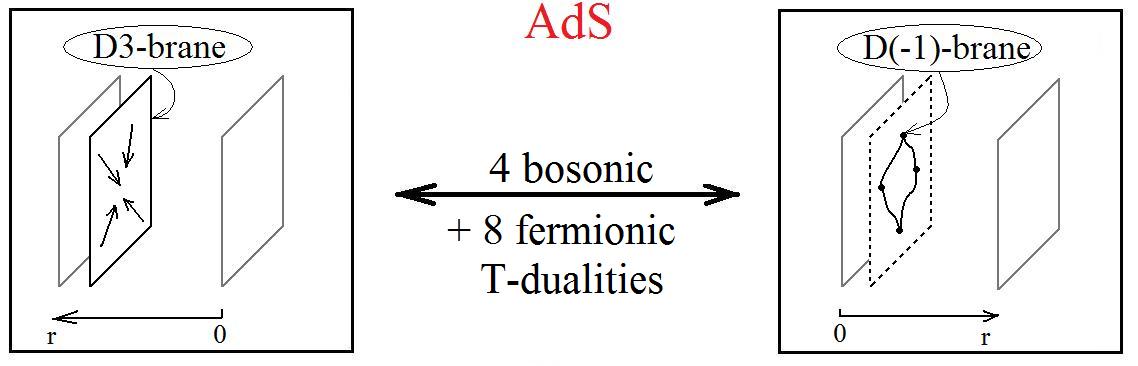}
\caption{Amplitude/Wilson loop correspondence in SYM (upper arrow) and its T-duality counterpart in string theory on $\5$.}
\label{corresp}
\end{center}
\end{figure}

In the SYM case the amplitude/Wilson loop correspondence has been explained by a combination of 4+8 T-dualities on the string theory side of the AdS/CFT correspondence. In particular, four ordinary T-dualities along the flat directions of $AdS_5$ \cite{Alday:2007hr} and eight fermionic T-dualities \cite{Berkovits:2008ic} were required to achieve precise self-duality of the $\5$ background. The AdS metric stays the same after the four bosonic T-dualities if one redefines the radial coordinate as $r'=\frac{R^2}{r}$, where $R$ is the AdS radius. The eight extra fermionic T-dualities are needed in order to make the entire background invariant, namely, they bring the RR fluxes and the dilaton back to their initial state (which has been disturbed by the bosonic T-dualities). While the background is invariant, a string theory configuration that corresponds to a scattering amplitude in SYM is mapped into a configuration related to a Wilson loop (more precisely a configuration consisting of a set of D(-1)-branes with strings stretched between them), hence the duality between the two on the field theory side. The whole setup is illustrated by the figure~\ref{corresp}.

This range of ideas is currently being applied to the other known instance of AdS/CFT correspondence, namely, the correspondence between the type IIA string theory on $\4$ background and the 3-dimensional Chern-Simons-matter theory (ABJM) \cite{Aharony:2008ug}. In particular, Yangian invariance \cite{Bargheer:2010hn, Beisert:2010gn, Bargheer:2011mm} and dual superconformal symmetry \cite{Huang:2010qy, Gang:2010gy, Lipstein:2011ej} of scattering amplitudes in ABJM theory have been observed. Recently dual superconformal symmetry has been observed for loop amplitudes \cite{Chen:2011vv, Bianchi:2011dg, Bianchi:2011fc}. There are hopes that these structures in ABJM theory will find a string theory explanation in terms of a set of bosonic and fermionic T-dualities, in analogy with the $\5$ case. This is further supported by comparison of the ABJM and SYM scattering amplitudes \cite{Wiegandt:2011uu}. 

However, it has been shown that this approach to dual superconformal symmetry cannot be straightforwardly reproduced in the $\4$ case \cite{Adam:2009kt, Grassi:2009yj, Hao:2009hw}. It is clearly impossible to achieve self-duality in the 3+6 setup (which would be a straightforward mimicking of the $\5$ case) because three ordinary T-dualities would take us from IIA to IIB theory. There has been a proposal \cite{Bargheer:2010hn} based on the superalgebra arguments that the correct set of T-dualities to perform in this case would be a `3+3+6' set: three flat $AdS_4$ T-dualities, three $\cp3$ T-dualities, and six fermionic T-dualities. Furthermore, the authors of \cite{Huang:2010qy} have established the existence of dual superconformal symmetry of the tree-level ABJM scattering amplitudes in case when the dual superspace includes three coordinates corresponding to complexified isometries
of $\mathbb{C}P^3$. Nevertheless, Adam,~Dekel,~and Oz have shown \cite{Adam:2010hh} that this combination of T-dualities is singular. The calculation in \cite{Adam:2010hh} has been done in the supercoset $\mathrm{OSp}(6|4)/(\mathrm{SO}(2,1)\times\mathrm{U}(3))$ realization of the sigma-model, and this was conjectured to be a possible cause of the problem. Since the coset is obtained by means of a partial gauge fixing the sigma-model $\k$-symmetry and some string configurations are prohibited by this gauge-fixing, the supercoset might be a good description not of the full field theory but some truncation thereof, and dual superconformal symmetry might be lost. The issues that arise when fixing the $\k$-symmetry in this model have been reviewed and discussed in \cite{Sorokin:2011mj}. Recently an algebraic condition has been formulated, which can be used as a criterion to decide if a particular supercoset sigma-model can be self-dual or not \cite{Dekel:2011qw}.

In this chapter we present the complementary point of view on how does this singularity arise, working with the supergravity component fields as in the previous chapter. This helps to evade the problems with $\k$-symmetry, but the fact that we also find a singularity means that there is still some missing ingredient in the recipe of how to make the dual superconformal symmetry manifest on the string theory side.

For the sake of simplicity, and also following the conjecture made in \cite{Bargheer:2010hn} that the dilaton shifts coming from the bosonic and the fermionic T-dualities seem to cancel, we confine our attention to the transformation of the dilaton. This turns out to be sufficient to expose the singularity involved. The dilaton gets two additive contributions --- a negative one from the bosonic T-dualitites (\ref{fi}):
\be
\delta_B\phi = -\frac12 \log |\det g|
\ee
and a positive one from the fermionic dualities (\ref{dil-mult}):
\be\label{f}
\delta_F\phi = \frac12 \log |\det C|.
\ee
Here $\det g$ is determinant of the block in the metric tensor that incorporates the directions that have been dualized (adapted coordinates have been chosen).

In what follows we shall consider the transformation of the string coupling $e^\f$, which according to the above formulae changes as
\be
\label{result}
e^{2\f'} = e^{2\f} \,\frac{\det C}{\det g}. 
\ee
The main result will be that this transformation is not only singular, but it is also indeterminate, in the sense that {\it both} determinants in the above formula vanish. This is to be contrasted with the $\5$ case \cite{Berkovits:2008ic}, where the two detereminants are nonzero and cancel precisely, thus allowing for the self-duality.

In what follows we will give a very brief review of the ABJM theory and its gravity dual, then describe the supergravity background in more detail, proceed to its Killing vectors and spinors and the symmetry superalgebra that they generate, and finally show what is the problem with bosonic and fermionic T-dualities in this setup.

\section{The background}

Aharony, Bergman, Jafferis and Maldacena (ABJM) have constructed a $d=3$ $\mathcal{N}=6$ superconformal Chern-Simons-matter theory that is conjectured to provide an effective description of $N$ M2-branes located on a $\mathbb{C}^4/\mathbb{Z}_k$ singularity~\cite{Aharony:2008ug}. This was motivated by the earlier work by Bagger and Lambert~\cite{Bagger:2006sk, Bagger:2007jr, Bagger:2007vi} and Gustavsson~\cite{Gustavsson:2007vu, Gustavsson:2008dy} and has lead to a great surge of activity. For reviews see~\cite{Klebanov:2009sg, Klose:2010ki}.

ABJM theory has gauge group $U(N)\times U(N)$ with the factors having opposite Chern-Simons levels $k$ and $-k$. In the 't~Hooft limit 
\be
N\rightarrow\infty,\quad \l=\frac{N}{k} \gg 1\; \mathrm{fixed}
\ee
the weakly coupled gravity dual description is valid, either in terms of the $AdS_4\times S^7 /\mathbb{Z}_k$ solution of $d=11$ supergravity (when $k^5 \ll N$), or in terms of the $\4$ solution of IIA supergravity (when $k \ll N \ll k^5$). We will be dealing with the $d=10$ description. The characteristic spacetime radius is related to the 't~Hooft coupling by $R^2 = 4\pi\a' \sqrt{2\l}$, and the background is given by
\bsub
\begin{align}
ds^2 &= \frac{R^3}{k} \left( \frac14 ds^2_{AdS_4} + ds^2_{CP^3} \right),\\
e^{2\phi} &= \frac{R^3}{k^3},\\
\label{F4}F_4 &= \frac{3R^3}{8} \e_4,\\
\label{F2}F_2 &= kJ.
\end{align}
\esub
$ds^2_{AdS_4}$ is a unit radius $AdS_4$ metric, e.g. in the Poincar\'e patch:
\be
\label{ads}
ds^2_{AdS_4} = r^2 \left[-(dx^0)^2+(dx^1)^2+(dx^2)^2 \right]+\frac{dr^2}{r^2},
\ee
and the corresponding 4-form flux $F_4$ is proportional to the totally antisymmetric symbol $\e_4$ in 4 dimensions.

As regards the $\cp3$ part of the background (which is supported by the 2-form $F_2$, proportional to the K\"ahler form $J$), let us introduce several coordinate systems that will be useful in what follows. Complex projective space $\cp3$ is by definition a linear space of complex lines through the origin of $\mathbb{C}^4$. Since in the 't~Hooft limit we have $k\rightarrow\infty$, the factor $S^7/\mathbb{Z}_k$ of the d=11 background turns into the Hopf fibration $S^7/S^1$. This is how the projective space arises as a submanifold of the d=10 gravitational background. We may illustrate this with a diagram, where the arrows represent factorisation first with respect to dilations by a positive $R$ and then with respect to $S^1$. Together these correspond to taking the factor of $\mathbb{C}^4$ with respect to the complex dilation by a number $c = R e^{i\th}$:
\be
\mathbb{R}^8 \cong \mathbb{C}^4 \xrightarrow{dilations} S^{7} \xrightarrow{Hopf} \cp3 \cong S^7/S^1. 
\ee

The resulting $\cp3$ is a manifold of complex dimension three, possessing a K\"ahler metric. This can be represented either in complex or in real coordinates, and in what follows we will make use of both types of representation. We introduce the following coordinate systems on the $\cp3$ factor of the d=10 geometry.
\begin{itemize}
\item Fubini-Study coordinates $(z,\zb)$, where $\zb_\a$ are complex conjugates 
of $z^\a$, $\a = 1,2,3$. Line element has the well-known form
\be
\label{metric-z}
ds^2_{\cp3} = \frac{dz^\a d\zb_\a}{1+|z|^2} - \frac{z^\a \zb_\b 
dz^\b d\zb_\a}{(1+|z|^2)^2},
\ee
where $|z|^2 = z^\a \zb_\a$. The metric is evidently real, which makes it possible to introduce six real coordinates instead.
\item Starting from the real components of the Fubini-Study coordinates $z^\a = \r^\a e^{i\varphi^\a}$, we can introduce six real coordinates $(\m,\a,\th,\ps,\c,\f)$ as follows \cite{Hohenegger:2009as}:
\be\label{coords-real}\bn
\r^1 &= \tan\m \,\sin\a \,\sin\frac{\th}{2}, \quad &\varphi^1 &= \frac12\, (\ps - \f + \c), \\
\r^2 &= \tan\m \,\cos\a,                     \quad &\varphi^2 &= \frac12\, \c,              \\
\r^3 &= \tan\m \,\sin\a \,\cos\frac{\th}{2}, \quad &\varphi^3 &= \frac12\, (\ps + \f + \c).
\en\ee
It is convenient to work with the Killing spinors in these coordinates because of the simple representation of the vielbein forms:
\be\bn
e^1 &= d\m,\\
e^2 &= \sin\m \,d\a,\\
e^3 &= \frac12 \sin\m \,\sin\a \left( \cos\ps \,d\th + \sin\th \,\sin\ps \,d\f \right),\\
e^4 &= \frac12 \sin\m \,\sin\a \left( \sin\ps \,d\th - \sin\th \,\cos\ps \,d\f \right),\\
e^5 &= \frac12 \sin\m \,\sin\a \,\cos\a \left( \,d\ps + \cos\th \,d\f \right),\\
e^6 &= \frac12 \sin\m \,\cos\m \left( \,d\c + \sin^2 \a \,d\ps + \sin^2 \a \,\cos\th \,d\f \right).
\en\ee
Line element is simply $ds^2_{\cp3} = \d_{ab} e^a e^b$. We shall use Latin letters for the tangent-space components.
\item Finally, we introduce the complexified $\cp3$ background by means of the following coordinate transformation:
\be\label{coords-w}\bn
w^\a &= z^\a,\\
\wb_\a &= \frac{\zb_\a}{1+|z|^2}.
\en\ee
These are six independent complex coordinates ($\wb_\a \neq (w^\a)^*$). The line element takes the simple form:
\be
\label{metric-w}
ds^2_{\cp3} = dw^\a d\wb_\a + \wb_\a \wb_\b dw^\a dw^\b.
\ee
\end{itemize}

Finally, the K\"ahler form $J$ in (\ref{F2}) has the simplest representation in the latter coordinates:
\be
\label{kahler}
J = -2i \,dw^\a \wedge d\wb_a.
\ee
Transforming it to the real coordinates, we get 
\be
\label{F2-real}
\bn
J = &-d\m \wedge (d\ps +d\f \cos\th) \sin 2\m \,\sin^2 \a -d\m \wedge d\c \sin 2\m \\
&- d\a \wedge (d\ps + d\f \cos\th) \sin^2 \m \,\sin 2\a + d\th \wedge d\f \sin^2 \m \,\sin^2 \a \,\sin \th.
\en
\ee
This looks much simpler when written in tangent-space components:
\be
\label{F2-tan}
J_{ab} = e^\m_a e^\n_b J_{\m\n} = 
\left(
	\begin{array}{cccccc}
		& & & & & -2\\
		& & & &-2 &\\
		& & &-2 & &\\
		& &2 & & & \\
		&2 & & & & \\
		2& & & & & 									
	\end{array}
\right).
\ee

\section{Killing vectors}
The six isometries that should be T-dualized are the shifts of three flat $AdS_4$ directions and three internal ($\cp3$) isometries. The contribution of the $AdS_4$ T-dualities can be trivially read off from (\ref{ads}), and it is nonsingular:
\be
\d\f = -3 \log r.
\ee
Therefore from now on we shall only be concerned with internal isometries.

The isometry algebra of $\cp3$ is $\mathfrak{su}(4)$, which is 15-dimensional. None of these isometries commute with any of the supersymmetries, which is the reason for complexifying the Killing vectors. We use the complexified Killing vectors of $\cp3$ as given in \cite{Hoxha:2000jf}:
\be\label{pope}
\bn
K^\a &= {T_0}^\a + {T_\b}^\a z^\b - {T_0}^0 z^\a - {T_\b}^0 z^\b z^\a,\\
K_\a &= -{T_\a}^0 - {T_\a}^\b \zb_\b + {T_0}^0 \zb_\a + {T_0}^\b \zb_\b \zb_\a,
\en
\ee
for a vector field
\be
K = K^\a \,\frac{\p}{\p z^\a} + K_\a \,\frac{\p}{\p \zb_\a}.
\ee
There are precisely 15 independent parameters ${T_A}^B$, $A,B = 0,\ldots,3$ because they are subject to the constraint ${T_A}^A = 0$.

We shall consider the three complex Killing vectors that result from keeping ${T_0}^\a$ in (\ref{pope}):
\be
\label{KV}
K_{(\a)} = \frac{\p}{\p z^\a} + \zb_\a \zb_\b \frac{\p}{\p \zb_\b},\quad \a=1,2,3.
\ee
These three vectors commute with each other and by transforming them to the real coordinates (\ref{coords-real}) one can check that they are of the form $a+ib$, where $a$ and $b$ are ordinary real Killing vectors of $\cp3$:
\bsub
\label{KV-real}
\be\bn
K_{(1)} = &\frac12 \,e^{-\frac{i}{2} (\ps-\f+\c)} \left( \sin\a \sin\frac{\th}{2} \frac{\p}{\p\m} + \cot\m \cos\a \sin\frac{\th}{2} \frac{\p}{\p\a} + \frac{\cot\m \cos\frac{\th}{2}}{\sin\a} \frac{\p}{\p\th}\right.
\\
&\left. -i \,\frac{\cot\m}{\sin\a \sin\frac{\th}{2}} \frac{\p}{\p\ps} +i \,\frac{\cot\m}{\sin\a \sin\frac{\th}{2}} \frac{\p}{\p\f} +2i \,\tan\m \sin\a \sin\frac{\th}{2} \frac{\p}{\p\c}\right),
\en
\ee
\be\bn
K_{(2)} = \frac12 \,e^{-\frac{i}{2} \c} &\left[ \cos\a \,\frac{\p}{\p\m} - \cot\m \sin\a \,\frac{\p}{\p\a} +2i \,\frac{\cot\m}{\cos\a} \,\frac{\p}{\p\ps}\right. 
\\
&\left. -2i \left( \frac{\cot\m}{\cos\a} - \frac{\cos\a}{\cot\m} \right) \frac{\p}{\p\c}\right],
\en
\ee
\be\bn
K_{(3)} = &\frac12 \,e^{-\frac{i}{2} (\ps+\f+\c)} \left( \sin\a \cos\frac{\th}{2} \,\frac{\p}{\p\m} + \cot\m \cos\a \cos\frac{\th}{2} \,\frac{\p}{\p\a} -2 \,\frac{\cot\m \sin\frac{\th}{2}}{\sin\a} \,\frac{\p}{\p\th}\right.
\\
&\left. -i \,\frac{\cot\m}{\sin\a \cos\frac{\th}{2}} \,\frac{\p}{\p\ps} -i \,\frac{\cot\m}{\sin\a \cos\frac{\th}{2}} \,\frac{\p}{\p\f} +2i \,\tan\m \sin\a \cos\frac{\th}{2} \,\frac{\p}{\p\c}\right).
\en
\ee
\esub

Note that alternatively one could also use the three vector fields corresponding to ${T_\a}^0$, which are complex conjugates of the vectors (\ref{KV}), or those resulting from keeping ${T_\a}^\a$ (no sum). These two subgroups of complexified symmetry superalgebra are also commuting.

Now we can reveal the reason for the introduction of the $(w,\wb)$ coordinates in (\ref{coords-w}). Transforming the vectors (\ref{KV}) to these coordinates one discovers that they are acting as shifts:
\be
K_{(\a)} = \frac{\p}{\p w^\a},
\ee
which enables us to calculate $\det g$ in (\ref{result}). For this purpose, we read off the metric tensor from the expression for the interval in $(w,\wb)$ coordinates (\ref{metric-w}):
\be
\label{metric}
g_{\m\n} = \left(
	\begin{array}{cc}
		\begin{array}{ccc}
			\wb^1 \wb^1 & \wb^1 \wb^2 & \wb^1 \wb^3\\
			\wb^2 \wb^1 & \wb^2 \wb^2 & \wb^2 \wb^3\\
			\wb^3 \wb^1 & \wb^3 \wb^2 & \wb^3 \wb^3\\						
		\end{array}	& 
		\begin{array}{ccc}
			  &   &  \\
			  & 1/2 &  \\
			  &   &  \\						
		\end{array} \\
		\begin{array}{ccc}
			  &   &  \\
			  & 1/2 &  \\
			  &   &  \\						
		\end{array} &
		\begin{array}{ccc}
			  &   &  \\
			  & 0 &  \\
			  &   &  \\						
		\end{array} \\
	\end{array}\right).
\ee
The upper-left block here corresponds to the $dw\,dw$ term in the interval. Rescaling of the string coupling under the three T-dualities with respect to $K_{(1,2,3)}$ is given by the determinant of this block, which is identically zero. Now we can rewrite (\ref{result}) as
\be
e^{2\f'} = e^{2\f} \,\frac{\det C}{0}. 
\ee
This is clearly a singularity, and now we proceed to showing that the numerator in this formula vanishes as well.

\section{Killing spinors}

In order to get an expression for the matrix $C$ (\ref{2a-c}) we need to know the Killing spinors $\e,\eh$. These can be found as solutions to the equations
\bsub
\begin{align}
\label{Ksp1}&\left( \slashed{F}_2  - \frac13 \slashed{F}_4 \G^{11} \right) \mathrm{E} = 0,\\
\label{Ksp2}&\nabla_M \mathrm{E} = \frac{e^\phi}{8} \left( \slashed{F}_2 \G_M \G^{11} - \slashed{F}_4 \G_M \right) \mathrm{E},
\end{align}
\esub
which are conditions that supersymmetry variations of the type IIA fermions vanish. Supersymmetry parameter $\mathrm{E}$ is a Majorana spinor, while $\e$ and $\eh$ are its Majorana-Weyl components, which can be obtained by applying the projections $\frac12 (1\pm \G^{11})$. We use the notation $\slashed{F}_n = \frac{1}{n!} F_{M_1\ldots M_n} \G^{M_1\ldots M_n}$. Note that the free index $M$ in (\ref{Ksp2}) is a curved index.

Original derivation of the Killing spinors of $\cp3$ can be found in \cite{Nilsson:1984bj}, \cite{Hohenegger:2009as}, and \cite{Hikida:2009tp}. Here we shall briefly overview the derivation for the sake of consistency with our notation and conventions. We decompose the spinor parameter $\mathrm{E} = \k \otimes \h$ into the product of the $SO(1,3)$ and $SO(6)$ spinors $\k$ and $\h$. With the corresponding decomposition of the gamma-matrices (for details see appendix \ref{app-gamma}), the first Killing spinor equation (\ref{Ksp1}) becomes
\be
\left( 1\otimes \frac12 F_{ij} \b^{ij} \right)\,\left(\k\otimes\h\right) = \left( 1\otimes 2 \b^7 \right)\,\left(\k\otimes\h\right).
\ee
We see that $\k$ is unconstrained, while the equation for $\h$ can be rewritten as follows:
\be
\label{eigen}
Q \,\b^7 \h = -2 \,\b^7 \h,
\ee
where $Q = \frac12 F_{ij} \b^{ij} \b^7$. Evaluating this matrix operator using the tangent-space components of the 2-form (\ref{F2-tan}) shows that indeed there is a~$-2$ eigenvalue, whose degeneracy is~$6$. The corresponding 6-parameter eigenspinor has the form
\be\label{eta}
\h = \left(\begin{array}{cccccccc}-f_1 & f_2 & f_3 & -f_1 & f_4 & -f_5 & -f_6 & f_4\end{array}\right)^T.
\ee
The exact functional dependence of the parameters $f_i$ on spacetime coordinates is fixed by the second Killing spinor equation (\ref{Ksp2}).

Performing the same decomposition as above and using (\ref{eigen}) we arrive at the following equations for $\k$ and $\h$:
\be
\left(\p_{\underline\m} + \frac14 \,\w_{\underline\m,\r\l} \,\a^{\r\l}\right)\k = \a_{\underline\m} \,\a^5\k,\\
\ee
\be
\left(\p_{\underline i} + \frac14 \,\w_{\underline i,kl} \,\b^{kl}\right)\h = \frac{i}{2} \b_{\underline i} \,\h - \frac{i}{4} F_{\underline i\,j} \,\b^j \b^7\,\h,
\ee
where we underline the world indices and our convention for the spin connection is 
\be
\w_{\underline A,BC} = \frac12 \,e^{\underline D}_B \,e^{\underline E}_C \left( \W_{\underline {ADE}} -\W_{\underline {DEA}}  +\W_{\underline {EAD}} \right),
\ee
\be
\W_{\underline{ABC}} = \p_{\left[\underline{A}\right.} e^D_{\left.\underline{B}\right]} \,e^E_{\underline{C}} \,\h_{DE}.
\ee

The $AdS_4$ Killing spinor equation is easy to solve and the solution $\k$ is 4-parametric:
\be\label{kappa}
\k =\left(
\begin{array}{c}
\k_1 r^{-1/2} \\ \k_2 r^{-1/2} \\ r^{1/2} \left[ -\k_2 (x^0-x^1) + \k_1 x^2 +\k_3 \right] \\ r^{1/2} \left[ \k_1 (x^0+x^1) - \k_2 x^2 +\k_4 \right]
\end{array}
\right).
\ee
Solving the equations for $\h$ is more tedious, but it can be done analytically. The solution is very bulky and is therefore given in the appendix \ref{app-spinors}. The overall result is that the $AdS_4$ part of the Killing spinor $\k$ is 4-parametric, while the $\cp3$ part is 6-parametric. Thus there are $24$ independent Killing spinors in the $\4$ background.

\section{Symmetry superalgebra}

To every supergravity solution is associated the superalgebra of its symmetries, where the even part is formed by the isometries represented by the Killing vectors, while the odd part is the algebra of unbroken supersymmetries given by the Killing spinors. This has the natural interpretation in the superspace picture where an unbroken supersymmetry is a shift invariance in a fermionic direction, i.e. essentially an isomtery as in the bosonic case. The superalgebra multiplication is a commutator (anticommutator) for the elements from within the even (odd) subalgebra. If one considers `commutation' of a Killing vector and a Killing spinor, then one is lead to the notion of a spinorial Lie derivative~\cite{FigueroaO'Farrill:1999va}, as we will see below.

We now need to establish which Killing spinors to use for the T-duality transformation. As long as we have chosen the three isometries generated by (\ref{KV}), the choice of the fermionic symmetries is dictated by the requirement that together they form a commuting subalgebra of the symmetry superalgebra. Bosonic generators (\ref{KV}) of this subalgebra are commuting; our next step will be to choose the fermionic generators (Killing spinors) that commute with these three vectors and finally we shall check the anticommutation of the selected supersymmetries among themselves.

First of all recall that apart from (\ref{KV}) our T-duality setup includes three bosonic dualities along the flat directions of $AdS_4$. Looking at the $AdS_4$ part of the Killing spinor (\ref{kappa}) we see that we must set $\k_{1,2} = 0$ for the product $\k\otimes\h$ to be invariant under the shifts of $x^{0,1,2}$. So what happens to the $\cp3$ part of the Killing spinor?

From the explicit expressions of the $\cp3$ spinors (appendix \ref{app-spinors}) it is not easy to tell what are their transformation properties under the shifts generated by the vectors (\ref{KV}). Therefore we calculate the Lie derivatives of our $\cp3$ Killing spinor fields with respect to the Killing vectors~\cite{FigueroaO'Farrill:1999va}. Lie derivative of a spinor $\h$ with respect to a vector $K$ is given by
\be
\mathcal{L}_K \h = K^i \nabla_i \h + \frac12 \nabla_{\left[i\right.} K_{\left.j\right]} \,\frac12 \b^{ij} \h,
\ee
where of course the covariant derivatives of a vector and of a spinor are taken with respect to the Christoffel and spin connections, correspondingly:
\bsub
\be
\nabla_i K_j = \p_i K_j - \G^k_{ij} K_k,
\ee
\be
\nabla_i \h = \p_i \h - \frac14 \w_{i,jk} \b^{jk} \h.
\ee
\esub

Using the expressions for $K_{(1,2,3)}$ (\ref{KV-real}) and for $\h_{1,\ldots,6}$ ((\ref{eta}) and appendix \ref{app-spinors}, where the spinor $\h_i$ results from keeping only the parameter $h_i = 1$ and setting all the rest to zero), one finds the following algebra:
\bsub
\be\bn
\mathcal{L}_{K_{(1)}} \h_1 &= -\frac{1}{2} (\h_3 - i\h_4 + i\h_5 - \h_6),\\
\mathcal{L}_{K_{(1)}} \h_2 &= -\frac{i}{2} (\h_3 - i\h_4 + i\h_5 - \h_6),\\
\mathcal{L}_{K_{(1)}} \h_3 &=  \frac{1}{4} (\h_1 + i\h_2               ),\\
\mathcal{L}_{K_{(1)}} \h_4 &= -\frac{i}{4} (\h_1 + i\h_2               ),\\
\mathcal{L}_{K_{(1)}} \h_5 &=  \frac{i}{4} (\h_1 + i\h_2               ),\\
\mathcal{L}_{K_{(1)}} \h_6 &= -\frac{1}{4} (\h_1 + i\h_2               ),\\
\en\ee
\be\bn
\mathcal{L}_{K_{(2)}} \h_1 &= 0,\\
\mathcal{L}_{K_{(2)}} \h_2 &= 0,\\
\mathcal{L}_{K_{(2)}} \h_3 &= -\frac{i}{2} (\h_4 - i\h_6               ),\\
\mathcal{L}_{K_{(2)}} \h_4 &=  \frac{i}{2} (\h_3 + i\h_5               ),\\
\mathcal{L}_{K_{(2)}} \h_5 &=  \frac{1}{2} (\h_4 - i\h_6               ),\\
\mathcal{L}_{K_{(2)}} \h_6 &=  \frac{1}{2} (\h_3 + i\h_5               ),\\
\en\ee
\be\bn
\mathcal{L}_{K_{(3)}} \h_1 &= -\frac{i}{2} (\h_3 + i\h_4 + i\h_5 + \h_6),\\
\mathcal{L}_{K_{(3)}} \h_2 &=  \frac{1}{2} (\h_3 + i\h_4 + i\h_5 + \h_6),\\
\mathcal{L}_{K_{(3)}} \h_3 &=  \frac{i}{4} (\h_1 + i\h_2               ),\\
\mathcal{L}_{K_{(3)}} \h_4 &= -\frac{1}{4} (\h_1 + i\h_2               ),\\
\mathcal{L}_{K_{(3)}} \h_5 &= -\frac{1}{4} (\h_1 + i\h_2               ),\\
\mathcal{L}_{K_{(3)}} \h_6 &=  \frac{i}{4} (\h_1 + i\h_2               ).\\
\en\ee
\esub
It is easy to see that there are three linear combinations of the Killing spinors that are invariant under the action of all three vectors:
\be
\h_1 + i\h_2,\quad \h_3 + i\h_5,\quad \h_4 - i\h_6.
\ee
Tensor multiplying these with the two $AdS_4$ spinors ($\k_3, \k_4 \neq 0$ in (\ref{kappa})) we get the six Killing spinors, which is precisely the number needed for the T-duality. Thus the symmetry superalgebra constraints unambiguously fix the fermionic directions to be T-dualized.

It remains to make sure that the corresponding supersymmetries anticommute. The constraint on the spinor $\mathrm{E} = \k\otimes\h$ is given in the appendix (\ref{2a-constraint}) and can be checked straightforwardly. For multiple supersymmetries one has to generalize this to the matrix constraint
\be
\bar{\mathrm{E}}_i \,\G^\m \mathrm{E}_j = 0, \quad i,j = 1,\ldots,6.
\ee

\section{Fermionic T-duality}

Finally we are in a position to calculate the matrix $C_{ij}$, $i,j=1,\ldots,6$:
\be
\p_\m C_{ij} = i\bar{\mathrm{E}}_i \,\G_\m \G^{11}\mathrm{E}_j,
\ee
which is a generalisation of (\ref{2a-c}) for the case of multiple T-dualities. These equations turn out to be consistent, and the solution is (up to integration constants)
\be\label{C2}
C_{\4} = \left(
			\begin{array}{cc}
				0 & \begin{array}{ccc}
							0 & a & b\\
						   -a & 0 & c\\
						   -b & -c & 0
					   \end{array}\\
				\begin{array}{ccc}
							0 & -a & -b\\
						    a & 0 & -c\\
						    b & c & 0
					   \end{array} & 0
			\end{array}
		\right) 
%		+
%		\left(
%			\begin{array}{cccccc}
%				k_{11} & & &\ldots & & k_{16}\\
%				\\
%				\vdots\\
%				\\
%				\\
%				k_{61} & & &\ldots & & k_{66}\\
%			\end{array}
%		\right)
,
\ee
where
\bsub
\label{abc}
\begin{align}
a &= -2\, r\, e^{-\frac{i}{2} (\ps +\c)} \sin{2\m} \sin{\a} \left[ \cos{\frac12 (\th +\f)} + i \sin{\frac12 (\th -\f)} \right],\\
b &=  2\, r\, e^{-\frac{i}{2} (\ps +\c)} \sin{2\m} \sin{\a} \left[ i \cos{\frac12 (\th -\f)} + \sin{\frac12 (\th +\f)} \right],\\
c &= -2\, r\, e^{-\frac{i}{2} \c} \sin{2\m} \cos{\a}.
\end{align}
\esub

The important point to notice here is that the determinant of the matrix (\ref{C2}) is identically zero, irrespective of the values (\ref{abc}). This is the second singularity, which manifests itself in the numerator of the formula (\ref{result}). 

The vanishing of $\det C$ in the present case is to be contrasted with the $\5$ case \cite{Berkovits:2008ic}, where the $C$-matrix has the same algebraic structure (symmetric matrix with off-diagonal antisymmetric blocks). However since in this setup one does $4$ bosonic ($AdS$) dualities and $8$ fermionic ones, $C_{\5}$ is now an $8 \times 8$ matrix:
\be\label{C3}
C_{\5} = \left(
			\begin{array}{cc}
				0 & \begin{array}{cccc}
							0 & a & b & c\\
						   -a & 0 & d & e\\
						   -b & -d & 0 & f\\
						   -c & -e & -f & 0
					   \end{array}\\
				\begin{array}{cccc}
							0 & -a & -b & -c\\
						    a & 0 & -d & -e\\
						    b & d & 0 & -f\\
							c & e & f & 0
					   \end{array} & 0
			\end{array}
		\right)
%		 +
%		\left(
%			\begin{array}{cccccccc}
%				k_{11} & & & \ldots & & & & k_{18}\\
%				\\
%				\\
%				\vdots\\
%				\\
%				\\
%				\\
%				k_{81} & & &\ldots & & & & k_{88}\\
%			\end{array}
%		\right)
,
\ee
where the entries are given by 
\bsub
\begin{align}
a &= 2 R r \sin y^1 \left(\cos y^2 - i \sin y^2 \cos y^3\right),\\
b &= 2 R r \left(i \cos y^1 + \sin y^1 \ldots \sin y^{5}\right),\\
c &= -2 R r \sin y^1 \sin y^2 \sin y^3 \left( \cos y^4 - i \sin y^4 \cos y^{5}\right),\\
d &=  2 R r \sin y^1 \sin y^2 \sin y^3 \left( \cos y^4 + i \sin y^4 \cos y^{5}\right),\\
e &= 2 R r \left(-i \cos y^1 + \sin y^1 \ldots \sin y^{5}\right),\\
f &= -2 R r \sin y^1 \left(\cos y^2 + i \sin y^2 \cos y^3\right).
\end{align}
\esub
Here $r$, as before, is the $AdS$ radial coordinate, $R$ is the $AdS$ radius and the variables $\{y^1,\ldots,y^5\}$ are the standard coordinates on $\mathbb{S}^5$:
\be
ds^2 = (dy^1)^2 + \sin^2 y^1 \left\{ (dy^2)^2 + \sin^2 y^2 \left[ (dy^3)^2 + \ldots\right]\right\}.
\ee
In the $\5$ case, not only is $\det C_{\5}$ nonvanishing, but for these particular values of the entries it can be simplified to
\be
\det C_{\5} = (2 R r)^8.
\ee
This is precisely cancelled by the $4$ $AdS$ dualities.

\section{Summary}

In this chapter we have observed an indeterminate transformation of the dilaton in the $\cp3$ background under a set of bosonic and fermionic T-duality transformations:
\be
e^{2\f'} = e^{2\f} \,\frac{0}{0}. 
\ee
This is to be contrasted with the $\5$ example, where an analogous set of dualities leave the background invariant. 

In the case at hand the degeneracy of the matrix (\ref{C2}) as opposed to (\ref{C3}) is due to their block structure with antisymmetric blocks (an odd-dimensional antisymmetric matrix has zero determinant). In a different setup the structure of the $C$-matrix may be different, as we have seen in the previous chapter \ref{ch:paper1}. We postpone the discussion of the possible reasons and ways around this problem to the conclusion (chapter \ref{ch:conc}).
\chapter{Generalized geometry from D1-brane worldvolume theory}
\label{ch:branes}

\section{Introduction}
Soon after the discovery of $d=11$ supergravity~\cite{Cremmer:1978km} it was observed that upon compactification on tori of various dimensions the theory exhibits lots of global continuous symmetries~\cite{Cremmer:1978ds, Cremmer:1979up, Julia:1980gr, Julia:1982fs, ThierryMieg:1980hq, Cremmer:1980gs}. The symmetries of each compactification (at least up to the dimension 8) can be described by a certain Lie group, whose dimensionality grows as one increases the dimensionality of compactification. The origin and in fact the very existence of these symmetries was obscure from the compactified supergravity action point of view, hence they were commonly referred to as ``hidden'' symmetries. Nowadays they usually go by the name of M-theory dualities or U-duality. The string or M-theoretic interpretation of the lower-dimensional U-duality groups is that duality is properly formulated in terms of the worldvolume of the corresponding extended objects, and it manifests itself in the low-energy effective theory (supergravity) as the U-duality symmetries~\cite{Hull:1994ys}. U-duality groups that arise when $d$ dimensions of 11-dimensional supergravity are compactified are given in Table \ref{table1} ($D$ is the dimension of the representation in which the spin one spacetime fields transform).
\begin{table}[hdtp]
\begin{center}
\begin{tabular}{ccc}
\hline\hline
$d$ & group & $D$ \\ \hline
%1 & $\mathbb{R}$ & 1 \\ \hline
2 & $\mathrm{GL}(2,\mathbb{R})$ & 3 \\ %\hline
3 & $\mathrm{SL}(3,\mathbb{R})\times \mathrm{SL}(2,\mathbb{R})$ & 6 \\% \hline
4 & $\mathrm{SL}(5,\mathbb{R})$ & 10 \\% \hline
5 & $\mathrm{SO}(5,5)$ & 16 \\ %\hline
6 & $\mathrm{E}_{6(+6)}$ & 27 \\ %\hline
7 & $\mathrm{E}_{7(+7)}$ & 56 \\ %\hline
8 & $\mathrm{E}_{8(+8)}$ & - \\ \hline\hline
\end{tabular}
\caption{U-duality groups for compactification of 11-dimensional supergravity on $\mathbb{T}^d$ and their representations.}
\label{table1}
\end{center}
\end{table}

\vspace{-0.7cm}
The U-duality groups contain both T and S-duality symmetries: for the compactifications of type II supergravity on a $d$-torus the U-duality group always contains $\mathrm{SL}(2,\mathbb{R})$ and $\mathrm{O}(d,d;\mathbb{R})$ as its subgroups. These are broken by quantum effects to the discrete $\mathrm{SL}(2,\mathbb{Z})$ and $\mathrm{O}(d,d;\mathbb{Z})$ where from the string-theoretic viewpoint $\mathrm{SL}(2,\mathbb{Z})$ is the S-duality group, while $\mathrm{O}(d,d;\mathbb{Z})$ is the standard T-duality group.

The generalized geometry framework allows a possibility of making the U-duality symmetries manifest on the Lagrangian level. This can be achieved if one combines the supergravity background fields into objects that transform covariantly under the U-dualities. This has been achieved initially on the T-duality level by employing Hitchin's generalized geometry \cite{Hitchin:2004ut, Hitchin:2005in, Gualtieri:2003dx}, where one augments a $d$-dimensional tangent space at each point with a cotangent space. The coordinates of the latter describe string winding modes, and the resulting theory is essentially formulated for a fibrewise sum $$TM \oplus T^*M.$$ The corresponding generalized metric is a $2d \times 2d$ matrix:
\be\label{gen-metric-T}
\mathbf{M} = \left(\begin{array}{cc} g_{\m\n} - b_{\m\a}g^{\a\b}b_{\b\n} & b_{\m\a}g^{\a\n} \\ -g^{\m\a}b_{\a\n} & g^{\m\n} \end{array}\right).
\ee
It unifies the metric and the $b$-field precisely in the way required to make the theory covariant under the T-duality group $O(d,d;\mathbb{Z})$ \cite{Duff:1989tf, Tseytlin:1990nb, Tseytlin:1990va, Hull:2004in, Hull:2006qs, Hull:2006va, Hull:2009mi, Thompson:2010sr}.

To give a U-duality symmetric formulation one would need to work with M-theory fundamental objects \cite{Berman:2007bv} and their winding modes, rather than fundamental strings. The corresponding generalized geometries have been considered in \cite{Hull:2007zu, Pacheco:2008ps}. A U-duality covariant action of $d=11$ supergravity has been constructed recently in~\cite{Berman:2010is}. The single U-duality covariant object unifying the metric and the 3-form potential of $d=11$ supergravity was in that case the metric on the generalized tangent bundle $$TM \oplus \L^2 T^*M,$$ where fibrewise summation is assumed:
\be\label{gen-metric-M}
\mathbf{M} = \left(\begin{array}{cc} g_{\m\n} + C_{\m\a\b}g^{\a\b,\g\d}C_{\g\d\n} & C_{\m\a\b}g^{\a\b,\s\t} \\ g^{\s\t,\a\b}C_{\a\b\m} & g^{\m\n,\s\t} \end{array}\right),
\ee
where $g^{\a\b,\g\d} = \frac12 (g^{\a\g} g^{\b\d} - g^{\a\d} g^{\b\g})$. In $d=4$ the generalized tangent space is 10-dimensional: $$\mathrm{dim}\, T_x M + \mathrm{dim}\, \L^2 T^*M = 4+6 = 10,$$ which is in line with the dimension of $\mathrm{SL}(5,\mathbb{R})$ representation given in Table \ref{table1}. The fibre at each point of the target spacetime is a direct sum of a tangent space and of the space of 2-forms. The latter correspond to the winding modes of the M2-brane, which is the only M-theory object relevant for compactifications on tori of dimension up to four (for $d>4$ M5-brane wrappings become relevant \cite{Berman:2011pe}). In a similar manner we can derive the representations of the U-duality groups in other dimensions by augmenting the tangent space at a point $x$ with $\L^2 T_x^*M$ for $d>1$, or $\L^2 T_x^*M \oplus \L^5 T_x^*M$ for $d>4$, which corresponds to the possibility that the M-theory branes may wrap the corresponding compactified spaces.  The results are shown in Table \ref{table2}\footnote{Much related work on string and M-theory generalized geometry has appeared after this thesis has been submitted, in particular \cite{Jeon:2011cn, Copland:2011yh, Andriot:2011uh, Thompson:2011uw, Hohm:2011dv, Albertsson:2011ux, Coimbra:2011nw, Hohm:2011cp, Kan:2011vg, Aldazabal:2011nj, Jeon:2011vx, Berman:2011kg, Berman:2011cg, Berman:2011jh, West:2011mm, Copland:2011wx, Hohm:2011nu, Jeon:2011sq}}.
\begin{table}[hdtp]
\begin{center}
\begin{tabular}{ccccc}
\hline\hline
$d$ & $\dim T_x M$ & $\dim \L^2 T_x^*M$ & $\dim \L^5 T_x^*M$ & total \\ \hline
1 & 1 & --- & --- & 1 \\ %\hline
2 & 2 & 1 & --- & 3  \\ %\hline
3 & 3 & 3 & --- & 6   \\ %\hline
4 & 4 & 6 & --- & 10   \\ %\hline
5 & 5 & 10 & 1 & 16  \\ %\hline
6 & 6 & 15 & 6 & 27 \\ \hline\hline
\end{tabular}
\caption{M-theory brane wrapping coordinates for $d$ compactified dimensions.}
\label{table2}
\end{center} 
\end{table}

In principle, there should be a way to extract this information from 10-dimensional string theory, provided one considers not only the fundamental strings, but also the nonperturbative states of the theory, i.e. the D-branes. In IIB string theory for compactifications on $\mathbb{T}^d$ with $d<6$ we may postulate a generalized tangent space of the form
\be\label{genspace1}
(T_x \oplus \L^1 T_x^* \oplus \L^1 T_x^* \oplus \L^3 T_x^* \oplus \L^5 T_x^* \oplus \L^5 T_x^*) M,
\ee
which takes care of the wrappings for F1-strings, D1-, D3-, and D5-branes, and NS5-branes. This space has dimensions precisely matching those shown in the Table \ref{table2} (note that there is a mismatch of one because of the difference between the $d=11$ M-theory and $d=10$ string theory), see Table \ref{table3}.
\begin{table}[hdtp]
\begin{center}
\begin{tabular}{cccccc}
\hline\hline
$d$ & $T_x M$ & $ 2 [\L^1 T_x^*M]$ & $\L^3 T_x^*M$ & $ 2 [\L^5 T_x^*M]$ & total \\ \hline
1 & 1 & 2  & --- & --- & 3    \\ %\hline
2 & 2 & 4  & --- & --- & 6    \\ %\hline
3 & 3 & 6  &  1  & --- & 10   \\ %\hline
4 & 4 & 8  &  4  & --- & 16   \\ %\hline
5 & 5 & 10 &  10 &  2  & 27   \\ \hline\hline
\end{tabular}
\caption{IIB string theory objects wrapping coordinates for $d$ compactified dimensions.}
\label{table3}
\end{center} 
\end{table}

IIA theory comes eqiupped with a generalized tangent space of the form
\be\label{genspace2}
\mathbb{R} \oplus (T_x \oplus \L^1 T_x^* \oplus \L^2 T_x^* \oplus \L^4 T_x^* \oplus \L^5 T_x^*) M,
\ee
where we take into account the wrapping coordinates of the F1-string, D2- and D4-branes, NS5-brane, and an extra summand of $\mathbb{R}$ is responsible for the dual coordinate of a D0-brane (a zero-form, i.e. a scalar). Once again the dimensions of this space (Table \ref{table4}) are in line with the representations of M-theory duality.
\begin{table}[hdtp]
\begin{center}
\begin{tabular}{cccccccc}
\hline\hline
$d$ & $\mathbb{R}$ & $T_x M$ & $\L^1 T_x^*M$ & $\L^2 T_x^*M$ & $\L^4 T_x^*M$ & $\L^5 T_x^*M$ & total \\ \hline
1 & 1 & 1 & 1 & --- & --- & --- & 3    \\ %\hline
2 & 1 & 2 & 2 &  1  & --- & --- & 6    \\ %\hline
3 & 1 & 3 & 3 &  3  & --- & --- & 10   \\ %\hline
4 & 1 & 4 & 4 &  6  &  1  & --- & 16   \\ %\hline
5 & 1 & 5 & 5 &  10 &  5  &  1  & 27   \\ \hline\hline
\end{tabular}
\caption{IIA string theory objects wrapping coordinates for $d$ compactified dimensions.}
\label{table4}
\end{center} 
\end{table}

\vspace{-0.7cm}
This leads us to conjecture that if one encodes all the dynamics of the string theory objects listed above in a single structure similar to the generalized metrics (\ref{gen-metric-T}, \ref{gen-metric-M}), then this may lead to a possibility of reformulation of 10-dimensional IIA/B supergravity in a U-duality covariant way. The T-duality-covariant generalized metric (\ref{gen-metric-T}) unifies the metric with the $b$-field; since we know that U-duality groups include T and S-dualities, and the latter mix NSNS and RR fields, we expect the type II string theory U-duality-covariant generalized metric to combine NSNS and RR fields. As we will see, this is achieved quite naturally by considering the D-brane worldvolume actions.

We proceed by briefly recapitulating the approach of~\cite{Duff:1990hn} to derivation of the T-duality-covariant generalized metric of a fundamental string (\ref{gen-metric-T}), and then extend the result to include the D-string. The hope is that eventually a single S-duality covariant result can be written down, describing both strings simultaneously. It is likely that an S-duality covariant description of a $(p,q)$-string will be needed in order to achieve such a goal. One then would need to proceed with including the higher dimensional D-branes into the game, in order to complete the spaces (\ref{genspace1}), (\ref{genspace2}).

\section{Generalized metric for an F1-string}

Taking the fundamental string to be the only object in the theory we see that the dual winding coordinates are 1-forms, and hence the relevant generalized tangent space is $$T_x M \oplus T_x^*M.$$ This is classical generalized geometry in the sense of Hitchin~\cite{Hitchin:2004ut}, i.e. a simple tangent bundle is replaced by a fibrewise sum of tangent and cotangent bundles. The metric for such a space arises naturally if one considers the worldvolume equations of motion and Bianchi identities and writes them down in a unified fashion. This task is simplified by making use of several alternative Lagrangian descriptions of a string, as we will now show. In this section we closely follow~\cite{Duff:1990hn}. The Lagrangian we begin with is given by (\ref{p-brane}) for $p=1$:
\be\label{L}
L = \frac12 \sqrt{-\g} \g^{ij} \p_i x^\m \p_j x^\n g_{\m\n} + \frac12 \e^{ij} \p_i x^\m \p_j x^\n b_{\m\n}.
\ee
We introduce the following notation:
\be\label{defs}\bn
&\mathfrak{F}^{i\m} = \sqrt{-\g} \g^{ij} \p_j x^\m,\quad &\tilde{\mathfrak{F}}^{i\m} = \e^{ij} \p_j x^\m,\\
&\mathfrak{G}^i_{\m} = \sqrt{-\g} \g^{ij} \p_j y_{\m},\quad &\tilde{\mathfrak{G}}^i_\m = \e^{ij} \p_j y_{\m}.
\en\ee
In this chapter $i,j$ are the worldvolume indices, running over the two values $0,1$, while $\m,\n$ are spacetime indices, see appendix~\ref{actions}. Equations of motion for $x^\m$ can then be written as
\be\label{x-eom}
\p_i \left( \mathfrak{F}^{i\n} g_{\m\n} + \tilde{\mathfrak{F}}^{i\n} b_{\m\n} \right) = 0,
\ee
while the Bianchi identities involving $x^\m$ follow from (\ref{defs}):
\be\label{bianchi}
\p_i \tilde{\mathfrak{F}}^{i\m} = 0.
\ee
The nonlinear action (\ref{p-brane}) can equivalently be represented by any of the two first order actions, which describe the system in the two dual frames. These are written in terms of the dynamical fields $F^\m_i$, $x^\m$, and a dual 1-form coordinate $y_{\m}$:
\be\bn
L_x = -&\frac12 \sqrt{-\g} \g^{ij} F_i^\m F_j^\n g_{\m\n} - \frac12 \e^{ij} F_i^\m F_j^\n b_{\m\n} +\\
+ &\p_i x^\m \left( \sqrt{-\g} \g^{ij} F_j^\n g_{\m\n} + \e^{ij} F_j^\n b_{\m\n} \right),
\en\ee
\be\bn
L_y = &\frac12 \sqrt{-\g} \g^{ij} F_i^\m F_j^\n g_{\m\n}+ \frac12 \e^{ij} F_i^\m F_j^\n b_{\m\n} +\\
+& \p_i y_{\m} \e^{ij} F_j^\m.
\en\ee
The $F_i^\m$ equation of motion for $L_x$ is $$F_i^\m = \p_i x^\m.$$ Substituting this back into the Lagrangian we get precisely $L$~(\ref{L}), thus establishing classical equivalence of the actions built out of $L_x$ and $L$. Of course, the $x^\m$ equation of motion for $L_x$
\be\label{x-eom-2}
\p_i (\sqrt{-\g} \g^{ij} F_j^\n g_{\m\n} + \e^{ij} F_j^\n b_{\m\n}) = 0 
\ee
also reduces to the one that follows from $L$ (\ref{x-eom}) upon substitution of the $F_i^\m$ field equation.

As regards $L_y$, it is a dual Lagrangian: the equation for $y_\m$ encoded in $L_y$ is what used to be the Bianchi identity in the $L_x$ description (\ref{bianchi}):
\be
\e^{ij} \p_i F_j^\m = 0\quad \Leftarrow \quad F_j^\m = \p_j x^\m.
\ee
To ensure that the relation between Bianchi identities and field equations is symmetric with respect to $L_x$ and $L_y$, consider the equation for $F_i^\m$ from $L_y$:
\be\label{uravnenie}
\e^{ij} \p_j y_\m = \sqrt{-\g} \g^{ij} F_j^\n g_{\m\n} + \e^{ij} F_j^\n b_{\m\n}.
\ee
Comparing this to (\ref{x-eom-2}), we see that the latter, which is the $x^\m$ equation following from $L_x$, implies the Bianchi identity for $L_y$:
\be
\delta_x L_x = 0\quad \Rightarrow \quad \e^{ij} \p_i \p_j y_\m \equiv 0.
\ee

We now proceed to show that all of the information given above can be summarized concisely, and that this results in a generalized metric structure. As one can check, the equation (\ref{uravnenie}) has the following solution:
\be\label{reshenie}
F_i^\m = p^{\m\n} \frac{1}{\sqrt{-\g}} \g_{ij} \e^{jk} \p_k y_\n + q^{\m\n} \p_i y_\n,
\ee
where $p_{\m\n}$ (symmetric) and $q^{\m\n}$ (antisymmetric) are defined by the relations
\be\bn
p_{\m\n} &= g_{\m\n} - b_{\m\a} g^{\a\b} b_{\b\n},\\
p_{\m\n} q^{\a\n} &= b_{\m\n} g^{\a\n},
\en\ee
and $p^{\m\n}$ is defined to be the inverse of $p_{\m\n}$: $p_{\m\a} p^{\a\n} = \d^\n_\m$.

Recalling the definitions (\ref{defs}) we notice that we can rewrite (\ref{uravnenie}, \ref{reshenie}) as
\be
\tilde{\mathfrak{G}}^i_\m = g_{\m\n} \mathfrak{F}^{i\n} + b_{\m\n} \tilde{\mathfrak{F}}^{i\n},
\ee
and
\be
\tilde{\mathfrak{G}}^i_\m = p_{\m\n} \mathfrak{F}^{i\n} + b_{\m\n} g^{\n\a} \mathfrak{G}^i_\a.
\ee
From these two we can express $\tilde{\mathfrak{F}}$ in terms of $\mathfrak{F,G}$. As a result we arrive at the following relationship between the worldvolume 1-forms ($\star$ is the worldvolume Hodge operator):
\be
\left( \begin{array}{c}\star dy \\ \star dx \end{array} \right) \sim
\left( \begin{array}{c}\tilde{\mathfrak{G}}\\ \tilde{\mathfrak{F}}\end{array} \right) =
\mathbf{M} \left( \begin{array}{c}\mathfrak{F}\\ \mathfrak{G}\end{array} \right) \sim
\mathbf{M} \left( \begin{array}{c} dx\\ dy \end{array} \right).
\ee
The generalized metric $\mathbf{M}$ is given by
\be\label{f1-gen-metric}
\mathbf{M} = \left(\begin{array}{cc} p_{\m\n} & b_{\m\r} g^{\r\n}\\ -g^{\m\r} b_{\r\n} & g^{\m\n} \end{array}\right).
\ee

\section{Generalized metric for a D1-brane}

We start with a Howe-Tucker form of the D1-brane action (\ref{d1-pol}) and repeat the manipulations of the previous section. Our initial Lagrangian is
\be\bn
L &= \frac12 e^{-\f} \sqrt{-H} H^{ij} \left[ \p_i x^\m \p_j x^\n (g_{\m\n} + b_{\m\n}) + 2\pi\a' (dA)_{ij} \right] +\\&+ \frac12 \e^{ij} \p_i x^\m \p_j x^\n C_{\m\n} + \frac12 \e^{ij} \mathcal{F}_{ij} C,
\en\ee
where in addition to the fundamental string worldvolume fields $\g_{ij}$ and $x^\m$ we have the worldvolume gauge field potential $A_i$, spacetime RR fields $C_{\m\n}$ and $C$ (the RR scalar), and an antisymmetric rank 2 worldvolume field that together with $\g_{ij}$ comprises $H_{ij}$ (for details see appendix \ref{actions}). We can rewrite the Lagrangian in the first order form by introducing an extra worldvolume field $F_i^\m$:
\be\bn
L_x =& -\frac12 e^{-\f} \sqrt{-H} H^{ij} F_i^\m F_j^\n (g_{\m\n} + b_{\m\n}) + \pi\a' (dA)_{ij} (e^{-\f} \sqrt{-H} H^{ij} + \e^{ij} C) -\\&- \frac12 \e^{ij} F_i^\m F_j^\n (C_{\m\n} + b_{\m\n} C) +\\&+ \p_i x^\m \left[  e^{-\f} \sqrt{-H} H^{ij} F_j^\n (g_{\m\n} + b_{\m\n}) + \e^{ij} F_j^\n (C_{\m\n} + b_{\m\n} C) \right].
\en\ee
This Lagrangian transforms precisely into $L$ upon substitution of the $F_i^\m$ field equation $F_i^\m = \p_i x^\m$. The dual Lagrangian is given by
\be\bn
L_y &= \frac12 e^{-\f} \sqrt{-H} H^{ij} F_i^\m F_j^\n (g_{\m\n} + b_{\m\n}) + \pi\a' (dA)_{ij} (e^{-\f} \sqrt{-H} H^{ij} + \e^{ij} C) +\\&+ \frac12 \e^{ij} F_i^\m F_j^\n (C_{\m\n} + b_{\m\n} C) + \p_i y_\m \e^{ij} F_j^\m,
\en\ee
where a dual D1-brane wrapping coordinate $y_\m$ was introduced. The situation with the field equations for $L_x$ being the Bianchi identities for $L_y$ and vice versa is exactly as in the fundamental string case. In particular, the $y_\m$ equation of motion implies $F_i^\m = \p_i x^\m$, and the equation of motion itself then turns into the Bianchi identity for $x^\m$.

In our search of the generalized metric we now need to consider the $F_i^\m$ equation of motion for $L_y$ together with its solution. The field equation is
\be\label{ur}
\e^{ij} \p_j y_\m = e^{-\f} \sqrt{-H} H^{ij} \p_j x^\n (g_{\m\n} + b_{\m\n}) + \e^{ij} \p_j x^\n (C_{\m\n} + b_{\m\n} C),
\ee
which can be solved by
\be\label{resh}
\p_i x^\m = \frac{e^{-\f}}{\sqrt{-H}}\, p^{\m\n} H_{ij} \e^{jk} \p_k y_\n + q^{\m\n} \p_i y_\n,
\ee
where now
\be
p_{\m\n} = e^{-2\f} (g+b)_{\m\n} - (C_{\m\a} + b_{\m\a} C) (g+b)^{-1\,\a\b} (C_{\b\n} + b_{\b\n} C),
\ee
\be\label{id}
q^{\m\a} p_{\a\n} = -(g+b)^{-1\,\m\a} (C_{\a\n} + b_{\a\n} C),
\ee
and as before $p_{\m\a} p^{\a\n} = \d^\n_\m$. Note, however, that both $p_{\m\n}$ and $q^{\m\n}$ have now lost their (anti)symmetry properties. Nevertheless, one can still find a simple expression for a product with reversed order $p_{\m\a}q^{\a\n}$, which will be important in what follows:
\be\label{id2}
p_{\m\a} q^{\a\n} = -(C_{\m\a} + b_{\m\a} C) (g+b)^{-1\,\a\n}.
\ee

From (\ref{resh}) we find that
\be\label{resh-2}
\e^{ij} \p_j y_\m = e^\f p_{\m\n} \sqrt{-H}H^{ij} \p_j x^\n - e^\f p_{\m\n} q^{\n\r} \sqrt{-H}H^{ij} \p_j y_\r.
\ee
Comparing this to (\ref{ur}) yields
\be\bn
\e^{ij} \p_j x^\n (C_{\m\n} + b_{\m\n} C) = &-e^\f \sqrt{-H} H^{ij} \p_j x^\r (C_{\m\n} + b_{\m\n} C) (g+b)^{-1\,\n\b} (C_{\b\r} + b_{\b\r} C) -\\&- e^\f p_{\m\n} q^{\n\r} \sqrt{-H} H^{ij} \p_j y_\r \\= &-e^\f \sqrt{-H} H^{ij} \p_j x^\r (C_{\m\n} + b_{\m\n} C) (g+b)^{-1\,\n\b} (C_{\b\r} + b_{\b\r} C) -\\&+ e^\f \sqrt{-H} H^{ij} \p_j y_\r (C_{\m\n} + b_{\m\n} C) (g+b)^{-1\,\n\r}.
\en\ee
We multiply this by $(C_2 + b_2 C)^{-1}$ which yields
\be\bn
\e^{ij} \p_j x^\m = &-e^\f \sqrt{-H} H^{ij} \p_j x^\r (g+b)^{-1\,\m\n} (C_{\n\r} + b_{\n\r} C) -\\&+ e^\f \sqrt{-H} H^{ij} \p_j y_\r (g+b)^{-1\,\m\r}.
\en\ee
The latter, together with (\ref{resh-2}), is sufficient to write down the generalized metric:
\be\label{GM}
\left(\begin{array}{c} \tilde{\mathfrak{G}}^i_\m \\ \tilde{\mathfrak{F}}^{i\m} \end{array}\right) =
e^\f \left(\begin{array}{cc} p_{\m\r} & -p_{\m\n} q^{\n\r} \\ q^{\m\n}p_{\n\r} & (g+b)^{-1\,\m\r} \end{array}\right) \left(\begin{array}{c} \mathfrak{F}^{i\r} \\ \mathfrak{G}^i_\r \end{array}\right),
\ee
where the products $p_{\m\n} q^{\n\r}, q^{\m\n}p_{\n\r}$ are given by (\ref{id}), (\ref{id2}), and the Gothic characters are the same as in (\ref{defs}) with $\g$ traded for $H$.

Note that due to the presence of the antisymmetric rank two worldvolume field in the D-brane action the matrix in (\ref{GM}) is no longer symmetric, thus no longer {\it a metric}. If we split it into symmetric and antisymmetric parts, then one can call the symmetric part the generalized metric:
\be\label{GM2}
\mathbf{M}_s = e^\f \left(\begin{array}{cc} p'_{\m\n} & (C_2 + b C)_{\m\a} (g+b)_s^{-1\,\a\n} \\ -(g+b)_s^{-1\,\m\a} (C_2 + b C)_{\a\n} & (g+b)_s^{-1\,\m\n} \end{array}\right),
\ee
where 
\be
p'_{\m\n} = e^{-2\f} g_{\m\n} - (C_{\m\a} + b_{\m\a} C) (g+b)_s^{-1\,\a\b} (C_{\b\n} + b_{\b\n} C),
\ee
and
\be
(g+b)_s^{-1\,\m\n} = (g+b)^{-1\,\m\a} g_{\a\b} (g+b)^{-1\,\b\n}
\ee
is the symmetric part of $(g+b)^{-1\,\m\n}$.

One can obtain the generalized metric (\ref{GM2}) from the fundamental string generalized metric (\ref{f1-gen-metric}) by replacing:
\be\label{repl}\bn
g^{\m\n} &\rightarrow e^\f (g+b)_s^{-1\,\m\n},\\
g_{\m\n} &\rightarrow e^{-\f} (g+b)_{s\,\m\n} = e^{-\f} g_{\m\n},\\
b_{\m\n} &\rightarrow C_{\m\n} + b_{\m\n}C.
\en\ee
The new metric 
\be\label{open}
(g+b)_s^{-1\,\m\n}
\ee
is the open string metric \cite{Seiberg:1999vs}. Note that in our interpretation it only appears with the upper indices, whereas $g_{\m\n}$ simply gets multiplied by $e^{-\f}$. In other words, a proper inverse for the contravariant open string metric (\ref{open}) would be 
\be
g_{\m\n} - b_{\m\a}g^{\a\b}b_{\b\n},
\ee
rather than $(g+b)_{s\,\m\n} = g_{\m\n}$ as in (\ref{repl}).

Note that the matrix in (\ref{GM}) also has an antisymmetric part:
\be
\mathbf{M}_a = e^\f \left(\begin{array}{cc} p''_{\m\n} & (C_2 + b C)_{\m\a} (g+b)_a^{-1\,\a\n} \\ -(g+b)_a^{-1\,\m\a} (C_2 + b C)_{\a\n} & (g+b)_a^{-1\,\m\n} \end{array}\right),
\ee
where 
\be
p''_{\m\n} = e^{-2\f} b_{\m\n} - (C_{\m\a} + b_{\m\a} C) (g+b)_a^{-1\,\a\b} (C_{\b\n} + b_{\b\n} C).
\ee
This corresponds to the anticommutativity parameter of \cite{Seiberg:1999vs}. One could speculate that some kind of anticommutativity arises in the generalized geometry for D-branes. Within the generic logic of this chapter this is quite an unexpected and interesting by-product, worth to be studied separately.
\chapter{Conclusion}
\label{ch:conc}

In this thesis we studied several aspects of string theory dualities from the worldvolume action perspective, and how they manifest themselves in the effective low energy supergravity theories. This study has taken the form of both purely theoretical and ``applied'' investigation (to the extent to which modern string theory may be applied). On the theoretical side we considered the fermionic T-duality transformation in type IIB supergravity (chapter \ref{ch:paper1}) and some aspects of U-duality in 10-dimensional string theory (chapter \ref{ch:branes}); on the applied side we studied an application of fermionic T-duality in the AdS/CFT approach to scattering amplitude dualities in $d=3$ $\mathcal{N}=6$ gauge theory (chapter \ref{ch:paper2}).

In the chapter \ref{ch:paper1} we have demonstrated the basic features of fermionic T-duality transformation by generating a few supersymmetric solutions of complexified supergravity. In this study we observe several interesting properties, some of which are quite unexpected. We have seen that fermionic T-duality does not commute with bosonic T-duality, because new components of the RR forms that arise after the duality transformation depend more on the structure of the Killing spinors, rather than Killing vectors and the corresponding spacetime symmetries. In retrospect this should not be a surprise since it is known that supersymmetries and isometries do not commute either. One can also think of examples where T-duality breaks supersymmetry (at the level of supergravity).

Several examples of real fermionic T-dual backgrounds have been found in the pp-wave case. These are however trivially related to the pp-wave itself (i.e. one can talk of a self-duality in a certain sense\footnote{After this thesis was submitted, the work investigating self-duality of a general class of pp-wave backgrounds has been completed \cite{Bakhmatov:2011aa}.}). Fermionic T-duals in general have complex RR backgrounds because of the constraint that the supersymmetries being dualized belong to the abelian subgroup of the symmetry supergroup; for that reason it would be interesting to consider the possibility of relaxing the abelian constraint in order to have a manifestly real nonabelian fermionic T-duality transformation.

Following this in the chapter \ref{ch:paper2} it was shown that under the combination of bosonic and fermionic T-dualities in the directions given by the three trivial $AdS_4$ isometries, three complexified $\cp3$ isometries and six complexified supersymmetries the transformation of the dilaton is indeterminate:
\be
e^{2\f'} = e^{2\f} \,\frac{0}{0}. 
\ee
This provides an alternative point of view on T-dualizing $\4$ background that has been done recently by Adam, Dekel, and Oz~\cite{Adam:2010hh} in the supercoset formulation of the sigma-model. The authors of~\cite{Adam:2010hh} who have also encountered a singularity suggest as one of the possible explanations the $\k$-symmetry gauge fixing that is used to obtain the coset model~\cite{Gomis:2008jt}. This may break dual superconformal invariance of the corresponding field theory since certain string configurations cannot be represented after the gauge fixing. The analysis presented here uses a supergravity description of the superstring, which certainly does not have this truncation, and yet the singularity persists. It is yet to be understood what makes singular transformations possible, and in particular what is the role of complexification of the fermionic symmetries that is obligatory for doing fermionic T-duality.

Perhaps a way to eliminate this $\frac{0}{0}$ ambiguity would be to consider a deformed $\4$ background, the deformation being parameterized by some $\l$, such that the dependence on the deformation parameter $e^{2\f'} = f(\l)\,e^{2\f}$ would have a well-defined limit as one removes the deformation $\lim_{\l\rightarrow 0} f(\l)$. In order to ensure that taking this limit sends us back to the initial background one would also require that the $\l$-deformation commutes with T-duality.

Most likely such a deformation would require giving the dilaton some nontrivial coordinate dependence. The dilaton equation of motion in our conventions is 
\be
R = 4(\p\f)^2 - 4\nabla^2 \f
\ee
(for a vanishing $B$-field). If we keep the dilaton constant, then the requirement that the $AdS$ part of the geometry be preserved will only allow for the deformations of the $\mathbb{C}P^3$ part that preserve $R=0$, which is problematic. One can also consider the Killing spinor equation (\ref{Ksp1}), which in the ABJM background reduces to the eigenspinor condition (\ref{eigen}). If one were to deform the RR 2-form, the eigenspinor condition would be broken, and for some supersymmetry to be preserved one would have to introduce the dilaton into the game. With nontrivial dilaton the equation (\ref{eigen}) gets modified to
\be
\left[ k \b^i \p_i \f - e^\f \left( Q + 2 \right) \right] \b^7 \h= 0,
\ee
where, as before, $Q = \frac12 F_{ij} \b^{ij} \b^7$, and we have absorbed the numerical factors that depend on the supergravity conventions into the constant $k$. An appropriate relative normalization of $F_2$ and $F_4$ is also assumed. It is possible that the dilaton field with nontrivial dependence on the internal manifold could allow for some supersymmetry to be preserved under the deformation.

A candidate recipe for the deformation is the TsT transformation \cite{Lunin:2005jy, Berman:2007tf}, which gives the beta-deformed $\4$ theory described in \cite{Imeroni:2008cr}. In order for the Killing vectors to be preserved under the beta-deformation, one may carry out the beta-deformation with respect to these Killing vectors. Therefore if we beta-deform the $\4$ background using the directions (\ref{KV}), we can then use the same Killing vectors for the T-duality. However the $dw\,dw$ block in (\ref{metric}) is not affected by such a beta-deformation, which means that the corresponding determinant is still zero. Thus the use of the TsT transformation for the deformation purposes in our setup is problematic.

Finally, in the chapter \ref{ch:branes} we have outlined the small step towards the remote aim of reformulation of type II supergravity in a U-duality covariant manner. Introducing the D1-brane into the game has lead to the generalized metric depending explicitly on the RR fields of type IIB theory. This is a natural feature for the theory that respects S-duality, whereby the NSNS and RR fluxes get mixed.

Thinking of the S-duality relation between the F1 and D1-strings we note, however, that a field redefinition such as (\ref{repl}) does not boil down just to an S-duality transformation. It is plausible that one could achieve a covariant formulation of the generalized geometry if one starts from the covariant $(p,q)$-string action formalism \cite{Townsend:1997kr, Cederwall:1997ts, Cederwall:1998xr, Bergshoeff:2006gs, Bandos:2000hw}.

Needless to say that a lot is still to be done on the way to the full U-duality covariant description of type II supergravity. One should carry out the procedure described in this chapter for the full spectrum of string theory objects as shown in (\ref{genspace1}), (\ref{genspace2}). Then it will be required to unify all of these in a framework of a unique generalized metric, such that the supergravity action could be written in terms of a single covariant object.

Aside from the main goal of the chapter we have found an antisymmetric contribution to the generalized metric of a D1-brane. This may be a manifestation of (generalized?) spacetime noncommutativity \cite{Seiberg:1999vs} and deserves further investigation.

\appendix

\chapter{IIB supergravity conventions}
\label{app:2b}

We will give the relevant action and equations of motion for IIB
supergravity so that all our conventions are transparent. Our metric signature is mostly plus, $(-\,+\ldots +)$; antisymmetric Levi-Civita tensor is defined with $\e_{0\ldots 9} = 1$.
Apart from the metric, which is represented by $g_{mn}$, the bosonic
field content of type IIB supergravity is given by two real scalars,
dilaton $\f$ and RR scalar $C_{(0)}$, two real
antisymmetric second-rank tensors $b$ and $C_{(2)}$ and a fourth-rank real
tensor $C_{(4)}$, whose field stregth $F_{(5)}=dC_{(4)}$ is self-dual:
\be
F_{m_1\ldots m_5} =  \frac{1}{5!} \e_{m_1\ldots m_5 n_1\ldots n_5} F^{n_1\ldots n_5}.
\ee
From string theory point of view, the fields $C_{(0)}, C_{(2)}$, and $C_{(4)}$ are
potentials of the RR fields $F_{(n+1)} = dC_{(n)}$. Three remaining fields
$g, b$, and $\f$ belong to the NSNS sector of type IIB superstring.

The action of type IIB supergravity in the string frame is a sum of three terms
\be
\label{action}
S = S_{NSNS} + S_{RR} + S_{CS},
\ee
where
\begin{align}
S_{NSNS} &= \frac{1}{2\k^2}\int d^{10}x \sqrt{|g|} \,e^{-2\f} \left[R+4(\p\f)^2-\frac12\frac{1}{3!}{H_{(3)}}^2\right],\\
S_{RR} &= -\frac{1}{4\k^2}\int d^{10}x \sqrt{|g|} \left[{F_{(1)}}^2 + \frac{1}{3!}{\tilde F_{(3)}}\!^2 +  \frac12\frac{1}{5!} {\tilde F_{(5)}}\!^2 \right],\\
S_{CS} &= - \frac{1}{4\k^2}\int C_{(4)}\wedge H_{(3)}\wedge F_{(3)}.
\end{align}
Here $H_{(3)} = db$ is the field strength of the NSNS antisymmetric tensor field, and we use a common notation ${F_{(n)}}^2 = F_{\m_1\ldots \m_n} F_{n_1\ldots  n_n} g^{m_1  n_1} \ldots g^{m_n  n_n}$. Modified field strengths $\tilde F_{(n)}$ are used in $S_{RR}$, and only there:
\bsub
\label{mod}
\begin{align}
\tilde F_{(3)} &= F_{(3)} - C_{(0)} H_{(3)},\\
\tilde F_{(5)} &= F_{(5)} - \frac12 C_{(2)} \wedge H_{(3)} + \frac12 b_{(2)} \wedge F_{(3)}.
\end{align}
\esub
Note that these reduce to ordinary $F_{(n)}$ if the $b$-field is zero.

The equations of motion of the two scalars in the
theory (\ref{action}) are the simplest. The dilaton equation reads
\be
\label{phi}
R = 4(\p\f)^2 - 4\nabla^2 \f + \frac12\frac{{H_{(3)}}^2}{2},
\ee
and the RR scalar field equation is
\be
\label{C_0}
\nabla^2 C_{(0)} + \frac{1}{3!} H_{(3)} \tilde F_{(3)} = 0.
\ee

The equations for $B_2, C_{(2)}$, and $C_{(4)}$ are respectively (note that the first two equations have been simplified by substitution of the third one):
\begin{align}
\label{B_2}
\nabla_m&\left[e^{-2\f} H - C_{(0)} \tilde F\right]^{a b m} \nonumber\\
& = \frac12\frac{1}{3!} \tilde F^{a b m n l} F_{ m n l} - \frac{1}{2\sqrt{|g|}}\frac{1}{5!}\frac{1}{3!} \e^{ a b m_1\ldots  m_5  n_1\ldots  n_3} \tilde F_{ m_1\ldots  m_5} F_{ n_1\ldots  n_3};
\end{align}
\begin{align}
\label{C_2}
\nabla_m &\tilde F^{ a b m} \nonumber\\
& = - \frac12\frac{1}{3!} \tilde F^{ a b m n l} H_{ m n l} + \frac{1}{2\sqrt{|g|}}\frac{1}{5!}\frac{1}{3!} \e^{ a b m_1\ldots  m_5  n_1\ldots  n_3} \tilde F_{ m_1\ldots  m_5} H_{ n_1\ldots  n_3};
\end{align}
\be
\label{C_4}
\nabla_m \tilde F^{ m n_1\ldots  n_4} = \frac{1}{\sqrt{|g|}}\frac{1}{3!}\frac{1}{3!}\e^{ n_1\ldots  n_4  l_1\ldots  l_3  r_1\ldots  r_3} H_{ l_1\ldots  l_3} F_{ r_1\ldots  r_3}.
\ee

Finally the Einstein equations, after simplifying by substitution
of the Ricci scalar as given by the dilaton equation (\ref{phi}) are:
\be
\label{G}
R_{ m n} + 2\nabla_m\nabla_n\f = \frac14 H_{ m a b} {H_n}^{ a b} + \frac{e^{2\f}}{2} \left[ T_{ m n}^{(1)} + T_{m n}^{(\tilde 3)} + \frac12 T_{mn}^{(\tilde 5)} \right],
\ee
where
\begin{align}
T_{mn}^{(1)} &= \p_m C \p_n C - \frac12 g_{mn} (\p C)^2,\\
T_{mn}^{(\tilde 3)} &= \frac12 \tilde F_{mab}  {\tilde F_n}^{~\,ab} - \frac12 g_{mn} \frac{1}{3!}\tilde {F_{(3)}}^2 ,\\
T_{mn}^{(\tilde 5)} &= \frac{1}{4!} \tilde F_{ m a_1\ldots  a_4}  {\tilde F_n}^{~\,a_1\ldots a_4}
\end{align}
(the ${\tilde F_{(5)}}\!^2$ term in the 5-form energy-momentum is identically zero since $\tilde F_{(5)} = \star \tilde F_{(5)}$).

The supergravity field equations, which we have derived here, simplify
considerably in the case of zero $b$-field, as is relevent for D-brane
solutions. For the dilaton, RR scalar, $b$, $C_{(2)}$, $C_{(4)}$, and $g$ we have correspondingly
\begin{align}
\label{_phi} &R = 4(\p\f)^2 - 4\nabla^2 \f,\\
\label{_C_0} &\nabla^2 C_{(0)} = 0,\\
\label{_B_2} &\nabla_m\left(C_{(0)} F\right)^{ a b m} = -\frac12\frac{1}{3!} F^{ a b m n l} F_{ m n l}+ \frac{1}{2\sqrt{|g|}}\frac{1}{5!}\frac{1}{3!} \e^{ a b m_1\ldots  m_5 n_1\ldots  n_3} F_{ m_1\ldots  m_5} F_{ n_1\ldots  n_3},\\
\label{_C_2} &\nabla_m F^{ a b m} = 0,\\
\label{_C_4} &\nabla_m F^{ m n_1\ldots  n_4} = 0,\\
\label{_G} &R_{ m n} + 2\nabla_m\nabla_n\f = \frac{e^{2\f}}{2} \left[ T_{mn}^{(1)} + T_{mn}^{(3)} + \frac12 T_{mn}^{(5)} \right].
\end{align}

\chapter{Clifford algebra and supersymmetry}
\label{gamma}

\section{$d=10$ conventions}
\label{gamma10d}

We work with the real 32 by 32 representation for the gamma-matrices of $\mathbb{R}^{9,1}$, which exists due to the isomorphism $\Cl(9,1)~\cong~\mathrm{Mat}(\mathbb{R},32)$. It is convenient to exploit the periodicity property of the Clifford algebras
\be
\Cl(9,1) \cong \Cl(1,1) \otimes \Cl(8,0)
\ee
to construct the gamma-matrices as tensor products of $\{\s_1, i\s_2\}$, which are the gamma-matrices of $Cl(1,1)$ with the following symmetric $\{\S_1,\ldots,\S_8\}$, which are the gamma-matrices of $8$-dimensional Euclidean space:
\be
\begin{array}{ccccccc}
\S^1 = \sigma_2 & \otimes & \sigma_2 & \otimes & \sigma_2 & \otimes & \sigma_2, \\
\S^2 = \sigma_2 & \otimes & 1             & \otimes & \sigma_1 & \otimes & \sigma_2, \\
\S^3 = \sigma_2 & \otimes & 1             & \otimes & \sigma_3 & \otimes & \sigma_2, \\
\S^4 = \sigma_2 & \otimes & \sigma_1 & \otimes & \sigma_2 & \otimes & 1,\\
\S^5 = \sigma_2 & \otimes & \sigma_3 & \otimes & \sigma_2 & \otimes & 1,\\
\S^6 = \sigma_2 & \otimes & \sigma_2 & \otimes & 1             & \otimes & \sigma_1, \\
\S^7 = \sigma_2 & \otimes & \sigma_2 & \otimes & 1             & \otimes & \sigma_3, \\
\S^8 = \sigma_1 & \otimes & 1             & \otimes & 1              & \otimes & 1,
\end{array}
\ee
and $\S^9=\S^1\cdot\ldots\cdot\S^8 = \s_3 \otimes  1 \otimes 1 \otimes 1$, which is a chirality operator in 8D.
In particular, the representation we use is:
\be\label{gamma-m}
\begin{aligned}
\G^0 = i\sigma_2 \otimes \mathbb{1}_{16} & = \left( \begin{array}{cc}
                                                                                    0 & \mathbb{1}_{16}\\ -\mathbb{1}_{16} & 0
                                                                                    \end{array}\right), \qquad (\G^0)^2 = -1;\\
\G^i = \sigma_1 \otimes \S^i & = \left( \begin{array}{cc}
                                                                0 & \S^i \\ \S^i & 0
                                                            \end{array}\right), \qquad (\G^i)^2 = 1.
\end{aligned}
\ee

The 10-dimensional chirality operator is $\G^{10} = \G^0\cdot\ldots\cdot\G^9 = \s_3\otimes\mathbb{1}_{16}$. Spinors of definite chirality are defined as usual, $\G^{10}\psi^\pm = \pm\psi^\pm$; they provide two inequivalent real 16-dimensional representations of $\Spin(9,1)$, $S_+$ and $S_-$. These are Majorana-Weyl spinors; we can also define $S_+ \oplus S_-$, which is a Majorana spinor (real 32 component) and $S_+ \otimes \mathbb{C}$ ($S_- \otimes \mathbb{C}$), which are Weyl spinors (complex 16 component) of positive (negative) chirality.

The ``small gamma'' $\g^\m$ matrices, which are used e.g. in the section \ref{sect}, are defined as off-diagonal 16 by 16 blocks of the $\G^\m$ matrices:
\be
\begin{aligned}
\G^\m =
    \left(
        \begin{array}{cc}
            0 & {\g^\m}^{\a\b}\\
            \g^\m_{\a\b} & 0\\
        \end{array}
    \right),
\end{aligned}
\ee
so that they are analogs of Pauli matrices in 4D. One can read off their values from (\ref{gamma-m}).
The $\g^\m$ matrices are symmetric and they satisfy a condition
\be
\g^\m_{\a\b} {\g^\n}^{\b\g} + \g^\n_{\a\b} {\g^\m}^{\b\g} = 2 \eta^{\m\n} \d_\a^\g.
\ee
Position of the spinor indices reflects the convention to denote the positive chirality spinors with $\psi^\a$ and the negative chirality spinors with $\chi_\a$. For example, action of a gamma-matrix on a Majorana spinor is given by
\be
\G^\m \Psi =
    \left(
        \begin{array}{cc}
            0 & {\g^\m}^{\a\b}\\
            \g^\m_{\a\b} & 0\\
        \end{array}
    \right)
    \left(
        \begin{array}{c}
            \psi^\b \\
            \chi_\b\\
        \end{array}
    \right) =
\left(
        \begin{array}{c}
            (\g^\m \chi)^\a\\
            (\g^\m \psi)_\a\\
        \end{array}
    \right),
\ee
and action on chiral (Majorana-Weyl or Weyl) spinors can be written by setting $\psi$ or $\chi$ to zero.

Since charge conjugation matrix in this representation can be taken to be $C=\G^0$:
\be
C \G^i C^{-1} = -{\G^i}^T,
\ee
the Lorentz-covariant spinor bilinear takes the form (using Majorana conjugation $\overline\Psi = \Psi^T C$):
\be
\begin{aligned}
\overline\Psi \G^\m \Phi &=
    \left(
        \begin{array}{cc}
            \psi^\a &   \chi_\a\\
        \end{array}
    \right)
    \left(
        \begin{array}{cc}
            0 & {1_\a}^\b\\
            -{1^\a}_\b & 0\\
        \end{array}
    \right)
    \left(
        \begin{array}{cc}
            0 & {\g^\m}^{\b\g}\\
            \g^\m_{\b\g} & 0\\
        \end{array}
    \right)
    \left(
        \begin{array}{c}
            \phi^\g \\
            \varphi_\g\\
        \end{array}
    \right) =\\
&=\psi^\a \g^\m_{\a\b} \phi^\b - \chi_\a {\g^\m}^{\a\b} \varphi_\b.
\end{aligned}
\ee
For chiral spinors, such as the supersymmetry parameters of IIB supergravity, this bilinear reduces to $\psi^\a\g^\m_{\a\b} \phi^\b$ (in the case of positive chirality). This type of 16-component spinor bilinear is used, e.g. in the formula (\ref{BM:constr}).

Killing spinor equations result from requiring that the supersymmetry variations of the fermions vanish. The fermions in type IIB supergravity are the doublets of gravitini and dilatini, which have opposite chirality. We take the dilatini $\l,\hat\l$ to have negative chirality. The supersymmetry parameters $\e,\eh$ are of the same (positive) chirality as the gravitini $\psi_\m,\hat\psi_\m$. Supersymmetry variations in the two-component formalism are:
\bsub
\label{susy-vars}
\begin{align}
\d\psi_m &= \nabla_m\e - \frac14 \slashed{H}_m\e - \frac{e^\f}{8} \left( \slashed{F}_{(1)} + \slashed{F}_{(3)} + \frac12 \slashed{F}_{(5)} \right) \G_m \eh, \\
\d\hat\psi_m &= \nabla_m\eh + \frac14 \slashed{H}_m \eh + \frac{e^\f}{8} \left( \slashed{F}_{(1)} - \slashed{F}_{(3)} + \frac12 \slashed{F}_{(5)} \right) \G_m \e,
\end{align}
\begin{align}
\d\l &= \slashed{\p}\f\,\e -\frac12 \slashed{H} \e + \frac{e^\f}{2} \left( 2\slashed{F}_{(1)} + \slashed{F}_{(3)} \right) \eh,\\
\d\hat\l &= \slashed{\p}\f\,\eh +\frac12 \slashed{H} \eh - \frac{e^\f}{2} \left( 2\slashed{F}_{(1)} - \slashed{F}_{(3)} \right) \e,
\end{align}
\esub
where
\begin{align}
\slashed{F}_{(n)} &= \frac{1}{n!} F_{m_1\ldots m_n} \G^{m_1\ldots m_n},\\
\slashed{H}_m &= \frac12 H_{mnr} \G^{nr}.
\end{align}
Sometimes it is more convenient to derive and solve the Killing spinor equations in terms of the single complex gravitino, dilatino and supersymmetry parameter, defined as
\be
\Psi_m = \psi_m + i \hat\psi_m, \quad\Lambda = \l + i \hat\l, \quad\varepsilon = \e + i\eh.
\ee
The above transformations can be rewritten in the complex notation as
\be
\d\Psi_m = \nabla_m \varepsilon -\frac14 \slashed{H}_m \varepsilon^* + \frac{i e^\f}{8} \left(  \slashed{F}_{(1)} + \frac12 \slashed{F}_{(5)} \right) \G_m \varepsilon - \frac{i e^\f}{8} \slashed{F}_{(3)} \G_m\varepsilon^*,
\ee
\be
\d\Lambda =  \slashed{\p}\f\,\varepsilon -\frac12 \slashed{H} \varepsilon^* -i e^\f \slashed{F}_{(1)} \varepsilon + \frac{i e^\f}{2} \slashed{F}_{(3)} \varepsilon^*.
\ee

%\chapter{Mathematica algorithm}
%\label{app:mathematica}

\section{Gamma-matrices for $\cp3$}
\label{app-gamma}
For the purposes of working with type IIA supergravity, whose spinorial quantities are Majorana spinors of $\mathbb{R}^{9,1}$, we need a Majorana representation of the gamma-matrices. We can construct this as a product of Majorana representations in $1+3$ and in $6$ dimensions. This representation is used in chapter \ref{ch:paper2}

Our spacetime signature convention is $(- + \ldots +)$, hence the following four real anticommuting matrices $\a^\m$ furnish a Majorana representation in $D=1+3$:
\be
\begin{array}{ccccc}
\a^0 & = & \s_3 & \otimes & i\s_2,\\
\a^1 & = & \s_3 & \otimes & \s_1,\\
\a^2 & = & \s_3 & \otimes & \s_3,\\
\a^3 & = & \s_1 & \otimes & 1.
\end{array}
\ee
Volume element $\a^5 = \a^0 \ldots \a^3 = i\s_2 \otimes 1$ is also real and squares to $-1$.

We choose the six gamma-matrices of $6D$ Euclidean space to be
\be
\begin{array}{ccccccc}
\b^1 & = & 1    & \otimes & \s_2 & \otimes & \s_1,\\
\b^2 & = & 1    & \otimes & \s_2 & \otimes & \s_3,\\
\b^3 & = & \s_1 & \otimes & 1    & \otimes & \s_2,\\
\b^4 & = & \s_3 & \otimes & 1    & \otimes & \s_2,\\
\b^5 & = & \s_2 & \otimes & \s_1 & \otimes & 1,   \\
\b^6 & = & \s_2 & \otimes & \s_3 & \otimes & 1.
\end{array}
\ee
These are imaginary and we define the corresponding volume element to be real: $\b^7 = -\b^1\ldots\b^6 = i\s_2\otimes i\s_2\otimes i\s_2$.

Finally, the ten-dimensional real gamma-matrices $\G$ are the following products (for $\a = 0,\ldots,3$ and $i = 1,\ldots,6$):
\be
\begin{array}{ccccc}
\G^\m & = & \a^\m & \otimes & 1,\\
\G^{i+3}  & = & i\a^5 & \otimes & \b^i
\end{array}
\ee
Ten-dimensional chirality operator is $\G^{11} = \G^0 \ldots \G^9 = -\a^5 \otimes \b^7$. This representation is clearly not Weyl.

\chapter{$\cp3$ Killing spinors}
\label{app-spinors}

The components of the $\cp3$ factor (\ref{eta}) of the Killing spinor are given by the following:
\be\bn
f_1 = \frac{1}{2} &\left\{2 h_1 \cos\a \sin\frac{\c}{4} + 
	2 h_2 \cos\a \cos\frac{\c}{4} +\right.
	\\   
   &\sin\a \left[\left(h_3 \sin\frac{\f}{2}+h_4
   \cos\frac{\f}{2}\right) \sin\frac{1}{4} (2 \th +\c -2 \ps)+\left(h_3 \cos\frac{\f}{2}
   -h_4 \sin\frac{\f}{2}\right) \right.
   \\
   &\cos\frac{1}{4} (2 \th -\c +2 \ps)-\left(h_6
   \cos\frac{\f}{2}-h_5 \sin\frac{\f}{2}\right) \cos
   \frac{1}{4} (2 \th +\c -2 \ps)
   \\
   &+\left.\left.\left(h_6 \sin\frac{\f}{2}+h_5 \cos\frac{\f}{2}\right) 
   \sin\frac{1}{4} (2 \th -\c +2 \ps)\right]\right\},
\en\ee
\be\bn
f_4 = \frac{1}{2} &\left\{2 h_1 \cos\a \cos\frac{\c}{4} - 2 h_2 \cos\a \sin\frac{\c}{4} +\right.
	\\   
   &\sin\a \left[\left(h_3 \sin\frac{\f}{2}+h_4
   \cos\frac{\f}{2}\right) \cos\frac{1}{4} (2 \th +\c -2 \ps)+\left(h_3 \cos\frac{\f}{2}
   -h_4 \sin\frac{\f}{2}\right) \right.
   \\
   &\sin\frac{1}{4} (2 \th -\c +2 \ps)+\left(h_6
   \cos\frac{\f}{2}-h_5 \sin\frac{\f}{2}\right) \sin
   \frac{1}{4} (2 \th +\c -2 \ps)
   \\
   &-\left.\left.\left(h_6 \sin\frac{\f}{2}+h_5 \cos\frac{\f}{2}\right) 
   \cos\frac{1}{4} (2 \th -\c +2 \ps)\right]\right\},
\en\ee
\be\bn
f_2 = \frac{1}{2} &\Big\{\mathcal{A}_1\big[(\cos\a +1) \sin\m-(\cos\a -1) \cos\m\big] + \mathcal{B}_1\big[(\cos\a +1) \cos\m -(\cos\a -1) \sin\m\big]
   \\
   &-2
   \sin\a (\cos\m -\sin\m) \left(h_1 \cos\frac{\c}{4}-h_2
   \sin\frac{\c}{4}\right)\Big\},
\en\ee
\be\bn
f_3 = \frac{1}{2} &\Big\{\mathcal{A}_1\big[(\cos\a -1) \sin\m+(\cos\a +1) \cos\m\big] + \mathcal{B}_1\big[(\cos\a -1) \cos\m +(\cos\a +1) \sin\m\big]
   \\
   &+2
   \sin\a (\cos\m +\sin\m) \left(h_1 \cos\frac{\c}{4}-h_2
   \sin\frac{\c}{4}\right)\Big\},
\en\ee
\be\bn
f_5 = \frac{1}{2} &\Big\{\mathcal{A}_2\big[(\cos\a +1) \sin\m-(\cos\a -1) \cos\m\big]
	+
   \mathcal{B}_2\big[(\cos\a +1) \cos\m -(\cos\a -1) \sin\m\big]
   \\
   &-2
   \sin\a (\cos\m -\sin\m) \left(h_1 \cos\frac{\c}{4}-h_2
   \sin\frac{\c}{4}\right)\Big\},
\en\ee
\be\bn
f_6 = \frac{1}{2} &\Big\{\mathcal{A}_2\big[(\cos\a -1) \sin\m+(\cos\a +1) \cos\m\big] + \mathcal{B}_2\big[(\cos\a -1) \cos\m +(\cos\a +1) \sin\m\big]
   \\
   &+2
   \sin\a (\cos\m +\sin\m) \left(h_1 \cos\frac{\c}{4}-h_2
   \sin\frac{\c}{4}\right)\Big\},
\en\ee
where
\be\bn
\mathcal{A}_1 &=\left[\sin
   \frac{\ps}{2} \sin\frac{1}{4} (2 \th -\c) \left(h_3 \cos
   \frac{\f}{2}-h_4 \sin\frac{\f}{2}\right)-\sin
   \frac{\ps}{2} \cos\frac{1}{4} (2 \th -\c) \right.
\\   
   &\left(h_6 \sin
   \frac{\f}{2}+h_5 \cos\frac{\f}{2}\right)+\cos
   \frac{\ps}{2} \left(\cos\frac{1}{4} (2 \th +\c)
   \left(h_6 \cos\frac{\f}{2}-h_5 \sin\frac{\f}{2}\right)\right.
   \\
   &-\left.\left.\sin\frac{1}{4} (2 \th +\c) \left(h_3 \sin
   \frac{\f}{2}+h_4 \cos\frac{\f}{2}\right)\right)\right],\\
\mathcal{B}_1 &=  \left(\cos\frac{\ps}{2} \cos\frac{1}{4} (2 \th -\c
   ) \left(h_3 \cos\frac{\f}{2}-h_4 \sin\frac{\f
   }{2}\right)+\cos\frac{\ps}{2} \sin\frac{1}{4} (2 \th -\c) \right.
	\\   
   &\left(h_6 \sin\frac{\f}{2}+h_5 \cos\frac{\f
   }{2}\right)-\sin\frac{\ps}{2} \left(\cos\frac{1}{4} (2 \th
   +\c) \left(h_3 \sin\frac{\f}{2}+h_4 \cos\frac{\f
   }{2}\right)\right.
\\   
   &+\left.\left.\sin\frac{1}{4} (2 \th +\c) \left(h_6 \cos
   \frac{\f}{2}-h_5 \sin\frac{\f}{2}\right)\right)\right),
\en\ee
and $\mathcal{A}_2, \mathcal{B}_2$ are the same with the following substitution: 
\be\bn
\sin\frac{\ps}{2} &\rightarrow -\cos\frac{\ps}{2},\\
\cos\frac{\ps}{2} &\rightarrow \sin\frac{\ps}{2}.
\en\ee

\chapter{D-brane action formalisms}
\label{actions}

%\section{F-string and D-string actions}

The prototype of generalized geometry first appeared in~\cite{Duff:1990hn}, where a fundamental $p$-brane theory has been considered and a generalized metric on the space $$TM \oplus \L^p T^*M$$ has been constructed. Embedding of a $p$-brane with worldvolume coordinates $\x^i, i=0,\ldots,p$ into the spacetime is given by the functions $x^\m(\x), \m=0,\ldots,d$. Using the independent worldvolume metric $\g_{ij}$ we can write down the Howe-Tucker form of the action as given in~\cite{Duff:1990hn}:
\be\label{p-brane}
S_p = -\frac{T_p}{2}\int d^{p+1}\x \sqrt{-\g}\, \g^{ij} \p_i x^\m \p_j x^\n \,g_{\m\n} - \frac{T_p}{2} (1-p)\int d^{p+1}\x \sqrt{-\g} - T_p\int C_{(p+1)}, 
\ee
where $T_p$ is the $p$-brane tension and the last term takes care of the $p$-brane electric coupling to the spacetime antisymmetric tensor field $C_{\m\m_1\ldots \m_p}$:
\be
\int C_{(p+1)} = \int d^{p+1}\x\frac{1}{(p+1)!}\,\ \e^{i i_1 \ldots i_p} \p_{i} x^\m \p_{i_1} x^{\m_1}\ldots \p_{i_p} x^{\m_p} \,C_{\m\m_1\ldots \m_p}.
\ee
This action is suitable for the fundamental objects of superstring theory and M-theory. When $p=2$ (\ref{p-brane}) is precisely the worldvolume action of the M-theory M2-brane, and for $p=1$ we get a fundametal string (F1-brane) action. Both these objects have no worldvolume fields apart from $\g_{ij}$ and $x^\m$.

The first two terms of the action (\ref{p-brane}) are equivalent to a Nambu-Goto action. Namely, if one varies (\ref{p-brane}) with respect to the worldvolume metric and then substitutes the equation of motion back into the action, one gets the new action with a Nambu-Goto kinetic term:
\be
S_p \left[ \g_{ij}\rightarrow \p_i x^\m \p_j x^\n \,g_{\m\n} \right] = -T_p \int d^{p+1}\x \sqrt{-\det \left( \p_i x^\m \p_j x^\n \,g_{\m\n} \right)} - T_p\int C_{(p+1)}.
\ee
As a result, one can write the F1 action (\ref{p-brane}) as either
\be\label{f1-pol}
S_{F1} = -\frac{T_1}{2}\int d^2\x \sqrt{-\g}\, \g^{ij} \p_i x^\m \p_j x^\n \,g_{\m\n} - T_1\int b_2
\ee
or
\be\label{f1-ng}
S_{F1} = -T_1 \int d^2\x \sqrt{- \det \left(\p_i x^\m \p_j x^\n \,g_{\m\n}\right)} - T_1\int b_2.
\ee

In order to make this action suitable for the description of other extended objects, be it the M5-brane, or the D-branes, we need to include a coupling to the corresponding extra worldvolume fields. In particular, for the D$p$-brane we need to introduce a worldvolume gauge field, and the corresponding dynamics is now given by the DBI action
\be\label{dbi}
S_{DBI} = -T_{p}\int d^{p+1}\x \,e^{-\f} \sqrt{-\det \left( g_{ij}+\F_{ij} \right)} - T_p \int e^\F \C,
\ee
where $g_{ij}$ is the pullback to the worldvolume of the spacetime metric $g_{\m\n}$, $$\F_{ij} = \p_i x^\m \p_j x^\n \,b_{\m\n} + 2\pi\a'(dA)_{ij}$$ is the gauge invariant worldvolume field strength 2-form, and $\C$ is a formal sum of all the RR fields in the theory. In the Wess-Zumino term $\int e^\F \C$ it is assumed that only one RR field of appropriate rank is multiplying each term in the power series expansion of the exponential. For example, for the D1-brane of type IIB superstring theory $\C = C_{(0)} \oplus C_{(2)} \oplus \ldots$, and
\be\label{dbi-d1}
S_{D1} = -T_1\int d^2\x \,e^{-\f} \sqrt{-\det \left( g+\F \right)} - T_1 \int (C_{(2)} + \F C_{(0)}).
\ee

The kinetic term of the DBI action can also be recast in the Howe-Tucker form. We can use 
\be\label{1}
S = -\frac{T_p}{2} \int d^{p+1}\x\, e^{-\f} \sqrt{-H}\, H^{ij} \left(g_{ij} + \F_{ij}\right) - \frac{T_p}{2} (1-p) \int d^{p+1}\x\, e^{-\f} \sqrt{-H},
\ee
where $H_{ij} = \g_{ij} + a_{ij}$ is now a generic matrix ($H = \det H_{ij}$), so that for its inverse $H^{ij}$ the symmetric part $\g^{ij}$ can be interpreted as an auxilliary worldvolume metric as before, whereas the antisymmetric part $a^{ij}$ is a worldvolume antisymmetric tensor that contracts $\F_{ij}$:
\be
H^{ij} \left(g_{ij} + \F_{ij}\right) = \g^{ij} g_{ij} + a^{ij} \F_{ij}.
\ee
Note that $H^{ij}$ is defined to be the inverse of $H_{ij}$, which means that its symmetric and antisymmetric parts $\g^{ij}$ and $a^{ij}$ are not necessarily the inverses of $\g_{ij}$ and $a_{ij}$, respectively.

The equation of motion for $H_{ij}$ that follows from the above action is
\be
g_{ij} + \F_{ij} - \frac12 H_{ij} H^{kl} \left( g_{kl} + \F_{kl} \right) = \frac{1-p}{2} H_{ij}.
\ee
This has a solution $H_{ij} = g_{ij} + \F_{ij}$, which upon substitution into (\ref{1}) transforms it into the kinetic term of the DBI action (\ref{dbi}):
\be\bn
S\left[H_{ij}\rightarrow g_{ij} + \F_{ij}\right] &=\\= -\frac{T_p}{2} \int d^{p+1}\x\, &e^{-\f} \sqrt{-H} (p+1)- \frac{T_p}{2} \int d^{p+1}\x\, e^{-\f} \sqrt{-H} (1-p) =\\= -T_p \int d^{p+1}\x\, &e^{-\f} \sqrt{-\det \left( g_{ij} + \F_{ij}\right)}.
\en\ee

Therefore a D1-string also admits two alternative formulations:
\be\label{d1-pol}
S_{D1} =  -\frac{T_1}{2} \int d^2\x\, e^{-\f} \sqrt{-H}\, H^{ij} \left(g_{ij} + \F_{ij}\right) - T_1 \int (C_{(2)} + \F C_{(0)}).
\ee
or
\be\label{d1-dbi}
S_{D1} = -T_1\int d^2\x \,e^{-\f} \sqrt{-\det \left( g+\F \right)} - T_1 \int (C_{(2)} + \F C_{(0)}).
\ee

\bibliographystyle{utphys}
\bibliography{bib}
%\end{singlespace}

\end{document}